\documentclass[aps,twocolumn,prd,showpacs,showkeys,preprintnumbers,superscriptaddress,bibnotes,floatfix,longbibliography,notitlepage,nofootinbib,groupedaddress]{revtex4-2}

\pdfoutput=1
\usepackage{amsmath}
\usepackage{amsfonts}
\usepackage{amssymb}
\usepackage{mathrsfs}
\usepackage{graphicx}
\usepackage{color}
\usepackage{bbding}
\usepackage{tabularx}
\usepackage[usenames,dvipsnames,table]{xcolor}
\usepackage[normalem]{ulem}
\usepackage{makecell}
\usepackage{siunitx}
\usepackage{enumerate}
\usepackage{multirow}
\usepackage{makecell}
\usepackage{upgreek}
\usepackage{blindtext}

\usepackage{marvosym}
\definecolor{darkred}{rgb}{0.5,0,0}
\definecolor{darkblue}{rgb}{0,0,0.5}
\definecolor{firebrick}{rgb}{0.75,0.125,0.125}
\definecolor{darkgreen}{rgb}{0,0.5,0}
\usepackage[colorlinks=true,linkcolor=firebrick,citecolor=darkgreen,urlcolor=darkblue]{hyperref} 
\usepackage[capitalise]{cleveref}

\newcommand{\ie}{{\it i.e.}}

\newcommand{\eg}{{\it e.g.}}

\newcommand{\eq}{Eq.}

\newcommand{\fig}{Fig.}

\newcommand{\Refe}{Ref.}
\newcommand{\Refes}{Refs.}

\newcolumntype{L}{>{\centering\arraybackslash}m{5cm}}
\newcommand{\equ}[1]{\eq~(\ref{equ:#1})}
\newcommand{\figu}[1]{\fig~\ref{fig:#1}}
\newcommand{\orcid}[1]{\href{https://orcid.org/#1}{\includegraphics[width=10pt]{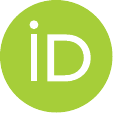}}}

\graphicspath{{figures/}{figures_appendix/}}

\begin{document}

\title{Near-future discovery of the diffuse flux of ultra-high-energy cosmic neutrinos}

\author{V\'ictor B.~Valera
\orcid{0000-0002-0532-5766}}
\email{vvalera@nbi.ku.dk}
\affiliation{Niels Bohr International Academy, Niels Bohr Institute,\\University of Copenhagen, DK-2100 Copenhagen, Denmark}

\author{Mauricio Bustamante
\orcid{0000-0001-6923-0865}}
\email{mbustamante@nbi.ku.dk}
\affiliation{Niels Bohr International Academy, Niels Bohr Institute,\\University of Copenhagen, DK-2100 Copenhagen, Denmark}

\author{Christian Glaser
\orcid{0000-0001-5998-2553}}
\email{christian.glaser@physics.uu.se}
\affiliation{Department of Physics and Astronomy, Uppsala University, Uppsala, SE-752 37, Sweden}

\date{\today}

\begin{abstract}
Ultra-high-energy (UHE) neutrinos, with EeV-scale energies, carry with them unique insight into fundamental open questions in astrophysics and particle physics.  For fifty years, they have evaded discovery, but maybe not for much longer, thanks to new UHE neutrino telescopes, presently under development.  We capitalize on this upcoming opportunity by producing state-of-the-art forecasts of the discovery of a diffuse flux of UHE neutrinos in the next 10--20 years.  By design, our forecasts are anchored in often-overlooked nuance from theory and experiment; we gear them to the radio array of the planned IceCube-Gen2 detector.  We find encouraging prospects: even under conservative analysis choices, most benchmark UHE neutrino flux models from the literature may be discovered within 10~years of detector exposure---many sooner---and may be distinguished from each other.  Our results validate the transformative potential of next-generation UHE neutrino telescopes.
\end{abstract}

\maketitle


\section{Introduction}
\label{section:introduction}

Ultra-high-energy (UHE) neutrinos, with energies in the EeV scale (1~EeV $\equiv 10^{18}$~eV), were first predicted in the late 1960s~\cite{Berezinsky:1969erk}, as a natural consequence~\cite{Greisen:1966jv, Zatsepin:1966jv} of the interaction of UHE cosmic rays (UHECRs), with comparable energies, and cosmological photon fields, like the cosmic microwave background.  They are the most energetic neutrinos expected to be produced from standard particle processes, at least 10--100 times more energetic than the TeV--PeV neutrinos discovered by the IceCube neutrino telescope~\cite{IceCube:2013cdw, IceCube:2013low, IceCube:2014stg, IceCube:2015qii, IceCube:2016umi,  IceCube:2020wum, IceCube:2021uhz} (there may be higher-energy neutrinos made in exotic processes~\cite{Berezinsky:2011cp, Ryabov:2016fac, Anchordoqui:2018qom, Creque-Sarbinowski:2022mex}, but we do not consider them).  UHE neutrinos provide unique insight into long-standing open problems in astrophysics---what are the most energetic astrophysical sources in the Universe---and particle physics---how do neutrinos, in particular, and fundamental physics, in general, behave at the highest energies~\cite{Ahlers:2018fkn, Ahlers:2018mkf, Ackermann:2019cxh, Ackermann:2019ows,  AlvesBatista:2019tlv, Arguelles:2019rbn, AlvesBatista:2021gzc, Abraham:2022jse, Ackermann:2022rqc, Adhikari:2022sve}.  Yet, despite efforts, they remain undiscovered; however, maybe not for much longer.

Over the last fifty years, UHE neutrinos have received considerable attention from experiment and theory.  Progress, while steady, has been challenging: past and present experiments have placed upper limits on their flux~\cite{IceCube:2018fhm, ANITA:2019wyx, ARIANNA:2019scz, PierreAuger:2019ens, ARA:2019wcf}, but even the tightest present-day limits~\cite{IceCube:2018fhm, PierreAuger:2019ens} leave vast swathes of the space of UHE neutrino flux models unconstrained; see \figu{benchmark_spectra}.  On the experimental front, the main challenge is that the flux of UHE neutrinos is possibly tiny~\cite{Aloisio:2009sj, Ahlers:2012rz}.  This motivates the need to build larger neutrino telescopes and explore more suitable detection strategies~\cite{Ackermann:2022rqc}.  On the theory front, the main challenge is that predictions of the UHE neutrino flux are uncertain because they depend on properties of UHECRs and their sources, which are known only uncertainly, such as the evolution with redshift of the source number density, the UHECR mass composition, the UHECR acceleration mechanism, including the maximum cosmic-ray energies achievable, and the neutrino production mechanism, among others; for details, see, \eg, \Refes~\cite{Kotera:2010yn, Heinze:2015hhp, Romero-Wolf:2017xqe, AlvesBatista:2018zui, Heinze:2019jou}.  This motivates the need to consider a host of competing, representative flux predictions~\cite{Fang:2013vla, Padovani:2015mba, Fang:2017zjf, Heinze:2019jou, Muzio:2019leu, Rodrigues:2020pli, Anker:2020lre, Muzio:2021zud}.  We tackle both challenges.

\begin{figure}[t!]
 \centering
 \includegraphics[width=\columnwidth]{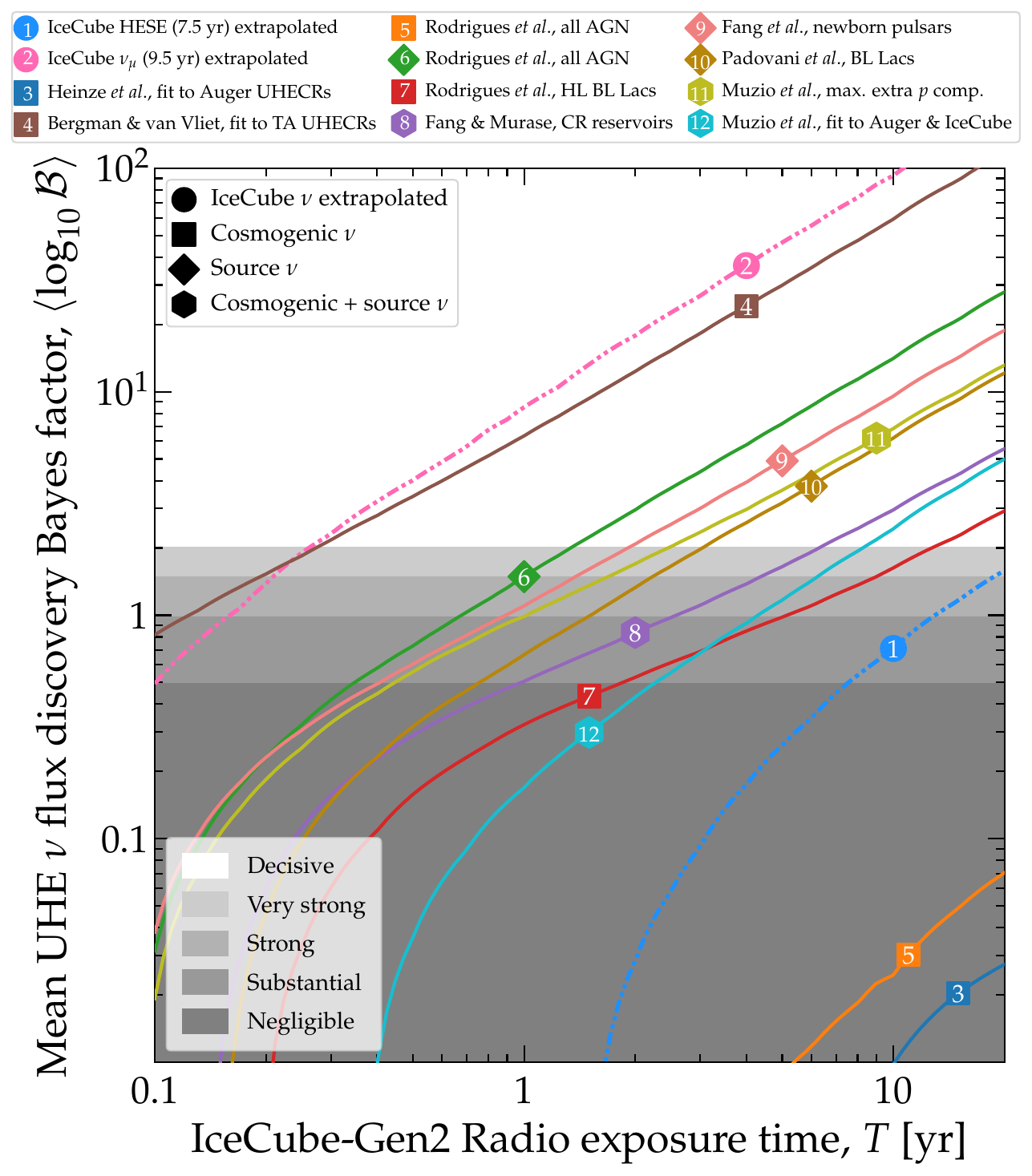}
 \caption{\label{fig:bayes_factor_hard}Discovery potential of benchmark diffuse ultra-high-energy (UHE) neutrino flux models 1--12~\cite{Fang:2013vla, Padovani:2015mba, Fang:2017zjf, Heinze:2019jou, Muzio:2019leu, Rodrigues:2020pli, Anker:2020lre, IceCube:2020wum,  Muzio:2021zud, IceCube:2021uhz} (\figu{benchmark_spectra}) in the radio array of IceCube-Gen2.  The background to discovery consists of atmospheric muons~\cite{Garcia-Fernandez:2020dhb, Hallmann:2021kqk}, for all models, plus the tentative UHE tail of the IceCube 9.5-year through-going $\nu_\mu$ flux~\cite{IceCube:2021uhz}, for models 3--12; see Section~\ref{sec:background}.  All analysis choices are baseline and conservative; see Table~\ref{tab:analysis_choices_base} and Section~\ref{sec:discovery_results_nominal_choices}.  {\it Decisive discovery may be achievable for most flux models after only a handful of years.}  See the main text, especially Sections~\ref{sec:discovery_stat_analysis} and \ref{sec:discovery_results_nominal_results}, for details.}
\end{figure}

Upcoming UHE neutrino telescopes, presently in different stages of planning, design, and prototyping, and built around different detection strategies, will have a real chance of discovering UHE neutrinos in the next 10--20 years, {\it even if their flux is low}~\cite{AlvesBatista:2021gzc, Ackermann:2022rqc, Adhikari:2022sve}.  We carve out this opportunity by providing the most detailed forecasts, to our knowledge, of the prospects of discovering a diffuse flux of UHE neutrinos.  Our results, even under conservative analysis choices, are encouraging.

To include realistic experimental nuance, we gear our forecasts to the radio array of IceCube-Gen2~\cite{IceCube-Gen2:2020qha} (``IceCube-Gen2 Radio" in our figures), the planned high-energy upgrade of IceCube, whose target UHE neutrino flux sensitivity is among the best~\cite{Ackermann:2022rqc}.  The array will instrument Antarctic ice with radio antennas that look for radio signals emitted by showers induced by UHE neutrinos~\cite{Askaryan:1961pfb, Zas:1991jv, Schroder:2016hrv}, a technique tested by ARA~\cite{ARA:2019wcf} and ARIANNA~\cite{ARIANNA:2019scz}  (and by ANITA~\cite{ANITA:2019wyx}, from the air).  RNO-G~\cite{RNO-G:2020rmc}, currently under deployment, will serve as a pathfinder for the radio array of IceCube-Gen2.  To make our forecasts comprehensive, we consider a large number of benchmark UHE neutrino flux models that span the full allowed space of models, in size and shape, from optimistic to pessimistic~\cite{Fang:2013vla, Padovani:2015mba, Fang:2017zjf, Heinze:2019jou, Muzio:2019leu, Rodrigues:2020pli, Anker:2020lre, IceCube:2020wum,  Muzio:2021zud, IceCube:2021uhz}.  

To produce our forecasts, we adopt the same flow of calculations as \Refe~\cite{Valera:2022ylt}.  For each UHE neutrino flux model, we propagate it through the Earth, computing neutrino interactions with matter along the way, and model its detection in the radio array of IceCube-Gen2.  We use the same state-of-the-art ingredients at every stage of the calculation as \Refe~\cite{Valera:2022ylt}: in the choice of diffuse UHE neutrino flux models (Section~\ref{sec:fluxes}), the UHE neutrino-nucleon cross section, the propagation of neutrinos through the Earth (Section~\ref{sec:propagation}), the neutrino detection, including the emission, propagation, and detection of radio signals in ice, and the neutrino and non-neutrino backgrounds (Section~\ref{sec:event_rates}).  See Section II of \Refe~\cite{Valera:2022ylt} for an overview.  Further, our forecasts account for random statistical fluctuations in the predicted event rates (Sections~\ref{sec:discovery_potential} and \ref{sec:model_separation}).  Below, we expand on all of the above.  

Figure~\ref{fig:bayes_factor_hard} shows our main results: the discovery prospects of the benchmark UHE neutrino flux models, computed under our baseline analysis choices, chosen to be largely conservative.  Because our statistical analysis is Bayesian, we report the flux discovery potential---and, later, the potential to tell apart  different flux models---via Bayes factors.  Figure~\ref{fig:bayes_factor_hard} reveals encouraging prospects: {\it conservatively, most benchmark UHE neutrino flux models may be discovered after only a handful of years}.  Later, we show that less conservative analysis choices, still well-motivated, lead to even better prospects. 

The overarching goal of our detailed forecasts is to help map the potential science reach that upcoming UHE neutrino telescopes will usher in in the next 10--20 years.  We make our forecasts realistic by factoring in nuance, from experiment and theory, that is often considered only partially, or not at all. We present our methods in considerable detail so that they can be readily adapted to produce forecasts for other upcoming UHE neutrino telescopes.  We hope that they help to assess and compare the complementary capabilities of competing designs.

This paper is organized as follows.  Section~\ref{sec:fluxes} presents the benchmark diffuse UHE neutrino flux models that we use in our forecasts.  Section~\ref{sec:propagation} sketches the effects of neutrino propagation inside Earth on them.  Section~\ref{sec:event_rates} introduces the method that we use to compute neutrino-induced event rates and the backgrounds that we consider.  Section~\ref{sec:discovery_potential} contains forecasts of the discovery potential of the benchmark flux models.  Section~\ref{sec:model_separation} contains forecasts of the separation between different benchmark flux models.  Section~\ref{sec:limits_improvements} outlines possible directions for future work.  Section~\ref{sec:summary} summarizes and concludes.


\section{Ultra-high-energy neutrinos}
\label{sec:fluxes}

\begin{figure*}[t!]
 \centering
 \includegraphics[width=\textwidth]{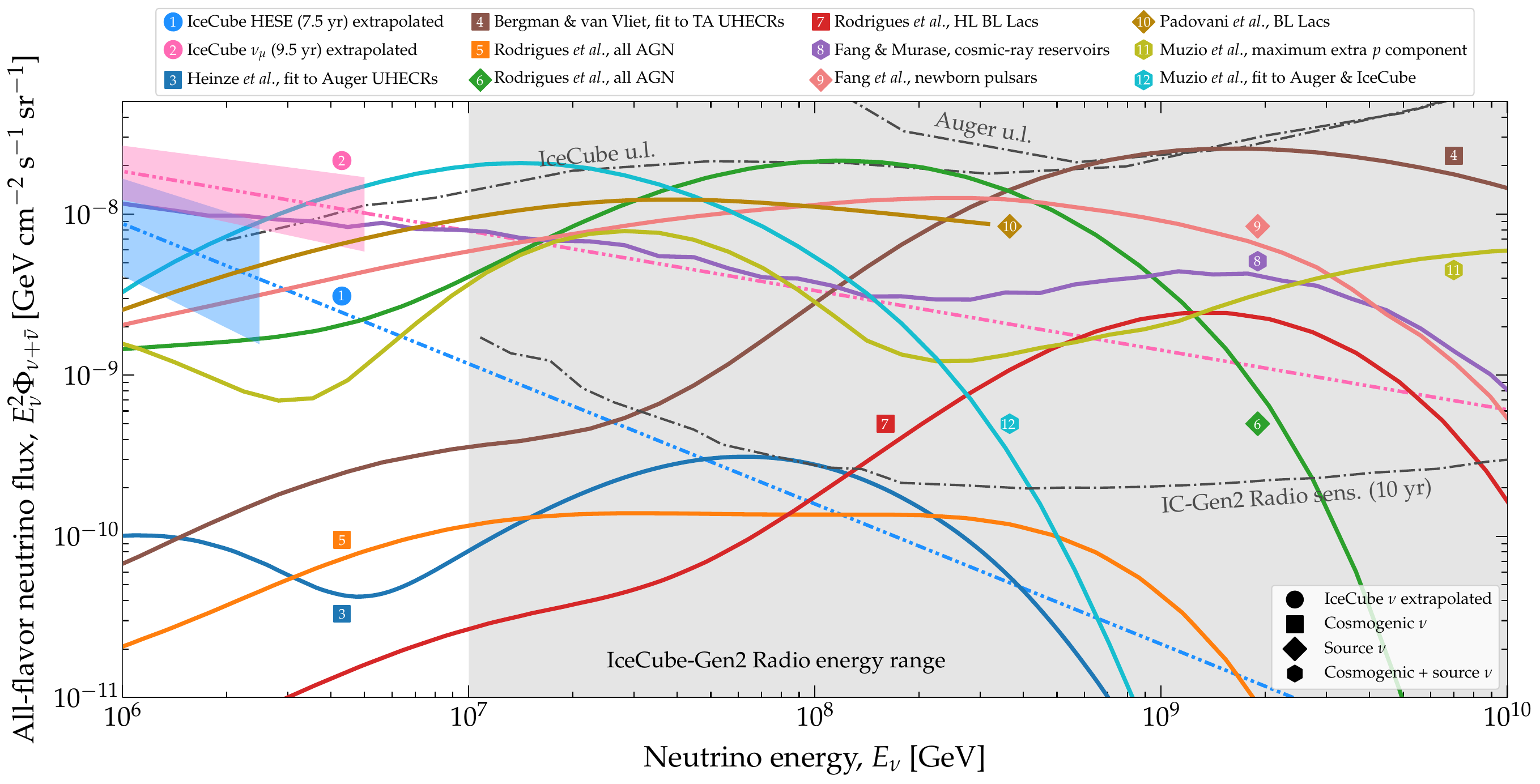}
 \caption{\label{fig:benchmark_spectra} Benchmark diffuse ultra-high-energy neutrino flux models~\cite{Fang:2013vla, Padovani:2015mba, Fang:2017zjf, Heinze:2019jou, Muzio:2019leu, Rodrigues:2020pli, Anker:2020lre, IceCube:2020wum,  Muzio:2021zud, IceCube:2021uhz} used here to assess the flux discovery capabilities of the radio array of IceCube-Gen2~\cite{IceCube-Gen2:2020qha} (``IceCube-Gen2 Radio").  These flux models are representative of the breadth of theoretical predictions in the literature.  The upper limits on the flux are from IceCube~\cite{IceCube:2018fhm} and the Pierre Auger Observatory~\cite{PierreAuger:2019ens}.  The shaded region indicates the approximate neutrino energy range to which the radio array of IceCube-Gen2 will be sensitive.   In this figure, fluxes are all-flavor, \ie, summed over all neutrino flavors, but our analysis treats individually the flux of each neutrino species, $\nu_e$, $\nu_\mu$, $\nu_\tau$, $\bar{\nu}_e$, $\bar{\nu}_\mu$, and $\bar{\nu}_\tau$.   See Fig.~6 in \Refe~\cite{Valera:2022ylt} for a breakdown of the flux of each neutrino species for each flux model.  See Section~\ref{sec:fluxes} for details.}
\end{figure*}

Ultra-high-energy neutrinos~\cite{Greisen:1966jv}, with energies above 100~PeV, are expected to be produced in the interaction of UHECRs~\cite{Zatsepin:1966jv, Berezinsky:1969erk}, with energies up to $10^{12}$~GeV, with matter and radiation, inside the UHECR sources ({\it source} neutrinos), outside them and en route to Earth ({\it cosmogenic} neutrinos), or both.   See \Refe~\cite{Ackermann:2022rqc} for a review.  

The interaction of UHECR protons on matter ($pp$) and radiation ($p\gamma$) produces a short-lived $\Delta(1232)$ resonance that decays into charged pions.  Upon decaying, they produce high-energy neutrinos, via $\pi^+ \to \mu^+ + \nu_\mu$, followed by $\mu^+ \to e^+ + \nu_e + \bar{\nu}_\mu$, and their charge-conjugated processes.  Each final-state neutrino carries, on average, 5\% of the energy of the parent proton.  En route to Earth, neutrino oscillations change the flavor composition of the flux, \ie, the relative content of $\nu_e$, $\nu_\mu$, and $\nu_\tau$ in it.  (Our benchmark UHE neutrino flux models below account for this change; more on this later.)

In realistic neutrino production models, including in some of our benchmark UHE neutrino flux models below, different production channels become accessible or dominant at different energies.  In $p\gamma$ interactions, neutrino production occurs via resonances heavier than $\Delta(1232)$ at intermediate energies, and via multi-pion production at high energies ~\cite{Mucke:1999yb,Hummer:2010vx, Morejon:2019pfu}.  In $pp$ interactions, the pion multiplicity changes with energy and affects the neutrino yield~\cite{Kelner:2006tc}.  The physical conditions inside the sources may affect neutrino production, too.  For instance, neutrino energies might be damped by strong magnetic fields that cool intermediate charged particles---protons, pions, muons---via synchrotron radiation~\cite{Waxman:1997ti, Waxman:1998yy, Winter:2013cla, Bustamante:2020bxp}, or by UHECR interactions in dense source environments~\cite{Boncioli:2016lkt, Biehl:2017zlw}.

For UHE neutrinos produced in $pp$ interactions, their energy spectrum is a power law that follows the power-law spectrum of the parent protons, and that may extend to low neutrino energies~\cite{Fang:2017zjf}.   For UHE neutrinos produced in $p\gamma$ interactions, their energy spectrum is determined by the spectra of the parent protons and photons.  Because the photon spectrum is typically peaked around a characteristic energy, the resulting neutrino energy spectrum is also peaked, at an energy set by the energy requirements to produce a $\Delta$ resonance.

Figure~\ref{fig:benchmark_spectra} shows the energy spectra of the benchmark UHE neutrino flux models 1--12~\cite{Fang:2013vla, Padovani:2015mba, Fang:2017zjf, Heinze:2019jou, Muzio:2019leu, Rodrigues:2020pli, Anker:2020lre, IceCube:2020wum,  Muzio:2021zud, IceCube:2021uhz} that we use in our forecasts below.  They span predictions from optimistic to pessimistic.  The wide variety in their size and shape is indicative of the present-day spread of the flux predictions available in the literature, and reflects large extant uncertainties in the properties of UHECRs and of their sources~\cite{Anchordoqui:2018qom, AlvesBatista:2019tlv}.   The benchmark flux models in \figu{benchmark_spectra} are the same ones that \Refe~\cite{Valera:2022ylt} used to forecast the measurement of the UHE neutrino-nucleon cross section.  Below, we only sketch the main features of the models; we defer to \Refe~\cite{Valera:2022ylt} for a detailed overview, and to the original \Refes~\cite{Fang:2013vla, Padovani:2015mba, Fang:2017zjf, Heinze:2019jou, Muzio:2019leu, Rodrigues:2020pli, Anker:2020lre, IceCube:2020wum,  Muzio:2021zud, IceCube:2021uhz} for full details.  

Our benchmark UHE neutrino flux models are grouped in four classes, depending on the origin of the flux:
\begin{enumerate}[(a)]
 \item
  {\bf UHE extrapolation of the IceCube neutrino flux (``IceCube $\nu$ extrapolated", {\Large $\bullet$}  models 1 and 2):}  These are unbroken extrapolations to ultra-high-energies of the power-law ($\propto E_\nu^{-\gamma}$) neutrino flux measured by IceCube in the TeV--PeV range.
  
  {\bf Flux model 1} (``IceCube HESE (7.5~yr) extrapolated") extrapolates the soft-spectrum flux ($\gamma = 2.87$) of the IceCube 7.5-year HESE analysis~\cite{IceCube:2020wum}.  
  
  {\bf Flux model 2} (``IceCube $\nu_\mu$ (9.5 yr) extrapolated") extrapolates the hard-spectrum flux ($\gamma = 2.37$) of the IceCube 9.5-year through-going $\nu_\mu$ analysis~\cite{IceCube:2021uhz}.  
  
  [In our forecasts below, we consider flux models 1 or 2, augmented with a high-energy cut-off (Section~\ref{sec:background_nu}), as a background to the discovery of the other flux models, 3--12; see Section~\ref{sec:discovery_stat_analysis}.  Section~\ref{sec:discovery_results_uhe_ic_tail} forecasts the discovery of flux models 1 and 2 themselves.]
 \item
  {\bf Models of cosmogenic neutrinos (``Cosmogenic $\nu$", $\blacksquare$ models 3--5, 7):}  These are models of cosmogenic neutrinos made either by a population of nondescript sources of UHECRs, or by known classes of potential UHECRs sources.
  
  {\bf Flux model 3}~\cite{Heinze:2019jou} (``Heinze {\it al.}, fit to Auger UHECRs") considers UHECRs produced by nondescript sources, and fits their flux and mass composition to recent UHECR observations by the Pierre Auger Observatory~\cite{Fenu:2017hlc, Bellido:2017cgf}.  (References~\cite{Romero-Wolf:2017xqe, AlvesBatista:2018zui} predict similar fluxes using similar procedures and data.)
  
  {\bf Flux model 4}~\cite{Anker:2020lre}) (``Bergman \& van Vliet, fit to TA UHECRs") is produced similarly to flux model 3, but using instead recent UHECR observations by the Telescope Array (TA)~\cite{Tsunesada:2017aaq, Bergman:2017ikd}.  (Reference~\cite{Bergman:2021djm} predicts a similar flux.)  Flux model 3 is significantly smaller than flux model 4 because Auger observations favor a heavier UHECR mass composition at the highest energies, and because the fit of the UHECR spectrum to Auger data favors a lower cosmic-ray maximum rigidity~\cite{Romero-Wolf:2017xqe, AlvesBatista:2018zui, Heinze:2019jou} than the fit to TA data.  
  
  {\bf Flux model 5}~\cite{Rodrigues:2020pli} (``Rodrigues {\it et al.}, all AGN") is the cosmogenic neutrino flux expected from the full population of active galactic nuclei (AGN), which are taken to be UHECR accelerators, including low- and high-luminosity BL Lacs and flat-spectrum radio quasars.  The resulting UHECR flux is fit to Auger data~\cite{Fenu:2017hlc}, and the associated cosmogenic neutrino flux satisfies the IceCube upper limit on the UHE neutrino flux~\cite{IceCube:2018fhm}.  We adopt the maximum allowed predicted cosmogenic neutrino flux from the entire AGN population (Fig.~2 in \Refe~\cite{Rodrigues:2020pli}).
  
  {\bf Flux model 7}~\cite{Rodrigues:2020pli} (``Rodrigues {\it et al.}, HL BL Lacs") isolates the contribution of high-luminosity (HL) BL Lacs to the cosmogenic neutrino flux of model 5.
 \item
  {\bf Models of UHE neutrinos made inside astrophysical sources (``Source $\nu$", \rotatebox[origin=c]{45}{$\blacksquare$} models 6, 9, 10):}  These are models based on more detailed descriptions of the physical properties of known UHECR and neutrino source classes.
  
  {\bf Flux model 6}~\cite{Rodrigues:2020pli} (``Rodrigues {\it et al.}, all AGN") is the counterpart source neutrino flux to the cosmogenic flux model 5.  We adopt the maximum allowed predicted source neutrino flux from the entire AGN population (Fig.~2 in \Refe~\cite{Rodrigues:2020pli}).
  
  {\bf Flux model 9}~\cite{Fang:2013vla} (``Fang {\it et al.}, newborn pulsars") is the neutrino flux predicted from newborn, fast-spinning pulsars with intense surface magnetic fields that may accelerate UHECRs in the pulsar wind.  UHECR $pp$ interactions on the surrounding supernova ejecta produce neutrinos.  We adopt the flux prediction from \Refe~\cite{Fang:2013vla} for which the number density of pulsars evolves with redshift following the star formation rate.  (We include only the contribution of neutrinos made inside the pulsar environment.)
  
  {\bf Flux model 10}~\cite{Padovani:2015mba} (``Padovani {\it et al.}, BL Lacs") is the neutrino flux produced by $p\gamma$ interactions inside the jets of BL Lacs, computed within the framework of the simplified view of blazars.  Following \Refe~\cite{IceCube:2016uab}, the ratio of the neutrino intensity to the gamma-ray intensity, a key parameter of the model~\cite{Padovani:2015mba}, is set to $Y_{\nu\gamma}= 0.13$ to satisfy the present IceCube upper limit on the UHE neutrino flux~\cite{IceCube:2018fhm}.
  
 \item
  {\bf Models of joint cosmogenic and UHE source neutrinos (``Cosmogenic + source $\nu$", $\rotatebox[origin=c]{90}{\HexaSteel}$ models 8, 11, 12):}  These are multi-messenger models that aim to explain the joint production of UHECRs and TeV--EeV neutrinos.
  
  {\bf Flux model 8}~\cite{Fang:2017zjf} (``Fang \& Murase, cosmic-ray reservoirs") is the flux of UHE neutrinos produced, via $pp$ and $p\gamma$ interactions, by UHECRs accelerated in the jets of radio-loud AGN embedded in galaxy clusters that act as cosmic-ray reservoirs, within a grand-unified multi-messenger model.  The predicted UHECR flux and mass composition are fit to Auger data~\cite{PierreAuger:2015fol} and the predicted TeV--PeV neutrino flux, to IceCube data~\cite{IceCube:2016umi, IceCube:2017zho}.
  
  {\bf Flux model 11}~\cite{Muzio:2019leu} (``Muzio {\it et al.}, maximum extra $p$ component") is the neutrino flux produced in $p\gamma$ interactions within the UFA15 multi-messenger framework~\cite{Unger:2015laa}, where the UHECR flux and mass composition are fit to Auger data.  The model includes a sub-dominant UHECR pure-proton component beyond $10^9$~GeV that enhances the UHE neutrino flux.  We adopt the maximum allowed neutrino flux from the joint single-mass UFA15 plus pure-proton components, computed using the {\sc Sybill 2.3c}~\cite{Fedynitch:2018cbl} hadronic interaction model (Fig.~9 in \Refe~\cite{Muzio:2019leu}).
  
  {\bf Flux model 12}~\cite{Muzio:2021zud} (``Muzio {\it et al.}, fit to Auger \& IceCube") is the neutrino flux produced in $p\gamma$ interactions within the UFA15 multi-messenger framework, and in $pp$ interactions of UHECRs in the source environment.  The UHECR flux and mass composition are fit to Auger data, and the neutrino flux is fit to the IceCube TeV--PeV neutrino flux~\cite{IceCube:2021rpz, IceCube:2020acn}.  We adopt the best-fit total neutrino flux, ``UHECR $\nu$" plus ``Non-UHECR $\nu$", from Fig.~1 in \Refe~\cite{Muzio:2021zud}).
\end{enumerate}

In each UHE neutrino flux model above, we treat individually the flux of each neutrino species, $\nu_e$, $\nu_\mu$, $\nu_\tau$, $\bar{\nu}_e$, $\bar{\nu}_\mu$, $\bar{\nu}_\tau$.  To compute the flavor composition at Earth, after oscillations, we follow the same detailed prescription as in \Refe~\cite{Valera:2022ylt}, based on recent values of the neutrino mixing parameters from the {\sc NuFit} 5.0~\cite{Esteban:2020cvm, NuFit_5.0} global fit to neutrino oscillation data.  See Section~\ref{sec:background_nu} for a sketch of our prescription (in the particular context of flux models 1 and 2 as background fluxes) and \Refe~\cite{Valera:2022ylt} for full details of the flavor composition of each flux model.  We maintain the individual treatment of the flux of each neutrino species during their propagation through the Earth (Section~\ref{sec:propagation}) and when computing their contribution to the predicted event rate (Section~\ref{sec:event_rates}).  However, we conservatively assume no capability to distinguish events made by different flavors in the radio array of IceCube-Gen2.


\section{Propagating neutrinos through Earth}
\label{sec:propagation}

Once UHE neutrinos arrive at the surface of the Earth, they propagate underground toward the detector, from all directions.  Because the neutrino-matter cross section grows with energy (see below), for UHE neutrinos interactions with matter underground are significant, and attenuate the flux of neutrinos that reaches the detector.  The attenuation is energy- and direction-dependent: the higher the energy and the longer the distance traveled by a flux of neutrinos inside the Earth, the stronger it is attenuated.  The attenuation is also flavor-dependent: $\nu_\tau$ are relatively less affected compared to $\nu_e$ and $\nu_\mu$.  In our forecasts, we account in detail for the in-Earth propagation of UHE neutrinos from the surface of the Earth to the radio array of IceCube-Gen2.  Below, we elaborate.

At neutrino energies above a few GeV, the leading neutrino interaction channel is neutrino-nucleon ($\nu N$) deep inelastic scattering (DIS)~\cite{CTEQ:1993hwr, Conrad:1997ne, Formaggio:2012cpf}.  In it, a neutrino scatters off of one of the partons, \ie, a quark or a gluon, inside a nucleon, $N$, \ie, a proton or a neutron.  The final-state parton promptly hadronizes into final-state hadrons, $X$.  A neutral-current (NC) DIS interaction, mediated by a $Z$ boson, produces in addition a final-state neutrino, \ie, $\nu_\alpha + N \to \nu_\alpha + X$ ($\alpha = e, \mu, \tau$).  A charged-current (CC) DIS interaction, mediated by a $W$ boson, produces in addition instead a final-state charged lepton, \ie, $\nu_\alpha + N \to l_\alpha + X$.  The $\nu N$ DIS cross section has been measured at sub-TeV neutrino energies by a variety of accelerator neutrino experiments~\cite{Mukhin:1979bd, Baranov:1978sx, Barish:1978pj, Ciampolillo:1979wp, deGroot:1978feq, Colley:1979rt, Morfin:1981kg, Baker:1982ty, Berge:1987zw, Anikeev:1995dj, Seligman:1997, Tzanov:2005kr, Wu:2007ab, Adamson:2009ju, Nakajima:2010fp, Abe:2013jth, Acciarri:2014isz, Abe:2014nox}; in the few-TeV range, by FASER~\cite{Arakawa:2022rmp} (and the upcoming FASER$\nu$~\cite{FASER:2019dxq}), and in the TeV--PeV range, using IceCube data~\cite{IceCube:2017roe, Bustamante:2017xuy, IceCube:2020rnc}.  At higher energies, the cross section is predicted~\cite{Gandhi:1995tf, Gandhi:1998ri, Cooper-Sarkar:2007zsa, Gluck:2010rw, Connolly:2011vc, Block:2014kza, Goncalves:2014woa, Arguelles:2015wba, Albacete:2015zra, Bertone:2018dse} and may be measured in upcoming UHE neutrino telescopes~\cite{Denton:2020jft, Huang:2021mki, Valera:2022ylt, Esteban:2022uuw}.

Computing the UHE $\nu N$ DIS cross section requires knowing the parton distribution functions in  protons and neutrons, which are measured in lepton-hadron collisions, and extrapolating them beyond the regime where they have been measured.  (Concretely, it requires extrapolating them to values of Bjorken-$x$---the fraction of nucleon momentum carried by the interacting parton---beyond the measured ones.)  At ultra-high energies, the NC and CC $\nu N$ cross sections, $\sigma_{\nu N}^{\rm NC}$ and $\sigma_{\nu N}^{\rm CC}$, respectively, grow roughly $\propto E_\nu^{0.36}$~\cite{Gandhi:1998ri}, are essentially equal for all flavors of $\nu_\alpha$ and $\bar{\nu}_\alpha$, and $\sigma_{\nu N}^{\rm NC} \approx \sigma_{\nu N}^{\rm CC}/3$.  Below, to produce our forecasts, we adopt the state-of-the-art BGR18 calculation of the $\nu N$ DIS cross sections~\cite{Bertone:2018dse} in the propagation and detection of neutrinos.  The BGR18 is built using recent experimental results and sophisticated next-to-leading-order calculations, including the major corrections described in Appendix B4 of \Refe~\cite{Bertone:2018dse}; for details, see \Refe~\cite{Bertone:2018dse} and\Refe~\cite{Gauld:2019pgt}, for a summary, see \Refe~\cite{Valera:2022ylt}.

In a DIS interaction, the final-state hadrons receive a fraction $y$---the inelasticity---of the neutrino energy, and the final-state lepton receives the remaining fraction $(1-y)$.  In each interaction, the value of $y$ is randomly sampled from a probability density that is proportional to the differential DIS cross sections, $d\sigma_{\nu N}^{\rm NC}/dy$ and $d\sigma_{\nu N}^{\rm CC}/dy$.  At the energies relevant for our work, the average value of $y$ is about 0.25~\cite{Gandhi:1995tf}.  However, because the distribution of values of $y$ has a large spread (see Fig.~4 in \Refe~\cite{Valera:2022ylt}), when propagating neutrinos through the Earth below (and also when computing the event rates that they induce, in Section~\ref{sec:event_rate_benchmarks}), we do it by using the distributions of $y$, separately for NC and CC DIS, rather than by using its average value.

Inside the Earth, NC interactions shift the UHE neutrino flux to lower energies, by regenerating lower-energy neutrinos, while CC interactions dampen the flux altogether, by replacing neutrinos with charged leptons.  The one exception is the CC interaction of $\nu_\tau$: in them, the final-state tauon may propagate for some distance inside the Earth before decaying and generating a new, high-energy $\nu_\tau$.  As a result of this ``$\nu_\tau$ regeneration," the flux of $\nu_\tau$ is less attenuated than that of $\nu_e$ and $\nu_\mu$.  

The severity of the effects of in-Earth propagation on the neutrino flux varies with neutrino energy, $E_\nu$, and direction, expressed via the zenith angle, $\theta_z$, measured from the South Pole, where IceCube-Gen2 will be located.  Higher energies and directions corresponding to longer path lengths inside the Earth yield more severe effects.  To illustrate this, we use a simplified calculation of the number of neutrino-induced events in the detector, $N_\nu^{\rm simp}$, similar to the one in \Refe~\cite{Bustamante:2017xuy}, \ie,
\begin{equation}
 \label{equ:event_rate_simple}
 N_\nu^{\rm simp} (E_\nu, \theta_z) 
 \propto
 \Phi_\nu (E_\nu)
 \sigma_{\nu N}(E_\nu) 
 e^{-L(\theta_z)/L_{\nu N}(E_\nu, \theta_z)} \;,
\end{equation}
where $\Phi_\nu$ is the neutrino flux at the surface of the Earth, $\sigma_{\nu N}$ is the $\nu N$ cross section (for this simplified calculation, it is the sum of NC and CC cross sections), $L(\theta_z) = \sqrt{\left( R_\oplus^2 - 2 R_\oplus d \right) \cos^2 \theta_z + 2 R_\oplus d} - (R_\oplus - d) \cos\theta_z$ is the distance traveled inside the Earth by a neutrino with incoming direction $\theta_z$, where $R_\oplus = 6371$~km is the radius of Earth, $d$ is the detector depth, approximately 200~m for the radio array of IceCube-Gen2, $L_{\nu N} \equiv (\sigma_{\nu N} n_N)^{-1}$ is the neutrino mean free path inside the Earth along this direction, and $n_N$ is the average number density of nucleons along this direction, based on knowledge of the internal matter density of Earth (more on this later).  (We use \equ{event_rate_simple} only for illustration; later we describe the  detailed calculation with which we produce our results.)

Equation~(\ref{equ:event_rate_simple}) accounts for flux attenuation during in-Earth propagation, via the exponential dampening term, but ignores the regeneration of lower-energy neutrinos.  Even so, it embodies essential features of the propagation and detection of high-energy and ultra-high-energy neutrinos. Upgoing neutrinos ($\cos \theta_z < 0$), \ie, neutrinos that reach the detector from below after traveling underground a distance of up to the diameter of the Earth, are more strongly attenuated than downgoing ($\cos \theta_z > 0$) and horizontal neutrinos ($\cos \theta_z \approx 0$).  For UHE neutrinos, the attenuation is so strong that virtually no upgoing neutrinos reach the detector (see Fig.~A2 in \Refe~\cite{Bustamante:2017xuy}), unless the neutrino flux at the surface is extraordinarily large; \eg, benchmark flux model 4 in Figs.~\ref{fig:benchmark_spectra}, \ref{fig:binned_events_all_benchmarks_vs_energy_dep}, and \ref{fig:binned_events_all_benchmarks_vs_costhzrec}.  This means that our forecasts below, which factor in the contribution of neutrinos from all directions, are driven primarily by downgoing and horizontal neutrinos.  

Further, \equ{event_rate_simple} shows that while flux attenuation is $\propto e^{-\sigma_{\nu N}}$, the rate of neutrino interactions in the detector is $\propto \sigma_{\nu N}$.  The interplay between these competing effects is accentuated at high energies, where the cross section is larger: a larger cross section makes the already tiny flux of upgoing neutrinos vanish, which has little marginal effect, but it appreciably increases the number of downgoing and horizontal neutrinos detected.

Finally, \equ{event_rate_simple} reveals important nuance in the rate of neutrino interactions, which is $\propto \Phi_\nu \sigma_{\nu N}$.  In a realistic setting, given that neither the UHE neutrino flux nor the UHE $\nu N$ cross section have been measured so far, or that they are known only uncertainly at best, the detection of a number of neutrinos $N_\nu^{\rm simp}$ implies a degeneracy between $\Phi_\nu$ and $\sigma_{\nu N}$~\cite{Hooper:2002yq, Hussain:2006wg, Borriello:2007cs, Hussain:2007ba, Connolly:2011vc}.  Reference~\cite{Valera:2022ylt} accounted for the uncertainty on $\Phi_\nu$ and $\sigma_{\nu N}$ when forecasting the potential of IceCube-Gen2 to measure the UHE $\nu N$ cross section.  Here, we account for the uncertainty on the UHE $\nu N$ cross section when forecasting the potential of IceCube-Gen2 to discover benchmark UHE neutrino flux models 1--12, and to distinguish between them.  As in \Refe~\cite{Valera:2022ylt}, we do so via the energy-independent scaling parameter $f_\sigma \equiv \sigma_{\nu N}/\sigma_{\nu N}^{\rm std}$, where $\sigma_{\nu N}^{\rm std}$ is the central BGR18 prediction~\cite{Bertone:2018dse}.  The nominal value is $f_\sigma = 1$, and the value of $f_\sigma$ is common to the NC and CC cross sections.  Values of $f_\sigma \neq 1$ scale the central BGR18 cross section up or down, but do not affect its energy dependence.

The effect of changing $f_\sigma$ on the flux attenuation is most evident in neutrinos that reach the detector from around the horizon, \ie, $80^\circ \lesssim \theta_z \lesssim 120^\circ$, for which flux attenuation is present but milder than for upgoing neutrinos.  Changing $f_\sigma$ affects the directional distribution of events induced by horizontal neutrinos: a larger value of $f_\sigma$ sharpens the drop in the event rate from horizontal to upgoing neutrinos, while a smaller value softens it.  These effects are intrinsic to our event-rate calculation in Section~\ref{sec:event_rate_benchmarks};  \Refe~\cite{Valera:2022ylt} illustrates them explicitly.  Below, as part of our statistical methods in Sections~\ref{sec:discovery_stat_analysis} and \ref{sec:model_separation_stat_analysis}, we allow the value of $f_\sigma$ to float in fits to predicted mock event rates in IceCube-Gen2.  When doing so, for a given test value of $f_\sigma$, we modify equally the cross section used in in-Earth propagation and in neutrino detection.
 
While \equ{event_rate_simple} is useful to understand the essential features of in-Earth propagation, in our forecasts we propagate neutrinos inside the Earth in a  more detailed manner, using the state-of-the-art Monte Carlo code {\sc NuPropEarth}~\cite{Garcia:2020jwr, NuPropEarth}.  {\sc NuPropEarth} propagates neutrinos accounting for the leading contribution from CC and NC $\nu N$ DIS, using the BGR18 cross sections, and for the subdominant contribution of other neutrino-matter interaction channels that, taken together, can attenuate the flux attenuation by up to an additional $10\%$~\cite{Garcia:2020jwr}.  It also includes $\bar{\nu}_e$ scattering on atomic electrons, via the Glashow resonance~\cite{Glashow:1960zz, IceCube:2021rpz}, $\nu_\tau$ regeneration, energy losses of intermediate tauons, and the regeneration of lower-energy neutrinos in NC interactions.  It takes the internal matter density profile of Earth to be that of the Preliminary Reference Earth Model~\cite{Dziewonski:1981xy}, given as a set of concentric layers of different densities and mass compositions.  Reference~\cite{Garcia:2020jwr} has a full description of {\sc NuPropEarth}; for a summary, see \Refe~\cite{Valera:2022ylt}.

Thus, to compute neutrino-induced event rates below, in Section~\ref{sec:event_rate_benchmarks}, we first propagate the fluxes of $\nu_e$, $\nu_\mu$, $\nu_\tau$, $\bar{\nu}_e$, $\bar{\nu}_\mu$, and $\bar{\nu}_\tau$ separately, from the surface of the Earth to the simulated surface of the radio array of IceCube-Gen2 (Section~\ref{sec:ic-gen2}), for multiple neutrino energies and across multiple directions.  Figures~10 and 11 in \Refe~\cite{Valera:2022ylt} illustrate the resulting neutrino fluxes at the detector.  (Later, when computing neutrino-induced event rates in Section~\ref{sec:event_rate_benchmarks}, the energy and angular dependence of the effective volume of IceCube-Gen2 that we use represents the detector response only, not the effect of in-Earth propagation on the neutrino fluxes.)


\section{Forecasting ultra-high-energy neutrino event rates}
\label{sec:event_rates}


\subsection{Overview of the experimental landscape}

In high-energy neutrino telescopes, neutrinos of TeV-scale energies and above interact with the detector medium---ice, air, rock---predominantly via $\nu N$ DIS; see Section~\ref{sec:event_rate_benchmarks}.  Final-state products interact with the medium, or decay, and initiate high-energy particle showers.  As a shower develops, charged particles within it emit electromagnetic radiation, in the optical, ultraviolet, or radio wavelengths, depending on the neutrino energy and on the medium where the shower develops.  Neutrino telescopes target this emission using a variety of techniques, which we overview below; for a comprehensive review, see~\Refes~\cite{Abraham:2022jse, Ackermann:2022rqc}.  From the properties of the detected electromagnetic emission, neutrino telescopes infer the neutrino energy, direction, and flavor, with varying degrees of precision.

Present TeV--PeV neutrino telescopes---IceCube~\cite{IceCube:2013cdw}, ANTARES~\cite{ANTARES:1999fhm}, Baikal NT-200~\cite{BAIKAL:1997iok}---instrument large bodies of water or ice to detect the optical Cherenkov light emitted by showers, initiated mostly by $\nu_e$ and $\nu_\tau$, and tracks, initiated mostly by $\nu_\mu$.  IceCube is the largest among them: it consists of about 1~km$^3$ of Antarctic ice instrumented by thousands of photomultipliers at depths of 1.5--2.5~km.  IceCube discovered~\cite{IceCube:2013cdw, IceCube:2013low} and regularly observes TeV--PeV cosmic neutrinos~\cite{IceCube:2020wum, IceCube:2021uhz}, but may not be large enough to either discover UHE neutrinos, of EeV-scale energies, whose predicted flux may conceivably be significantly smaller~\cite{Aloisio:2009sj, Ahlers:2012rz, Romero-Wolf:2017xqe, AlvesBatista:2018zui, Heinze:2019jou}, or to observe a large number of them.  Indeed, currently the most constraining upper limits on the flux of UHE neutrinos come from IceCube~\cite{IceCube:2018fhm} and the Pierre Auger Observatory~\cite{PierreAuger:2019ens}; see \figu{benchmark_spectra}.  Future in-ice and in-water optical neutrino telescopes---Baikal-GVD~\cite{Baikal-GVD:2020xgh}, the optical array of IceCube-Gen2~\cite{IceCube-Gen2:2020qha}, KM3NeT~\cite{KM3Net:2016zxf}, P-ONE~\cite{P-ONE:2020ljt}, TRIDENT~\cite{Ye:2022vbk}---will be as large or larger than IceCube, and so will have higher detection rates,  but will remain mainly sensitive in the TeV--PeV range.

In the search for UHE neutrinos, the main limitation of optical detection is the attenuation length of light in ice or water, of 100--200~m~\cite{IceCube:2013llx}, which forces optical neutrino telescopes to use relatively dense arrays of photomultipliers.  Scaling the arrays up to the size required to achieve sensitivity to a conceivably tiny UHE neutrino flux is technically and financially challenging.  Instead, for UHE neutrinos, a variety of alternative techniques exist that can monitor a larger detection volume using more sparse instrumentation.  They target the particles, light, and radio emission from the showers initiated by UHE neutrinos in the atmosphere, in ice, or from space~\cite{Abraham:2022jse, Ackermann:2022rqc}.  Large arrays of surface particle detectors, like Auger~\cite{PierreAuger:2019ens} and the proposed TAMBO~\cite{Romero-Wolf:2020pzh}, are sensitive to showers from Earth-skimming neutrinos~\cite{Fargion:1999se}.  Atmospheric imaging telescopes, like the Telescope Array~\cite{TelescopeArray:2019mzl}, the Cherenkov Telescope Array (CTA)~\cite{Fiorillo:2020xst}, under construction, and the proposed Trinity~\cite{Otte:2019knb} and Ashra NTA~\cite{Sasaki:2014mwa}, target the Cherenkov and fluorescence light from neutrino-initiated extensive air showers, from different vantage points on the surface.  The proposed POEMMA~\cite{POEMMA:2020ykm} satellites target Cherenkov emission from space, while the proposed ANDIAMO~\cite{Marinelli:2021upw} aims for acoustic neutrino detection in water. 

In recent years, the technique of radio-detection of UHE particles, including neutrinos, has matured.  Because the attenuation length of radio is roughly 1~km in ice and negligible in air~\cite{Schroder:2016hrv}, radio-based neutrino telescopes can monitor large detector volumes using sparse arrays of radio antennas.  We focus on them below.


\subsection{Radio-detection of UHE neutrinos}
\label{sec:radio_detection}

Reference~\cite{Zas:1991jv} first proposed using radio emission from showers initiated by high-energy neutrinos as a means to detect them.  In a dense, transparent, and dielectric medium, like ice, as the shower develops it accumulates an excess of electrons on the shower front that can reach 20--30\% over the number of electrons plus positrons~\cite{Zas:1991jv} at shower maximum, after which the charge imbalance fades away. The time-varying excess charge produces a nanosecond-long pulse, known as {\it Askaryan radiation}~\cite{Askaryan:1961pfb} with a frequency content of approximately $100$~MHz--$1$~GHz. For a comprehensive introduction to the in-ice radio-detection technique, see \Refe~\cite{Barwick:2022vqt}.

Pioneering experiments established the viability of the radio-detection of UHE neutrinos.  ANITA~\cite{ANITA:2019wyx} was a balloon-borne detector that targeted radio emission from extensive air showers.  ARA~\cite{ARA:2019wcf} and ARIANNA~\cite{ARIANNA:2019scz}, and RICE~\cite{Kravchenko:2006qc} before them, are underground antenna arrays in Antarctica that target the radio emission from neutrino-initiated showers in ice.  (They, and other radio detectors like AERA~\cite{Holt:2016rsp}, CODALEMA~\cite{Ardouin:2006nb}, LOPES~\cite{LOPES:2005ipv}, LOFAR~\cite{vanHaarlem:2013dsa}, and Tunka-Rex~\cite{Bezyazeekov:2015rpa}, also look for UHECRs that interact in the atmosphere.)

In spite of their larger effective volume, these experiments have not yet been able to discover EeV neutrinos.  Thus, a number of radio-based neutrino telescopes currently in planning---BEACON~\cite{Wissel:2020sec}, GRAND~\cite{GRAND:2018iaj}, the radio array of IceCube-Gen2~\cite{IceCube-Gen2:2020qha, Hallmann:2021kqk}, PUEO~\cite{PUEO:2020bnn}, RET~\cite{RadarEchoTelescope:2021rca}, RNO-G~\cite{RNO-G:2020rmc}, TAROGE~\cite{Chen:2021egw}---aim to do so by using larger detectors and refined techniques. 

The main advantage of the radio-detection technique is the long attenuation length of radio waves in ice: up to 1.5~km at the South Pole~\cite{Barwick:2005zz, Barrella:2010vs}, and roughly 1~km in Greenland~\cite{Avva:2014ena,Aguilar:2022kgi}, {\it vs.}~100--200~m for optical signals~\cite{IceCube:2013llx, IceCube-Gen2:2020qha}.  This makes it possible to build larger detectors by placing a smaller number of radio antennas sparsely distributed, covering a larger area, and reaching a flux sensitivity that would be technically and economically more demanding with an optical detector. 

Below, we gear our forecasts to the radio array of IceCube-Gen2, one of the detectors in an advanced stage of planning and that envisions one of the best target flux sensitivities~\cite{Ackermann:2022rqc}.  IceCube-Gen2~\cite{IceCube-Gen2:2020qha} will be located in Antarctica, at the same site as IceCube, and will include an extension of the optical array aimed at high-statistics measurements in the TeV--PeV range, and a new underground radio antenna array aimed at discovering EeV neutrinos.  RNO-G~\cite{RNO-G:2020rmc}, in Greenland, presently under construction, has an order-of-magnitude smaller sensitivity than foreseen for IceCube-Gen2, but will field-test its design features.  In our forecasts, we model in detail the propagation of radio signals in ice, and their detection in antennas with the capabilities envisioned in the IceCube-Gen2 baseline design~\cite{IceCube-Gen2:2020qha, Hallmann:2021kqk}.  This allows us to make forecasts that include experimental nuance.  


\subsection{IceCube-Gen2}
\label{sec:ic-gen2}

The planned design of IceCube-Gen2~\cite{IceCube-Gen2:2020qha} includes an underground radio array that spans a total surface area of $500$~km$^2$.  The baseline design of the array~\cite{Hallmann:2021kqk}, which we adopt for our work, consists of 313 stations, each containing a cluster of antennas, and separated by 1--2~km from each other.  Because the stations are located far apart from each other, they function largely as stand-alone detectors, \ie, they have nearly independent effective volumes, unlike the strings of digital optical modules used in optical detectors, a number of which typically need to be triggered by the same shower in order to claim detection.  Thus, the total effective volume of the radio array grows roughly linearly with the number of stations.

The radio stations contain shallow antennas, buried close to the surface, and deep antennas, buried up to 200~m in the ice~\cite{IceCube-Gen2:2020qha, Hallmann:2021kqk}.  The final design of the stations and of the array is still evolving.  Placing antennas deeper in the ice increases the sky coverage, whereas shallow antennas have a field of view of that is more concentrated around the horizon~\cite{Barwick:2022vqt}.  This is because signal trajectories bend downward in the upper 200~m of the ice sheet due to a changing refraction index that restricts the region of ice that can be effectively monitored.  In our forecasts, we adopt the current baseline design of 169 stations containing only shallow antennas and 144 stations containing both shallow and deep antennas~\cite{Hallmann:2021kqk}. 

Regarding the angular resolution of the detector, the ability to reconstruct the incoming direction of detected events is different for shallow and deep antennas.  Shallow antennas have good angular resolution, expected to be as good as $3^\circ$~\cite{Gaswint:2021smu, ARIANNA:2021pzm}, which translates into $2^\circ$ when projected on the zenith angle.  Deep antennas have, on average, worse angular resolution due to the more limited ability to measure the horizontal signal polarization component~\cite{RNO-G:2021zfm}.  Presently, the development of algorithms to reconstruct the energy and direction of detected events is still in an early stage; future improvements are possible. To reflect this, below we repeat our analysis for different assumptions of angular resolution.  For our baseline results, in Section~\ref{sec:discovery_results_nominal_results}, we assume a zenith angle resolution of $\sigma_{\theta_z} = 2^\circ$ for each radio station; see Table~\ref{tab:analysis_choices_base}.  (We assume a common resolution of shallow and deep antennas.)  This is optimistic, but not unrealistic given likely improvements in reconstruction methods foreseen for the next decade, especially via the use of deep learning~\cite{Glaser:2022lky}.  In Section~\ref{sec:discovery_results_impact_resolution}, we present  results for a poorer resolution of $\sigma_{\theta_z} = 5^\circ$ and $10^\circ$.  Angular resolution is especially important to break the degeneracy between flux and cross section described in  Section~\ref{sec:propagation}.  It is also key to discovering point sources of UHE neutrinos; see \Refes~\cite{Fang:2016hop, Fiorillo:2022ijt}.

\begin{figure*}[t!]
 \centering
 \includegraphics[width=\textwidth]{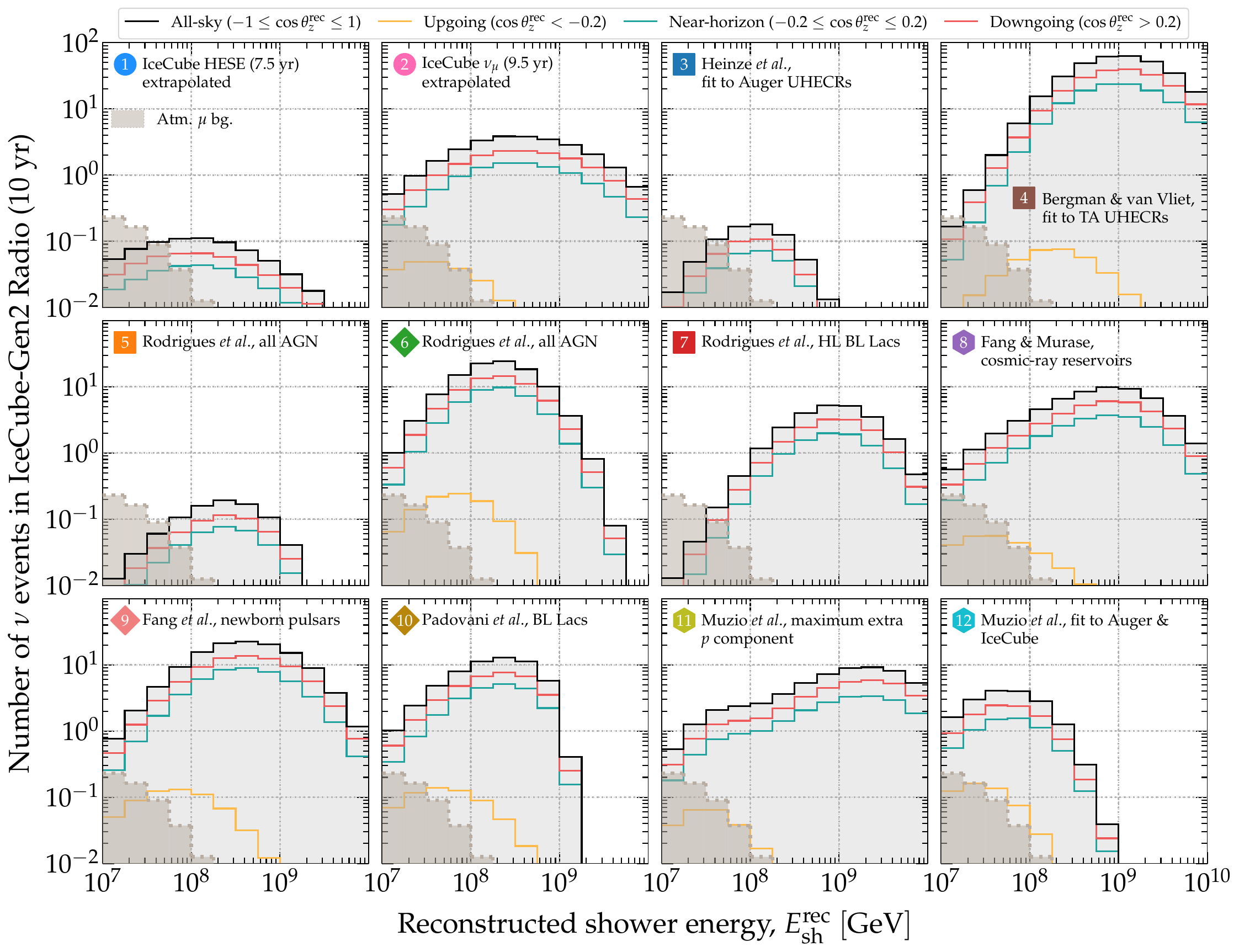}
 \caption{\label{fig:binned_events_all_benchmarks_vs_energy_dep}Mean distribution of events in reconstructed shower energy, $E_{\rm sh}^{\rm rec}$, expected in the radio array of IceCube-Gen2 after 10~years of exposure, for the benchmark UHE neutrino flux models 1--12 from \figu{benchmark_spectra}.  Figure \ref{fig:binned_events_all_benchmarks_vs_costhzrec} shows the corresponding distribution in reconstructed direction.  The neutrino-induced event rates are computed using the methods from Section~\ref{sec:event_rate_benchmarks} and, in this plot (and also  in Figs.~\ref{fig:binned_events_all_benchmarks_vs_costhzrec} and \ref{fig:binned_events_ic_hard}), by adopting our baseline analysis choices (see Table~\ref{tab:analysis_choices_base} and Section~\ref{sec:discovery_results_nominal_results}); in particular, the energy resolution is $\sigma_\epsilon = 0.1$ and angular resolution is $\sigma_{\theta_z} = 2^\circ$.  Figures~\ref{fig:bayes_factor_hard} and \ref{fig:confusion_hard} show, respectively, the associated baseline flux discovery potential and flux model separation.  Table~\ref{tab:event_rates_binned} shows the corresponding all-sky mean integrated event rates (however, to obtain our main results, in Sections~\ref{sec:discovery_potential} and \ref{sec:model_separation}, we use binned event rates).  We include the baseline background of atmospheric muons (see Section~\ref{sec:background_mu}), but not the background of the UHE tail of the high-energy IceCube neutrino flux (see Section~\ref{sec:background_nu}), though both enter our analysis; see Sections~\ref{sec:discovery_stat_analysis} and \ref{sec:model_separation_stat_analysis}. 
 }
\end{figure*}

Regarding the energy resolution of the detector, current estimates of the resolution of the shower energy are of a factor of two, \ie, a standard deviation of 0.3 on logarithmic scale, or better \cite{Anker:2019zcx,Gaswint:2021smu,Glaser:2022lky, Aguilar:2021uzt}, and as good as 30\%, \ie, a standard deviation of 0.1 on logarithmic scale, for certain conditions~\cite{Aguilar:2021uzt}.  For our baseline results, in Section~\ref{sec:discovery_results_nominal_results}, we assume an uncertainty of $\sigma_\epsilon = 0.1$ on the logarithm of the reconstructed shower energy; see Table~\ref{tab:analysis_choices_base}.  In Section~\ref{sec:discovery_results_impact_resolution}, we present results for a poorer energy resolution of $\sigma_\epsilon = 0.5$ and 1.0.  Energy resolution is especially important to distinguish between signal and background event distributions (see \figu{binned_events_all_benchmarks_vs_energy_dep}), and between predictions from alternative UHE neutrino flux models (Section~\ref{sec:model_separation}).

Later, we describe in detail how the predicted event rates in IceCube-Gen2 are affected by the angular and energy resolution of the detector, in connection to \equ{spectrum_rec}.

\begin{figure*}[t!]
 \centering
 \includegraphics[width=\textwidth]{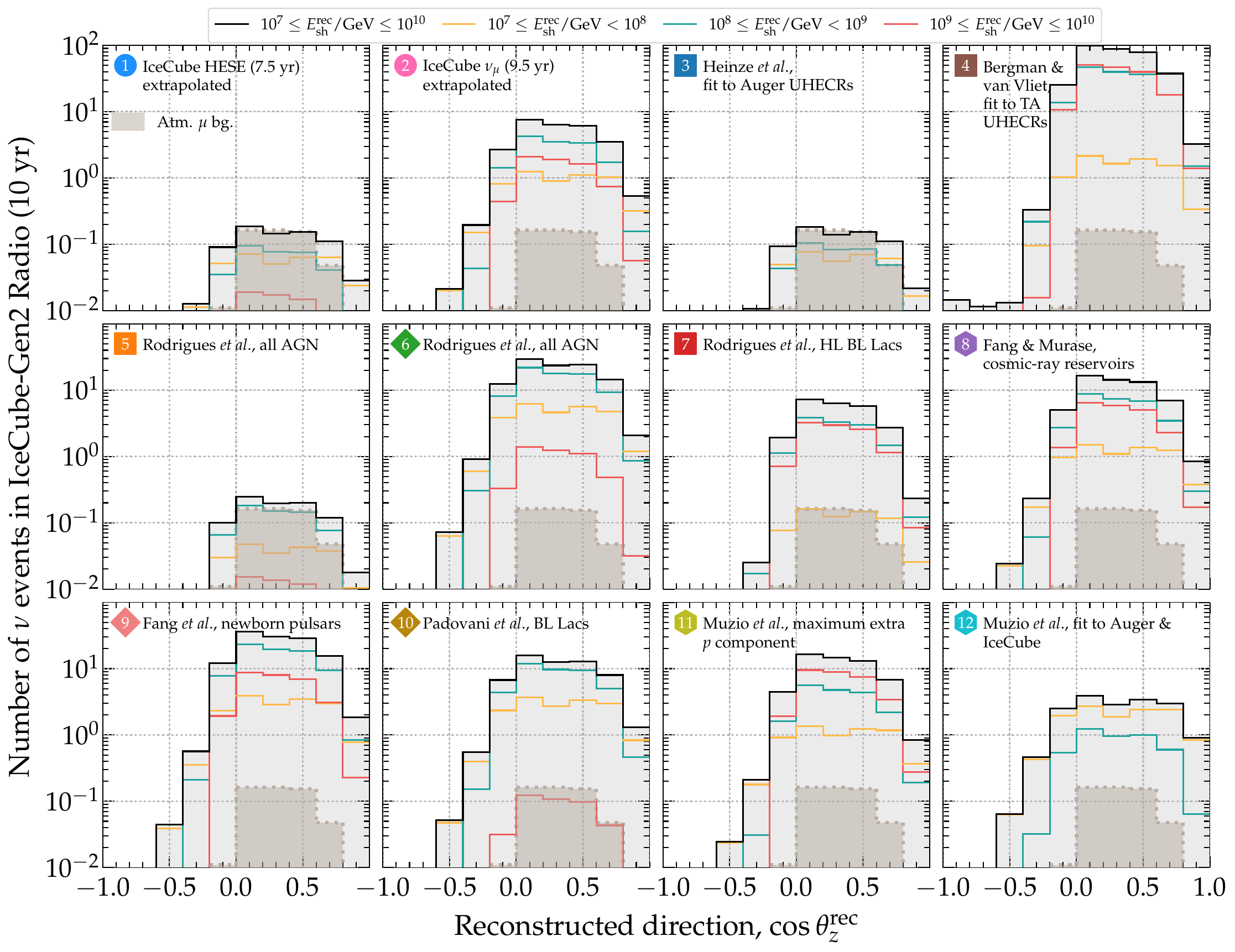}
 \caption{\label{fig:binned_events_all_benchmarks_vs_costhzrec}Mean distribution of events in reconstructed direction, $\cos \theta_{z}^{\rm rec}$, expected in the radio array of IceCube-Gen2 after 10~years of exposure, for the benchmark UHE neutrino flux models 1--12 from \figu{benchmark_spectra}.  Figure \ref{fig:binned_events_all_benchmarks_vs_energy_dep} shows the corresponding distribution in reconstructed shower energy.  The neutrino-induced event rates are computed using the methods from Section~\ref{sec:event_rate_benchmarks} and, in this plot (and also  in Figs.~\ref{fig:binned_events_all_benchmarks_vs_energy_dep} and \ref{fig:binned_events_ic_hard}), by adopting our baseline analysis choices (see Table~\ref{tab:analysis_choices_base} and Section~\ref{sec:discovery_results_nominal_results}); in particular, the energy resolution is $\sigma_\epsilon = 0.1$ and angular resolution is $\sigma_{\theta_z} = 2^\circ$.  Figures~\ref{fig:bayes_factor_hard} and \ref{fig:confusion_hard} show, respectively, the associated baseline flux discovery potential and flux model separation.  The uniform binning in this plot is only for illustration; the binning used in our statistical analysis is finer for events around the horizon ($80^\circ \leq \theta_z^{\rm rec} \leq 100^\circ$) and coarser for downgoing ($\theta_z^{\rm rec} < 80^\circ$) and upgoing directions ($\theta_z^{\rm rec} > 100^\circ$); it is described in Section~\ref{sec:discovery_results_nominal_choices}.  Table~\ref{tab:event_rates_binned} shows the corresponding all-sky mean integrated event rates (however, to obtain our main results, in Sections~\ref{sec:discovery_potential} and \ref{sec:model_separation}, we use binned event rates).  We include the baseline background of atmospheric muons (see Section~\ref{sec:background_mu}), but not the background of the UHE tail of the high-energy IceCube neutrino flux (see Section~\ref{sec:background_nu}), though both enter our analysis; see Sections~\ref{sec:discovery_stat_analysis} and \ref{sec:model_separation_stat_analysis}.}
\end{figure*}


\subsection{Computing event rates}
\label{sec:event_rate_benchmarks}

To forecast neutrino-induced event rates in the radio array of IceCube-Gen2, we follow the same methods introduced in \Refe~\cite{Valera:2022ylt}.  We sketch them below, and defer to \Refe~\cite{Valera:2022ylt} for details.  While we make our predictions particular to neutrino radio-detection IceCube-Gen2, our methods can be adapted to other neutrino telescopes, radio-based or otherwise; see Section~\ref{sec:limits_improvements}.

In-ice, radio-based neutrino telescopes, like ARA~\cite{ARA:2019wcf}, ARIANNA~\cite{ARIANNA:2019scz}, RNO-G~\cite{RNO-G:2020rmc}, and IceCube-Gen2~\cite{IceCube-Gen2:2020qha}, measure the energy deposited in the ice by particle showers that emit Askaryan radiation~\cite{Askaryan:1961pfb}.  In a shower initiated by a neutrino-nucleon ($\nu N$) DIS event, the shower energy, $E_{\rm sh}$, is a fraction of the parent neutrino energy, $E_\nu$.  The value of the fraction depends on the flavor of the interacting neutrino and on whether the interaction is NC or CC; we elaborate on this below.  See also Section~\ref{sec:propagation} for the effect of DIS in neutrino propagation inside Earth.

In the NC DIS interaction initiated by any flavor of $\nu_\alpha$ or $\bar{\nu}_\alpha$ ($\nu_\alpha + N \to \nu_\alpha + X$, $\alpha = e, \mu, \tau$), only the final-state hadrons, $X$, shower.  The final-state hadrons receive a fraction $y$---the inelasticity---of the parent neutrino energy, while the final-state neutrino, which escapes without interacting, receives the remaining fraction $1-y$; so, in this case, $E_{\rm sh} = y E_\nu$.  In the CC DIS interaction initiated by a $\nu_e$ or $\bar{\nu}_e$ ($\nu_e + N \to e + X$), the showers initiated by the final-state electron and hadrons both radiate, so the full neutrino energy is transmitted to the shower, \ie, $E_{\rm sh} = E_\nu$.  Finally, in the CC DIS interaction initiated by a $\nu_\mu$, $\nu_\tau$, $\bar{\nu}_\mu$, or $\bar{\nu}_\tau$ ($\nu_\alpha + N \to \alpha + X$, $\alpha = \mu, \tau$), the shower initiated by the final-state hadrons dominates the radiation; so, in this case, $E_{\rm sh} = y E_\nu$.  (There is an additional sub-dominant contribution to the shower rate, of up to 20\%, coming from showers initiated by the final-state muons and tauons~\cite{Garcia-Fernandez:2020dhb, Glaser:2021hfi}.  However, as in \Refe~\cite{Valera:2022ylt}, we do not account it in our simulations because it is computationally taxing to include.  This makes our forecasts below conservative.)   In summary, for a given shower energy $E_{\rm sh}$, the neutrino energy is
\begin{equation}
 \label{equ:energy_nu}
 E_{\nu_\alpha}^{i}(E_{\rm sh}, y)
 =
 \left\{
  \begin{array}{cl}
   E_{\rm sh}/y,                    &  {\rm for}~\nu_\alpha, i = {\rm NC} \\
   E_{\rm sh},                      &  {\rm for}~\nu_e~, i = {\rm CC}      \\
   E_{\rm sh}/y,                     &  {\rm for}~\nu_\mu~{\rm and}~\nu_\tau, i = {\rm CC}      \\
  \end{array}
 \right. \;.
\end{equation}
As during in-Earth neutrino propagation (Section~\ref{sec:propagation}), in each DIS interaction in the detector the value of $y$ is sampled at random from a probability distribution that is proportional to the energy-dependent differential $\nu N$ cross section, $d\sigma^{{\rm NC}}_{\nu_\alpha}/dy$ for NC interactions and $d\sigma^{{\rm CC}}_{\nu_\alpha}/dy$ for CC interactions; see \equ{spectrum_true} below.  In our forecasts, we use the inelasticity distribution built from the BGR18 UHE $\nu N$ cross section~\cite{Bertone:2018dse}; see Fig.~4 in \Refe~\cite{Valera:2022ylt}.

After a $\nu N$ DIS event, Askaryan radiation propagates through the ice, attenuating en route to the detector, and, upon reaching it, may or may not trigger the antennas, depending on the shower energy and direction, the characteristics of the antenna, and the size and geometry of the detector array.  We account for these features via dedicated Monte Carlo simulations of neutrino-induced shower production, propagation, and detection using the same state-of-the-art computational tools as the IceCube-Gen2 Collaboration, {\sc NuRadioMC}~\cite{Glaser:2019cws} and {\sc NuRadioReco}~\cite{Glaser:2019rxw}.  These simulations characterize the expected detector response; we describe it via the detector effective volumes, $V_{{\rm eff}, \nu_\alpha}^{\rm NC}$ and $V_{{\rm eff}, \nu_\alpha}^{\rm CC}$ below, which depend on the energy and direction of the shower. 

In our forecasts, we adopt the same simulated effective volumes introduced in \Refe~\cite{Valera:2022ylt}.  First, we simulate separately the effective volumes for NC and CC interactions of a shallow and a deep detector station; see Fig.~12 in \Refe~\cite{Valera:2022ylt}.  The shallow-station components are triggered by requiring a time-coincident high-and-low threshold crossing of two out of four LPDA antennas in an optimized trigger bandwidth~\cite{Glaser:2020pot}. The deep-station components are triggered by an interferometric phased array installed at a depth of 200~m~\cite{Allison:2018ynt}.  These are the trigger settings foreseen for IceCube-Gen2~\cite{Hallmann:2021kqk}.  We simulate the interactions, for NC and CC separately, in several cubic kilometers of ice surrounding the antennas, of a large number of neutrinos with different energies and from different directions, and the ensuing production and propagation of showers and Askaryan radiation.  We keep record of which showers trigger the antennas and are deemed as detected.  With this, we compute the effective volume as the fraction of showers that are detected times the simulated detector volume.  

Second, we scale up the effective volumes obtained for the single detector component up to the size of the array to obtain the full-detector volumes, $V_{{\rm eff}, \nu_\alpha}^{\rm NC}$ and $V_{{\rm eff}, \nu_\alpha}^{\rm CC}$, by multiplying the effective volume of a single component times their total number in the array.  To do that, we adopt the baseline array design of \Refe~\cite{Hallmann:2021kqk}, \ie, 144 hybrid stations, each containing a shallow component and a deep component, plus 169 shallow-only stations; see Fig.~13 in \Refe~\cite{Valera:2022ylt}.   Equation~(\ref{equ:spectrum_true}) shows the role of the effective volume in the calculation of event rates.  Unlike common practice, the energy and direction dependence of the effective volume that we use stem exclusively from the detector response, not from the neutrino propagation through the Earth.  We account for the latter in the neutrino flux that reaches the detector, $\Phi_{\nu_\alpha}^{\rm det}$, computed as described in Section~\ref{sec:propagation}.

The differential rate of showers induced by the NC and CC interactions of $\nu_\alpha$ is the convolution of the effective volume, the differential cross section, and the neutrino flux at the detector, \ie,
\begin{widetext}
\begin{equation}
 \label{equ:spectrum_true}
 \begin{split}
  \frac{d^2N_{\nu_\alpha}}{dE_{\rm sh} d\cos\theta_z}
  =  
  2 \pi T n_t
  \int_0^1 dy
  \biggl(
  &
  \frac{E_{\nu_\alpha}^{\rm NC}(E_{\rm sh}, y)}{E_{\rm sh}}
  V_{{\rm eff}, \nu_\alpha}^{\rm NC}(E_{\rm sh}, \cos\theta_z)
  \frac{d\sigma_{\nu_\alpha{\rm w}}^{\rm NC}(E_\nu, y)}{dy}
  \Phi^{\rm det}_{\nu_\alpha}(E_\nu,\cos\theta_z)_{E_\nu = E_{\nu_\alpha}^{\rm NC}(E_{\rm sh}, y)}
  \\
  &
  +~
  {\rm NC} \to \biggl. {\rm CC}
  \biggr)
   \;,
 \end{split}
\end{equation}
\end{widetext}
where $T$ is the detector exposure time, $n_t \equiv N_{\rm Av} \rho_{\rm ice} / M_{\rm ice}$ is the number density of water molecules in ice, $N_{\rm Av}$ is Avogadro's number, $\rho_{\rm ice} = 0.9168$~g~cm$^{-3}$ is the density of ice, and $M_{\rm ice} = 18.01528$~g~mol$^{-1}$ is the molar mass of water.  On the right-hand side of \equ{spectrum_true}, the term $E_{\nu_\alpha}^{\rm NC}/E_{\rm sh}$, and its CC equivalent, transforms the energy scale from neutrino energy to shower energy; it is given by \equ{energy_nu}.  The differential cross section is for neutrino DIS on one molecule of water (H$_2$O), \ie, $\sigma_{{\nu_\alpha} w}^{\rm NC} = 10 \sigma_{\nu_\alpha p}^{\rm NC} + 8 \sigma_{\nu_\alpha n}^{\rm NC}$, where $\sigma_{\nu_\alpha p}^{\rm NC}$ and $\sigma_{\nu_\alpha n}^{\rm NC}$ are the $\nu_\alpha p$ and $\nu_\alpha n$ BGR18 cross sections~\cite{Bertone:2018dse}, respectively, and similarly for CC interactions.  The event rate induced by $\bar{\nu}_\alpha$ is the same as for $\nu_\alpha$, \ie, \equ{spectrum_true}, but changing the cross section to $\sigma_{\bar{\nu}_\alpha {\rm w}}^{\rm NC}$, which, at ultra-high energies, is nearly equal to $\sigma_{\nu_\alpha {\rm w}}^{\rm NC}$ (also true for CC interactions; see Fig.~3 in \Refe~\cite{Valera:2022ylt}), and the flux to $\Phi_{\bar{\nu}_\alpha}^{\rm det}$.

Equation~(\ref{equ:spectrum_true}) computes the event rate in terms of the true shower energy, $E_{\rm sh}$, and the true shower direction, $\theta_z$.  We account for the limited energy and angular resolution of the detector by using energy and angular resolution functions, and by expressing the event rate in terms of reconstructed shower energy, $E_{\rm sh}^{\rm rec}$, and reconstructed direction, $\theta_z^{\rm rec}$.  The energy resolution function, $\mathcal{R}_{E_{\rm sh}}$, is a Gaussian probability density function of $\log_{10} E_{\rm sh}^{\rm rec}$, centered at the true shower energy, $\log_{10} E_{\rm sh}$, with a width $\sigma_{E_{\rm sh}} \equiv 10^{\sigma_\epsilon} E_{\rm sh}$, where $\epsilon \equiv \log_{10}(E_{\rm sh}^{\rm rec}/E_{\rm sh})$.  For our baseline results (Table~\ref{tab:analysis_choices_base}), we set $\sigma_\epsilon = 0.1$ as discussed above.  The angular resolution function, $\mathcal{R}_{\theta_z}$, is a Gaussian probability density function of $\theta_z^{\rm rec}$, centered at the true direction, $\theta_z$, with a width of $\sigma_{\theta_z}$.  For our baseline results (Table~\ref{tab:analysis_choices_base}), we set $\sigma_{\theta_z} = 2^\circ$ as discussed above.  In Section~\ref{sec:discovery_results_impact_resolution}, we show the impact on our results of varying the values of $\sigma_\epsilon$ and $\sigma_{\theta_z}$.  Reference~\cite{Valera:2022ylt} contains explicit definitions of the resolution functions.

Thus, the differential event rate of showers induced by $\nu_\alpha$, in terms of reconstructed energy and direction, is
\begin{widetext}
\begin{equation}
 \label{equ:spectrum_rec}
 \frac{d^2N_{\nu_\alpha}}
 {dE_{\rm sh}^{\rm rec} d\theta_{z}^{\rm rec}}
 =
 \int_{-1}^{+1} d\cos\theta_z
 \int_{0}^{\infty} dE_{\rm sh}
 \frac{d^2N_{\nu_\alpha}}{dE_{\rm sh} d\cos\theta_z} 
 \mathcal{R}_{E_{\rm sh}}(E_{\rm sh}^{\rm rec}, E_{{\rm sh}})~
 \mathcal{R}_{\theta_{z}}(\theta_{z}^{\rm rec}, \theta_z) \;.
\end{equation}
\end{widetext}
The CC interaction of $\nu_e$ and $\bar{\nu}_e$ dominates the event rate, since these are the two cases for which $E_{\rm sh} = E_\nu$.  The NC interaction of each species, and the CC interaction of $\nu_\mu$, $\nu_\tau$, $\bar{\nu}_\mu$, and $\bar{\nu}_\tau$ each contributes at roughly the same level.  For details, see Fig.~14 in \Refe~\cite{Valera:2022ylt}.  To be conservative in our forecasts, we do not assume that flavor identification will be possible, though there are promising early results~\cite{Stjarnholm:2021xpj, Glaser:2021hfi}.  Accordingly, we use only the total event rate induced by all flavors of $\nu_\alpha$ and $\bar{\nu}_\alpha$, \ie,
\begin{equation}
 \label{equ:spectrum_rec_all}
 \frac{d^2N_{\nu}}
 {dE_{\rm sh}^{\rm rec} d\theta_{z}^{\rm rec}}
 =
 \sum_{\alpha}^{e,\mu,\tau}
 \left(
 \frac{d^2N_{\nu_\alpha}}
 {dE_{\rm sh}^{\rm rec} d\theta_{z}^{\rm rec}}
 +
  \frac{d^2N_{\bar{\nu}_\alpha}}
 {dE_{\rm sh}^{\rm rec} d\theta_{z}^{\rm rec}}
 \right) \;.
\end{equation}
Below, as part of our analysis, we compute event rates in bins of reconstructed energy and direction; to do so, we integrate \equ{spectrum_rec_all} in $E_{\rm sh}^{\rm rec}$ and $\theta_z^{\rm rec}$ inside each bin.  For our baseline results, we use 12 bins of $E_{\rm sh}^{\rm rec}$, evenly distributed in logarithmic scale from $10^7$~GeV to $10^{10}$~GeV, and 13 bins of $\theta_z^{\rm rec}$, with a denser coverage around the horizon; see Table~\ref{tab:analysis_choices_base} and Section~\ref{sec:discovery_results_nominal_choices} for details.  

Figure~\ref{fig:binned_events_all_benchmarks_vs_energy_dep} shows the mean predicted energy distribution of events after 10~years of exposure in the radio array of IceCube-Gen2, for the benchmark flux models 1--12 introduced in Section~\ref{sec:fluxes} and \figu{benchmark_spectra}.  Because of the severe in-Earth attenuation of UHE neutrinos, the event rate is dominated by downgoing events and, to a lesser extent, by near-horizontal events.  For each flux model, the shape of the event energy distribution traces the shape of its corresponding neutrino energy spectrum from \figu{benchmark_spectra}.  Later, in Section~\ref{sec:model_separation}, this feature will allow us to distinguish between different flux models.  For all flux models, below $E_{\rm sh}^{\rm rec} = 10^7$~GeV the event rates dip because the effective volume decreases at low neutrino energies as a result of Askaryan radiation weakening.

Figure~\ref{fig:binned_events_all_benchmarks_vs_costhzrec} shows the corresponding mean predicted angular distribution of events for the benchmark flux models.  (The angular binning used in \figu{binned_events_all_benchmarks_vs_costhzrec} is for illustrative purposes only.  Our analysis uses a finer binning around the horizon; see Table~\ref{tab:analysis_choices_base} and Section~\ref{sec:discovery_results_nominal_choices}.)   Above the horizon, \ie, $\cos\theta_z^{\rm rec} > 0$, where in-Earth attenuation is small or negligible, the angular event distribution primarily traces the angular dependence of the effective volume.  At the horizon and below it, \ie, $\cos\theta_z^{\rm rec} < 0$, the angular distribution has a sharp cut-off due to  the strong in-Earth attenuation.  Only flux models with a large normalization, \ie, models 2, 4, 6--12, overcome the suppression and yield a handful of events below the horizon.  In contrast to the energy distribution of the events, where differences between flux models are evident, differences in the angular distribution between flux models are mild.

The event rates computed using Eqs.~(\ref{equ:spectrum_true})--(\ref{equ:spectrum_rec_all}) and shown in Figs.~\ref{fig:binned_events_all_benchmarks_vs_energy_dep} and \ref{fig:binned_events_all_benchmarks_vs_costhzrec} are the mean expected rates.  In a specific experimental observation, the number of events in each bin will be an integer value, and might deviate appreciably from the mean, especially when it has a low value.  Therefore, later, as part of our statistical analyses in Sections~\ref{sec:discovery_stat_analysis} and \ref{sec:model_separation_stat_analysis}, we account for random statistical fluctuations around the mean.

\begingroup
\squeezetable
\begin{table*}[t!]
 \begin{ruledtabular}
  \caption{\label{tab:event_rates_binned}Expected mean rates of neutrino-induced events in the radio array of IceCube-Gen2, after exposure time $T$, for the benchmark UHE diffuse neutrino flux models used in this analysis (see Section~\ref{sec:fluxes} and \figu{benchmark_spectra}), time needed for their decisive discovery, $T^{\rm disc}$, \ie, for the mean discovery Bayes factor $\langle \mathcal{B^{\rm disc}} \rangle > 100$ (see Section~\ref{sec:discovery_stat_analysis}), and mean number of events induced by them until the time of their decisive discovery, $N_\nu^{\rm disc}$.  Flux models with blank entries (--) are not expected to be discovered within 20~years.  Entries marked with an asterisk (*) signal that the background includes only atmospheric muons (see Sections~\ref{sec:background_mu} and \ref{sec:discovery_results_uhe_ic_tail}); unmarked entries include in addition the background from the UHE tail of the IceCube high-energy neutrino flux (see Section~\ref{sec:background_nu}).  Flux types are (Section~\ref{sec:fluxes}): extrapolation to ultra-high energies ({\Large $\bullet$}), cosmogenic ($\blacksquare$), source (\rotatebox[origin=c]{45}{$\blacksquare$}), and cosmogenic + source ($\rotatebox[origin=c]{90}{\HexaSteel}$).  Results in this table are obtained using our baseline analysis choices (Table~\ref{tab:analysis_choices_base} and Section~\ref{sec:discovery_results_nominal_choices}).  Figure~\ref{fig:bayes_factor_hard} shows the continuous evolution of $\langle \mathcal{B^{\rm disc}} \rangle$ with $T$; this table shows only snapshots.  Results for alternative analysis choices are in Sections~\ref{sec:discovery_results_impact_muon_bg}--\ref{sec:discovery_results_uhe_ic_tail}.  The event rates shown are all-sky, \ie, summed over all reconstructed directions, $-1 \leq \cos \theta_z^{\rm rec} \leq 1$, and grouped in a single bin of reconstructed shower energy, $10^7 \leq E_{\rm sh}^{\rm rec}/{\rm GeV} \leq 10^{10}$.  However, all-sky rates are only illustrative; the statistical analysis with which we compute $\langle \mathcal{B^{\rm disc}} \rangle$ uses instead binned event rates; see Section~\ref{sec:discovery_stat_analysis} for details.}
  \centering
  \renewcommand{\arraystretch}{1.3}
  \begin{tabular}{cccccccc}
  \multirow{2}{*}{\#} &
  \multirow{2}{*}{Type} &
  \multirow{2}{*}{UHE $\nu$ flux model} & 
  \multicolumn{3}{c}{All-sky integrated event rate, $N_\nu$} &
  \multicolumn{2}{c}{Decisive flux discovery ($\langle \mathcal{B}^{\rm disc} \rangle > 100$)} \\
  \cline{4-6}
  \cline{7-8}
  &
  &
  &
  $T = 1$~yr &
  $T = 3$~yr &
  $T = 10$~yr &
  $T^{\rm disc}~[{\rm yr}]$ &
  $N^{\rm disc}_\nu$\\
  \hline
  1 &
  {\Large $\bullet$} &
  IceCube HESE (7.5~yr) extrapolated~\cite{IceCube:2020wum} &
  0.07 &
  0.22 &
  0.73 &
  $> 20^\star$ &
  -- \\
  &
  &
  $\ldots$ with cut-off at $E_{\nu, {\rm cut}}^{\rm HE} = 500~{\rm PeV}$ &
  0.05 &
  0.14 &
  0.45 &
  $> 20^\star$ &
  -- \\
  &
  &
  $\ldots$ with cut-off at $E_{\nu, {\rm cut}}^{\rm HE} = 100~{\rm PeV}$ &
  0.02 &
  0.07 &
  0.22 &
  $> 20^\star$ &
  -- \\  
  &
  &
  $\ldots$ with cut-off at $E_{\nu, {\rm cut}}^{\rm HE} = 50~{\rm PeV}$ &
  0.01 &
  0.04 &
  0.13 &
  $> 20^\star$ &
  -- \\
  2 &
  {\Large $\bullet$} &
  IceCube $\nu_\mu$ (9.5~yr) extrapolated~\cite{IceCube:2021uhz} &
  2.69 &
  8.07 &
  26.90 &
  0.26* &
  0.70* \\
  &
  &
  $\ldots$ with cut-off at $E_{\nu, {\rm cut}}^{\rm HE} = 500~{\rm PeV}$ &
  1.02 &
  3.06 &
  10.20 &
  $1.05^\star$ &
  $1.07^\star$ \\
  &
  &
  $\ldots$ with cut-off at $E_{\nu, {\rm cut}}^{\rm HE} = 100~{\rm PeV}$ &
  0.35 &
  1.04 &
  3.47 &
  $4.97^\star$ &
  $1.74^\star$ \\  
  &
  &
  $\ldots$ with cut-off at $E_{\nu, {\rm cut}}^{\rm HE} = 50~{\rm PeV}$ &
  0.18 &
  0.53 &
  1.75 &
  $11.10^\star$ &
  $2.00^\star$ \\
  3 &
  $\blacksquare$ &
  Heinze {\it et al.}, fit to Auger UHECRs~\cite{Heinze:2019jou} &
  0.07 &
  0.21 &
  0.71 &
  $> 20$ &
  -- \\
  4 &
  $\blacksquare$ &
  Bergman \& van Vliet, fit to TA UHECRs~\cite{Anker:2020lre} &
  33.23  &
  99.70 &
  332.34 &
  0.28 &
  9.30 \\
  5 &
  $\blacksquare$ &
  Rodrigues {\it et al.}, all AGN benchmark~\cite{Rodrigues:2020pli} &
  0.09 &
  0.27 &
  0.89 &
  $> 20$ &
  -- \\
  6 &
  \rotatebox[origin=c]{45}{$\blacksquare$} &
  Rodrigues {\it et al.}, all AGN benchmark~\cite{Rodrigues:2020pli} &
  10.72 &
  32.15 &
  107.16 &
  1.31 &
  14.04 \\
  7 &
  $\blacksquare$ &
  Rodrigues {\it et al.}, HL BL Lacs~\cite{Rodrigues:2020pli} &
  2.42 &
  7.27 &
  24.24 &
  13.03 &
  31.53 \\
  8 &
  \rotatebox[origin=c]{90}{\HexaSteel} &
  Fang \& Murase, cosmic-ray reservoirs~\cite{Fang:2017zjf} &
  5.74 &
  17.22 &
  57.41 &
  6.16 &
  35.36 \\
  9 &
  $\blacksquare$ &
  Fang {\it et al.}, newborn pulsars~\cite{Fang:2013vla} &
  12.54 &
  37.61 &
  125.38 &
  1.89 &
  23.70 \\
  10 &
  $\blacksquare$ &
  Padovani {\it et al.}, BL Lacs~\cite{Padovani:2015mba} &
  5.79 &
  17.34 &
  57.85 &
  3.07 &
  17.78 \\
  11 &
  $\blacksquare$ &
  Muzio {\it et al.}, maximum extra $p$ component~\cite{Muzio:2019leu} &
  5.66 &
  19.97 &
  56.55 &
  2.48 &
  5.97 \\
  12 &
  \rotatebox[origin=c]{90}{\HexaSteel} &
  Muzio {\it et al.}, fit to Auger \& IceCube ~\cite{Muzio:2021zud} &
  1.71 &
  5.14 &
  17.12 &
  8.05 &
  13.77 
  \vspace{0.02cm} \\
  \hline
  \vspace{-0.30cm} \\
  -- &
  -- &
  Atmospheric muon background~(baseline) &
  0.05 &
  0.16 &
  0.54 &
  -- &
  --
  \end{tabular}
 \end{ruledtabular}
\end{table*}
\endgroup


\subsection{Backgrounds}
\label{sec:background}

Below, in Sections~\ref{sec:discovery_potential} and \ref{sec:model_separation}, we forecast the potential to discover the benchmark UHE neutrino flux models 1--12, and to distinguish between them, factoring in the contamination from background that may mimic the events induced by the flux models.  We account for two expected sources of background---atmospheric muons and the potential UHE tail of the IceCube neutrino flux---and comment on the pressing need to characterize a third likely source of background---air-shower cores.


\subsubsection{Atmospheric muons}
\label{sec:background_mu}

High-energy muons produced in the interaction of UHECRs in the atmosphere may trigger in-ice showers whose Askaryan radiation is expected to generate a small, but irreducible background for UHE neutrino searches~\cite{Garcia-Fernandez:2020dhb}.  We estimate the rate of muon-induced events in the radio array of IceCube-Gen2 using the hadronic interaction model {\sc Sybill 2.3c}~\cite{Fedynitch:2018cbl} and applying a surface veto that mitigates its effect by detecting the air shower that accompanies the muon.  This is the same prescription that was used to compute the muon background in \Refes~\cite{Valera:2022ylt, Fiorillo:2022ijt}.  In our forecasts below, the background of atmospheric muons affects all benchmark flux models 1--12.

Figures~\ref{fig:binned_events_all_benchmarks_vs_energy_dep} and \ref{fig:binned_events_all_benchmarks_vs_costhzrec} show the resulting energy and angular distribution of muon-induced events in the radio array of IceCube-Gen2.  They are concentrated at the lowest energies, $E_{\rm sh}^{\rm rec} \lesssim 10^8$~GeV, and in downgoing directions, $\cos \theta_z^{\rm rec} \gtrsim 0$, since muons cannot penetrate far inside Earth.  The irreducible all-sky integrated rate of muon-induced events above $10^8$~GeV, \ie, that cannot be vetoed by the surface veto, is lower than $0.1$ events per year.  Hence, atmospheric muons represent an obstacle only to the discovery of an UHE neutrino flux that is small and that peaks at low neutrino energies, \eg, benchmark flux models 1, 3, and 9; Section~\ref{sec:discovery_results} shows this in detail.  In Appendix~\ref{sec:impact_surface_veto} we comment on the effect on the flux discovery potential of not using a surface veto; in that case, the discovery potential is only degraded mildly.

The muon background shown in Figs.~\ref{fig:binned_events_all_benchmarks_vs_energy_dep} and \ref{fig:binned_events_all_benchmarks_vs_costhzrec} constitutes our baseline analysis choice; see Section~\ref{sec:discovery_results_nominal_choices}, and Tables~\ref{tab:analysis_choices_base} and \ref{tab:analysis_choices_alt} for a full list of analysis choices.  Below, we produce our main results using it.  In Section~\ref{sec:discovery_results_impact_muon_bg}, we show that even if our baseline muon-induced background was a significant underestimation of its true size, this would only erode mildly the prospects of discovering most of our benchmark UHE neutrino flux models.  (However, in Section~\ref{sec:discovery_results_impact_muon_bg}, we only change the normalization of the atmospheric muon flux, not the shape of its energy spectrum.  If the energy spectrum of atmospheric muons were to extend to higher energies than in our baseline prescription of it, the conclusions about its importance in our forecasts might change.)


\subsubsection{UHE tail of the IceCube high-energy neutrino flux}
\label{sec:background_nu}

Presently, using roughly ten years of data, IceCube has found that the diffuse flux of high-energy cosmic neutrinos that it measures spans the neutrino energy range from about 10~TeV to at least a few PeV~\cite{IceCube:2020wum, IceCube:2021uhz}.  In the PeV range, data is sparse because the neutrino energy spectrum falls steeply with energy.  As a result, it is presently unknown whether the flux measured at TeV--PeV energies extends to ultra-high energies, beyond 100~PeV and, if so, what the size and shape of its spectrum is at those energies.  

In \figu{benchmark_spectra}, benchmark UHE neutrino flux models 1 and 2 are straightforward UHE extrapolations of two IceCube TeV--PeV power-law flux measurements~\cite{IceCube:2020wum, IceCube:2021uhz}, without any high-energy suppression (more on this later).  They illustrate that if the UHE tail of the IceCube high-energy neutrino flux is large enough to trigger events in the radio array of IceCube-Gen2, it would constitute a background to the discovery of UHE neutrino flux models.  In our forecasts below, the background from the UHE tail of the IceCube high-energy neutrino flux affects benchmark flux models 3--12.  (In Section~\ref{sec:discovery_results_uhe_ic_tail}, we study separately the discovery of the UHE tail of the IceCube high-energy neutrino flux itself.)

Currently, the TeV--PeV neutrino flux seen by IceCube is described well as a simple power law $\propto E_\nu^{-\gamma}$.  Two properties of the TeV--PeV neutrino flux determine whether its UHE tail may be detectable in UHE neutrino telescopes: the value of the spectral index, $\gamma$, and whether the flux is further suppressed, relative to the simple power law, at or above the few-PeV scale.

Regarding the spectral index, its value depends on the set of IceCube events that is used to perform the fit.  The neutrino spectrum is harder ($\gamma \approx 2.37$) when derived from a fit to 9.5 years of through-going muon tracks~\cite{IceCube:2021uhz}, created by $\nu_\mu$ that interact outside the detector and make muons that cross it.  The spectrum is softer ($\gamma \approx 2.87$) when derived from a fit to 7.5 years of High Energy Starting Events (HESE), created by neutrinos of all flavors that interact inside the detector~\cite{IceCube:2020wum}.  In the TeV--PeV range, these results are compatible with each other within $1\sigma$~\cite{IceCube:2020wum}.  However, because a harder spectrum falls more slowly with energy, its UHE tail is more likely to be prominent and trigger events in the radio array of IceCube-Gen2.  In our forecasts, we explore different values of the spectral index (and of the corresponding normalization of the flux), motivated by IceCube results; see Sections~\ref{sec:discovery_results_nominal_choices} and \ref{sec:discovery_results_impact_neutrino_bg}, and Tables~\ref{tab:analysis_choices_base} and \ref{tab:analysis_choices_alt}.

Regarding the possible further suppression of the IceCube flux in the few-PeV range, it is unknown whether the flux extends beyond a few PeV as a simple power law, or whether it is suppressed by an exponential cut-off $e^{-E_\nu/E_{\nu, {\rm cut}}^{\rm HE}}$ at a cut-off energy $E_{\nu, {\rm cut}}^{\rm HE}$ of a few PeV or more.  A lower value of the cut-off energy implies a smaller UHE tail of the flux, and a lower contribution of it as a background in the radio array of IceCube-Gen2.  Currently, there is no significant evidence for the existence of a cut-off: a recent analysis using 7.5 years of HESE events~\cite{IceCube:2020wum} strongly disfavors the presence of an exponential cut-off below $370$~TeV and finds no substantial evidence for a cut-off above $1.6$~PeV.  However, the measurement is hampered by the paucity of events in the PeV range.  In our forecasts, we explore different possibilities for the value of the cut-off energy, representative of our present and possible future knowledge of it; see Sections~\ref{sec:discovery_results_nominal_choices} and \ref{sec:discovery_results_impact_prior_cut-off}, and Tables~\ref{tab:analysis_choices_base} and \ref{tab:analysis_choices_alt}.

We model the background flux of high-energy (HE) $\nu_\alpha + \bar{\nu}_\alpha$ as an exponentially suppressed power law, \ie,
\begin{equation}
 \label{equ:powerlaw_cutoff}
 \Phi_{\nu_\alpha+\bar{\nu}_\alpha}^{\rm HE} (E_\nu)
 =
 f_{\alpha, \oplus}
 \Phi_0^{\rm HE}
 \left( \frac{E_\nu}{100~\rm{TeV}}\right)^{-\gamma}
 e^{- \frac{E_\nu}{E_{\nu, {\rm cut}}^{\rm HE}}},
\end{equation}
where $f_{\alpha, \oplus}$ is the ratio of the flux of $\nu_\alpha + \bar{\nu}_\alpha$ to the all-flavor flux and $\Phi_0^{\rm HE}$ is the normalization of the all-flavor flux.  The flux shape in \equ{powerlaw_cutoff} is the same one used by searches for an exponential suppression performed by the IceCube Collaboration, \eg, in \Refes~\cite{IceCube:2020wum, IceCube:2021uhz}. Equation~(\ref{equ:powerlaw_cutoff}) makes the typical simplifying assumption that the fluxes of neutrinos of all flavors share common values of $\gamma$ and $E_{\nu, {\rm cut}}^{\rm HE}$, and that the flux of $\nu_\alpha$ and $\bar{\nu}_\alpha$ are equal, \ie, $\Phi_{\nu_\alpha} = \Phi_{\bar{\nu}_\alpha} = \Phi_{\nu_\alpha + \bar{\nu}_\alpha} / 2$, which is expected from neutrino production in $pp$ interactions~\cite{Kelner:2006tc} and, at high energies, in $p\gamma$ interactions~\cite{Mucke:1999yb, Hummer:2010vx, Morejon:2019pfu}.

For the flavor composition, $f_{\alpha, \oplus}$, in \equ{powerlaw_cutoff}, we adopt the canonical scenario where high-energy neutrinos come from the decay of pions produced in $pp$ and $p\gamma$ interactions in astrophysical sources (S).  Thus, at production, the flavor composition is approximately $\left( f_{e, {\rm S}}, f_{\mu, {\rm S}}, f_{\tau, {\rm S}} \right) = (1/3, 2/3, 0)$.  Oscillations en route to Earth change the flavor composition into $f_{\alpha, \oplus} = \sum_{\beta}^{e, \mu, \tau} P_{\beta\alpha} f_{\beta, {\rm S}}$, where $P_{\beta\alpha} \equiv \sum_{i=1}^3 \lvert U_{\alpha i} \rvert^2 \lvert U_{\beta i} \rvert^2$ is the average flavor-transition probability for $\nu_\alpha \to \nu_\beta$~\cite{Pakvasa:2008nx}, and $\mathbf{U}$ is the Pontecorvo-Maki-Nagawa-Sakata (PMNS) mixing matrix.  We evaluate the PMNS matrix using the present-day best-fit values of the mixing parameters from the {\sc NuFit} 5.0 global fit to oscillation data~\cite{Esteban:2020cvm, NuFit_5.0}.  This yields flavor ratios at Earth ($\oplus$) close to equipartition~\cite{Bustamante:2015waa, Song:2020nfh}, \ie,
\begin{equation}
 \left( f_{e,\oplus}, f_{\mu,\oplus}, f_{\tau,\oplus} \right) = (0.298:0.359:0.342) \;.
\end{equation}
These are the flavor ratios that we use to evaluate \equ{powerlaw_cutoff}.  (These are also the flavor ratios with which we build the benchmark UHE neutrino flux models that originally lacked detailed flavor composition; see \Refe~\cite{Valera:2022ylt} for details.)  We neglect uncertainties on $f_{\alpha,\oplus}$ that stem from uncertainties in the mixing parameters.  References~\cite{Song:2020nfh, Valera:2022ylt} showed that by the time that IceCube-Gen2 is operating, in the 2030s, precise measurement of the mixing parameters in upcoming oscillation experiments DUNE~\cite{DUNE:2020lwj}, Hyper-Kamiokande~\cite{Hyper-Kamiokande:2018ofw}, JUNO~\cite{JUNO:2015zny}, and the IceCube Upgrade~\cite{Ishihara:2019aao}, will have rendered the uncertainty on the predicted values of $f_{\alpha, \oplus}$ negligible.  (The issue of inferring the flavor composition at the neutrino sources is related, but separate~\cite{Bustamante:2019sdb, Song:2020nfh}.)

For the all-flavor flux normalization, $\Phi_0^{\rm HE}$, and the spectral index, $\gamma$, in \equ{powerlaw_cutoff}, we consider three possibilities based on their best-fit values reported in IceCube analyses (see also Section~\ref{sec:discovery_results_nominal_choices}, and Tables~\ref{tab:analysis_choices_base} and \ref{tab:analysis_choices_alt}):
\begin{itemize}
 \item
  {\bf Hard flux motivated by the 9.5-year through-going $\nu_\mu$ analysis~\cite{IceCube:2021uhz}:}  We set $\Phi_0^{\rm HE} = \Phi_{\nu_\mu+\bar{\nu}_\mu, 0} / f_{\mu, \oplus}$, where $\Phi_{\nu_\mu+\bar{\nu}_\mu, 0} = 1.44 \times 10^{-18}$~GeV$^{-1}$~cm$^{-2}$~s$^{-1}$~sr$^{-1}$ is the best-fit value of the normalization of the $\nu_\mu + \bar{\nu}_\mu$ flux in \Refe~\cite{IceCube:2021uhz}, and $\gamma = 2.37$.  This is the hardest background neutrino flux that we consider: it induces the largest background event rate.  To be conservative, we adopt it as our baseline analysis choice.  (Without a cut-off, \ie, for $E_{\nu, {\rm cut}}^{\rm HE} \to \infty$, this background matches flux model 2.)
 \item
  {\bf Intermediate flux motivated by the 9.5-year through-going $\nu_\mu$ analysis:}  We keep the same normalization as for the case of the hard flux above, but change the spectral index to $\gamma = 2.50$.
 \item
  {\bf Soft flux motivated by the 7.5-year HESE analysis~\cite{IceCube:2020wum}:}  We set $\Phi_0^{\rm HE} = 6.37 \times 10^{-18}$~GeV$^{-1}$~cm$^{-2}$~s$^{-1}$~sr$^{-1}$ and $\gamma = 2.87$.  This is the softest background neutrino flux that we consider: it induces the smallest background event rate.  (Without a cut-off, \ie, for $E_{\nu, {\rm cut}}^{\rm HE} \to \infty$, this background matches flux model 1.)
\end{itemize}
We make the reasonable assumption that, by the time that IceCube-Gen2 is operating the values of $\Phi_{0}^{\rm HE}$ and $\gamma$ will be known precisely from measurements in TeV--PeV neutrino telescopes~\cite{Ackermann:2022rqc}.  Thus, in our forecasts we neglect the uncertainty on their values, and use only their present-day best-fit values.  Below, Sections~\ref{sec:discovery_results_nominal_results} and \ref{sec:model_separation_results} show results for our baseline choice of a hard background high-energy neutrino flux; Section~\ref{sec:discovery_results_impact_neutrino_bg} and Appendix~\ref{sec:appendix_ic_uhe_tail_bg} show results for the two alternative choices. 

For the cut-off energy, $E_{\nu, {\rm cut}}^{\rm HE}$, in \equ{powerlaw_cutoff}, we assume that its value lies between $10^7$~GeV and $10^{12}$~GeV.  For our conservative baseline forecasts, we assume complete ignorance of its value in the statistical analysis to reflect the present-day scenario; see Section~\ref{sec:discovery_results_nominal_choices}  and Table~\ref{tab:analysis_choices_base}. The lack of knowledge of the size and shape of the UHE tail of the high-energy neutrino flux encumbers the discovery of benchmark UHE neutrino flux models 3--12 and the separation between them.  For our forecasts made with alternative analysis choices, we assume limited and precise knowledge of the value of $E_{\nu, {\rm cut}}^{\rm HE}$; see Section~\ref{sec:discovery_results_impact_prior_cut-off}.  There, we show that our current ignorance of $E_{\nu, {\rm cut}}^{\rm HE}$ erodes, but does not destroy, the potential to discover benchmark flux models.

Some of the benchmark UHE neutrino flux models we consider predict a sizable flux of neutrinos at $\lesssim 10$~PeV, around the high-energy end of current IceCube measurements; this is the case for models 6 and 8--12 (see \figu{benchmark_spectra}).  In these cases, adding the high-energy tail of the IceCube flux may naively seem to overshoot present-day IceCube flux measurements below 10~PeV.  We argue that this is not necessarily the case: in the 1--10~PeV energy range, present-day IceCube measurements are rather limited, with only a handful of events detected so far.  This is reflected in the fluxes inferred from the IceCube HESE and through-going muon analyses in \figu{benchmark_spectra}: they stop at about 2~PeV and 5~PeV, respectively, and their allowed bands are rather wide.  This allows for additional flux components to coexist with the high-energy tail of the IceCube flux, like the flux models that we study.

We compute event rates induced by the UHE tail of the IceCube high-energy neutrino flux using the same methods introduced in Section~\ref{sec:event_rate_benchmarks}.


\subsubsection{Air-shower cores}
\label{sec:background_cores}

In addition to the two sources of background described above, a likely third one is the background from air-shower cores~\cite{Besson:2021wmj, DeKockere:2022bto,ryan_rice_smith_2022_6785120}.  These are cores of particle showers initiated by cosmic-ray interactions in the atmosphere, that develop downwards and penetrate the ice, where they may trigger detectable Askaryan radiation.  Reflection layers in the deep ice~\cite{ryan_rice_smith_2022_6785120} may then reflect the radiation upwards, resulting in signals that mimic those expected from neutrinos.  Presently, the estimates of the size and shape of this background are uncertain.  (Reference~\cite{ryan_rice_smith_2022_6785120} contains early results on this front.)  Thus, we do not account for it in our forecasts below.  Nevertheless, as in \Refe~\cite{Valera:2022ylt}, we point out that characterizing the background of air-shower cores, and possibly minimizing its effect, is a pressing issue in assessing the science reach of upcoming in-ice UHE neutrino telescopes.


\section{Diffuse flux discovery potential}
\label{sec:discovery_potential}

Below, we answer the question of how long it would take to discover benchmark flux models 1--12 (see \figu{benchmark_spectra}) in the radio array of IceCube-Gen2.  The UHE neutrino diffuse flux discovery potential of the detector represents its ability to distinguish between a signal induced by a diffuse UHE neutrino flux model plus background, \ie, the {\it signal hypothesis}, {\it vs.}~a signal induced by background only, \ie, the {\it background-only hypothesis}.  We compute the discovery potential via a Bayesian statistical comparison of the two hypotheses, accounting for random statistical fluctuations in the predicted rate of detected events and for uncertainties in key analysis ingredients.  Later, in Section~\ref{sec:model_separation}, we use similar methods to distinguish between different neutrino flux models, \ie, between different signal hypotheses.

Because of the degeneracy between the neutrino flux and the $\nu N$ cross section (Section~\ref{sec:propagation}), and because both the event rates induced by the signal---\ie, one of the flux models (Section~\ref{sec:fluxes})---and the background---\ie, atmospheric muons and the UHE tail of the IceCube high-energy neutrino flux (Section~\ref{sec:background})---scale roughly linearly with the cross section [see \equ{event_rate_simple}], in our treatment below we account for the uncertainty on the cross section when computing the flux discovery potential.  Reference~\cite{Valera:2022ylt} used similar methods to address the related issue of simultaneously measuring the UHE neutrino flux and the UHE $\nu N$ cross section.  


\subsection{Statistical analysis}
\label{sec:discovery_stat_analysis}

For a given choice of UHE neutrino flux model, $\mathcal{M}_{\rm UHE}$, out of models 3--12 in \figu{benchmark_spectra}, and for a given choice of the background UHE tail of the IceCube high-energy neutrino flux (Section~\ref{sec:background_nu}), $\mathcal{M}_{\rm HE}$, we quantify the discovery potential on the basis of a likelihood function binned in reconstructed shower energy and direction.  For the signal hypothesis (s+bg), this is
\begin{equation}
 \label{equ:likelihood_s_bg}
 \mathcal{L}_{\mathcal{M}_{\rm UHE}, \mathcal{M}_{\rm HE}}^{({\rm s+bg})}(\boldsymbol\theta)
 =
 \prod_{i=1}^{N_{E_{\rm sh}^{\rm rec}}} \prod_{j=1}^{N_{\theta_z^{\rm rec}}}
 \mathcal{L}_{{\mathcal{M}_{\rm UHE}, \mathcal{M}_{\rm HE}, ij}}^{({\rm s+bg})}(\boldsymbol\theta) \;,
\end{equation}
and for the background-only hypothesis (bg), this is
\begin{equation}
 \label{equ:likelihood_bg}
 \mathcal{L}_{\mathcal{M}_{\rm HE}}^{({\rm bg})}(\boldsymbol\theta)
 =
 \prod_{i=1}^{N_{E_{\rm sh}^{\rm rec}}} \prod_{j=1}^{N_{\theta_z^{\rm rec}}}
 \mathcal{L}_{{\mathcal{M}_{\rm HE}, ij}}^{({\rm bg})}(\boldsymbol\theta) \;,
\end{equation}
In Eqs.~(\ref{equ:likelihood_s_bg}) and (\ref{equ:likelihood_bg}), $\boldsymbol\theta \equiv (\log_{10} f_\sigma, \log_{10} (E_{\nu, {\rm cut}}^{\rm HE} / {\rm GeV}))$ represents the free parameters on which the neutrino-induced event rate depends, \ie, the $\nu N$ cross section (Section~\ref{sec:propagation}), $f_\sigma \equiv \sigma_{\nu N}/\sigma_{\nu N}^{\rm std}$, and the cut-off energy of the background IceCube high-energy neutrino flux (Section~\ref{sec:background_nu}), $E_{\nu, {\rm cut}}^{\rm HE}$.  The number of bins of $E_{\rm sh}^{\rm rec}$ is $N_{E_{\rm sh}^{\rm rec}}$ and the number of bins of $\cos \theta_{z}^{\rm rec}$ is $N_{\theta_z^{\rm rec}}$.  (Because flux models 1 and 2 represent a particular realization of the background UHE tail of the IceCube high-energy neutrino flux---\ie, one where the cut-off energy $E_{\nu, {\rm cut}}^{\rm HE} \to \infty$, we forecast their discovery separately from models 3--12, in Section~\ref{sec:discovery_results_uhe_ic_tail}.  Specifically, for them we only consider as background the atmospheric muons.)

The total likelihood in Eqs.~(\ref{equ:likelihood_s_bg}) and (\ref{equ:likelihood_bg}) is the product of partial likelihoods over all bins of reconstructed energy and direction.  The partial likelihood in bin $ij$ compares the predicted (pred) average event rate in the radio component of IceCube-Gen2, computed following the procedure in Section~\ref{sec:event_rate_benchmarks}, {\it vs.}~a particular realization of the observed (obs) event rate, $N_{{\rm obs},ij}$.  To account for possibly low event rates, we take the partial likelihood to be Poissonian.  For the signal hypothesis, it is
\begin{equation}
 \label{equ:likelihood_partial_s_bg}
 \mathcal{L}_{{\mathcal{M}_{\rm UHE}, \mathcal{M}_{\rm HE}, ij}}^{({\rm s+bg})}(\boldsymbol\theta)
 = \frac{
 N_{{\rm pred},ij}^{({\rm s+bg})}(\boldsymbol\theta)
 ^{N_{{\rm obs},ij}}
 e^{-N_{{\rm pred},ij}^{({\rm s+bg})}(\boldsymbol\theta)}
 }
 {
 N_{{\rm obs},ij}!
 } \;,
\end{equation}
where the predicted event rate, $N_{{\rm pred}, ij}^{({\rm s}+{\rm bg})}$, is due to the UHE neutrino flux model being tested (Section~\ref{sec:fluxes}), $N_{\nu, ij}^{\mathcal{M}_{\rm UHE}}$, the background from the UHE tail of the IceCube high-energy neutrino flux (Section~\ref{sec:background_nu}), $N_{\nu, ij}^{\mathcal{M}_{\rm HE}}$, and the background of atmospheric muons (Section~\ref{sec:background_mu}), $N_{\mu, ij}$, \ie, 
\begin{equation}
 \label{equ:event_rate_pred_s_bg}
 N_{{\rm pred}, ij}^{({\rm s}+{\rm bg})}(\boldsymbol\theta) 
 =
 N_{\nu, ij}^{\mathcal{M}_{\rm UHE}}(f_\sigma)
 +
 N_{\nu, ij}^{\mathcal{M}_{\rm HE}}(f_\sigma, E_{\nu, {\rm cut}}^{\rm HE})
 +
 N_{\mu, ij} \;.
\end{equation}
Similarly, for the background-only hypothesis, the partial likelihood is
\begin{equation}
 \label{equ:likelihood_partial_bg}
 \mathcal{L}_{{\mathcal{M}_{\rm HE}, ij}}^{({\rm bg})}(\boldsymbol\theta)
 = \frac{
 N_{{\rm pred},ij}^{({\rm bg})}(\boldsymbol\theta)
 ^{N_{{\rm obs},ij}}
 e^{-N_{{\rm pred},ij}^{({\rm bg})}(\boldsymbol\theta)}
 }
 {
 N_{{\rm obs},ij}!
 } \;,
\end{equation}
where the predicted event rate, $N_{{\rm pred}, ij}^{({\rm bg})}$, is due solely to the background, \ie,
\begin{equation}
 \label{equ:event_rate_pred_bg}
 N_{{\rm pred}, ij}^{({\rm bg})}(\boldsymbol\theta) 
 =
 N_{\nu, ij}^{\mathcal{M}_{\rm HE}}(f_\sigma, E_{\nu, {\rm cut}}^{\rm HE})
 +
 N_{\mu, ij} \;.
\end{equation}
Thus, the likelihood function in \equ{likelihood_partial_s_bg} represents the probability that the observed event rate is due to the signal hypothesis, computed for a given UHE neutrino flux model $\mathcal{M}_{\rm UHE}$ out of models 3-12, and the likelihood function in \equ{likelihood_partial_bg} represents the probability that it is due to the background-only hypothesis.  Broadly stated, the UHE neutrino flux model will be more easily discoverable when the former is higher than the latter. 

When computing the likelihood, Eqs.~(\ref{equ:likelihood_partial_s_bg})--(\ref{equ:event_rate_pred_bg}), we sample the value of the observed event rate, $N_{{\rm obs},ij}$, at random from a Poisson distribution with central value equal to $N_{{\rm pred}, ij}^{({\rm s}+{\rm bg})}(\boldsymbol\theta^\star)$, where $\boldsymbol\theta^\star$ represents the set of real parameter values of the model, \ie, $f_\sigma = 1$ and $E_{\nu, {\rm cut}}^{\rm HE}$ fixed at a value between $10^7$ and $10^{12}$~GeV, with the specific value depending on the scenario we are testing.  To account for statistical fluctuations, we perform the sampling many times.  We elaborate on this below.

For the signal hypothesis, the posterior probability distribution associated to the likelihood in \equ{likelihood_s_bg} is
\begin{equation}
 \label{equ:posterior_s_bg}
 \mathcal{P}_{\mathcal{M}_{\rm UHE}, \mathcal{M}_{\rm HE}}^{({\rm s+bg})}(\boldsymbol\theta)
 =
 \frac{\mathcal{L}_{\mathcal{M}_{\rm UHE}, \mathcal{M}_{\rm HE}}^{({\rm s+bg})}(\boldsymbol\theta)
 \pi(\boldsymbol\theta)}
 {\mathcal{Z}_{\mathcal{M}_{\rm UHE}, \mathcal{M}_{\rm HE}}^{({\rm s+bg})}} \;,
\end{equation}
where $\pi(\boldsymbol\theta) \equiv \pi(\log_{10} f_\sigma) \pi(\log_{10}(E_{\nu, {\rm cut}}^{\rm HE}/{\rm GeV}))$ is the prior on the model parameters; we expand on them in Section~\ref{sec:discovery_results}.  The normalization factor in \equ{posterior_s_bg},
\begin{equation}
 \label{equ:evidence_s_bg}
 \mathcal{Z}_{\mathcal{M}_{\rm UHE}, \mathcal{M}_{\rm HE}}^{({\rm s+bg})}
 =
 \int d\boldsymbol\theta \mathcal{L}_{\mathcal{M}_{\rm UHE}, \mathcal{M}_{\rm HE}}^{({\rm s+bg})}(\boldsymbol\theta) \pi(\boldsymbol\theta) \;,
\end{equation}
is the statistical evidence, \ie, the likelihood fully marginalized over the space of model parameters.  For the background-only hypothesis, the posterior and evidence, $\mathcal{P}_{\mathcal{M}_{\rm HE}}^{({\rm bg})}$ and $\mathcal{Z}_{\mathcal{M}_{\rm HE}}^{({\rm bg})}$, are computed as in Eqs.~(\ref{equ:posterior_s_bg}) and (\ref{equ:evidence_s_bg}), but using $\mathcal{L}_{\mathcal{M}_{\rm HE}}^{({\rm bg})}$ instead, \ie,
\begin{equation}
 \label{equ:posterior_bg}
 \mathcal{P}_{\mathcal{M}_{\rm HE}}^{({\rm bg})}(\boldsymbol\theta)
 =
 \frac{\mathcal{L}_{\mathcal{M}_{\rm HE}}^{({\rm bg})}(\boldsymbol\theta)
 \pi(\boldsymbol\theta)}
 {\mathcal{Z}_{\mathcal{M}_{\rm HE}}^{({\rm bg})}} \;
\end{equation}
and
\begin{equation}
 \label{equ:evidence_bg}
 \mathcal{Z}_{\mathcal{M}_{\rm HE}}^{({\rm bg})}
 =
 \int d\boldsymbol\theta \mathcal{L}_{\mathcal{M}_{\rm HE}}^{({\rm bg})}(\boldsymbol\theta) \pi(\boldsymbol\theta) \;.
\end{equation}

We report the discovery potential of the UHE neutrino flux model $\mathcal{M}_{\rm UHE}$ via the ratio of the statistical evidence of the signal and background-only hypotheses, \ie, the {\it discovery} Bayes factor, 
\begin{equation}
\label{equ:bayes_s_bg}
 \mathcal{B}_{\mathcal{M}_{\rm UHE}, \mathcal{M}_{\rm HE}}^{\rm disc}
 =
 \frac{\mathcal{Z}_{\mathcal{M}_{\rm UHE}, \mathcal{M}_{\rm HE}}^{({\rm s}+{\rm bg})}}
 {\mathcal{Z}_{\mathcal{M}_{\rm HE}}^{({\rm bg})}} \;.
\end{equation}
The Bayes factor represents the preference for the signal hypothesis over the background-only hypothesis.  Following convention, we ascribe qualitative significance to its value using Jeffreys' table~\cite{jeffreys1998theory}: $10^0 \leq \mathcal{B} < 10^{0.5}$ represents negligible evidence for the signal hypothesis; $10^{0.5} \leq \mathcal{B} < 10^1$, moderate evidence; $10^1 \leq \mathcal{B} < 10^{1.5}$, strong evidence; $10^{1.5} \leq \mathcal{B} < 10^2$, very strong evidence; and $\mathcal{B} \geq 10^2$, decisive evidence.  In our discussion below we focus mainly on flux discovery with decisive evidence.

To compute the statistical evidence, we use {\sc UltraNest}~\cite{Ultranest}, an efficient Bayesian nested-importance sampler~\cite{Buchner:2014, Buchner:2017}.  {\sc UltraNest} integrates Eqs.~(\ref{equ:evidence_s_bg}) and (\ref{equ:evidence_bg}) numerically and reports the result of each as $\ln\mathcal{Z} \pm \Delta\ln\mathcal{Z}$, where $\Delta\ln\mathcal{Z}$ is the numerical error of the integral.  With it, we compute the numerical error on the Bayes factor in \equ{bayes_s_bg} as
\begin{eqnarray}
\label{equ:bayes_err}
 && 
 \Delta \ln\mathcal{B}_{\mathcal{M}_{\rm UHE}, \mathcal{M}_{\rm HE}}^{\rm disc}
 =
 \nonumber \\
 && \qquad
 \sqrt{
  \left( \Delta \ln \mathcal{Z}_{\mathcal{M}_{\rm UHE}, \mathcal{M}_{\rm HE}}^{({\rm s}+{\rm bg})} \right)^2
  +
  \left( \Delta \ln \mathcal{Z}_{\mathcal{M}_{\rm HE}}^{({\rm bg})} \right)^2
  }
  \;.
\end{eqnarray}

For a given choice of UHE neutrino flux model, $\mathcal{M}_{\rm UHE}$, and background high-energy neutrino flux, $\mathcal{M}_{\rm HE}$, we account for the effect of random statistical fluctuations in the observed event rate by repeating the above procedure $N_{\rm samples} = 10^4$ times.  Each time, we draw a different random realization of the distribution of the observed event rate across all bins, $N_{{\rm obs}, ij}$, as explained above.  For each random realization, we compute the discovery Bayes factor and its error as in Eqs.~(\ref{equ:evidence_s_bg}) and (\ref{equ:bayes_err}).

Below, when presenting results, the value of the Bayes factor that we report is always the average over all random realizations, $\langle \log_{10} \mathcal{B}_{\mathcal{M}_{\rm UHE}, \mathcal{M}_{\rm HE}}^{\rm disc} \rangle$.  Specifically, it is the weighted arithmetic mean of $\log_{10} \mathcal{B}_{\mathcal{M}_{\rm UHE}, \mathcal{M}_{\rm HE}}^{\rm disc}$ (equal to the weighted geometric mean of $\mathcal{B}_{\mathcal{M}_{\rm UHE}, \mathcal{M}_{\rm HE}}^{\rm disc}$), \ie,
\begin{eqnarray}
 \label{equ:bayes_factor_avg}
 &&
 \langle \log_{10} \mathcal{B}_{\mathcal{M}_{\rm UHE}, \mathcal{M}_{\rm HE}}^{\rm disc} \rangle
 = 
 \nonumber \\
 && \qquad
 \frac{1}{\ln 10}
 \frac{\sum_{i=1}^{N_{\rm samples}} w_i \ln \mathcal{B}_{\mathcal{M}_{\rm UHE}, \mathcal{M}_{\rm HE}, i}^{\rm disc}}
 {\sum_{i=1}^{N_{\rm samples}} w_i} \;,
\end{eqnarray}
where $\mathcal{B}_{\mathcal{M}_{\rm UHE}, \mathcal{M}_{\rm HE}, i}^{\rm disc}$ is the Bayes factor computed in the $i$-th random realization and the weight, $w_i$, is
\begin{equation}
 w_i = \left(\Delta\ln\mathcal{B}_{\mathcal{M}_{\rm UHE}, \mathcal{M}_{\rm HE}, i}^{\rm disc}\right)^{-2} \;,
\end{equation}
where $\Delta\ln\mathcal{B}_{\mathcal{M}_{\rm UHE}, \mathcal{M}_{\rm HE}, i}^{\rm disc}$ is the error on the Bayes factor computed in the same realization.  (Computing the weighted arithmetic mean of $\log_{10} \mathcal{B}_{\mathcal{M}_{\rm UHE}, \mathcal{M}_{\rm HE}}^{\rm disc}$, instead of the weighted arithmetic mean of $\mathcal{B}_{\mathcal{M}_{\rm UHE}, \mathcal{M}_{\rm HE}}^{\rm disc}$, avoids the bias that the mean Bayes factor would otherwise have towards non-representative large values resulting from the pull of the relatively few random realizations that yield significantly larger Bayes factors.)  In Appendix~\ref{sec:stat_sig_bayes_factor} we comment on the spread of the distribution of values of the Bayes factor obtained from the different random realizations of the observed event rate, and how much they may deviate from the mean value.

In some cases scenarios, the sum of the UHE tail of the IceCube high-energy neutrino flux plus the UHE flux model 3--12 may exceed the present-day experimental upper limits from IceCube and Auger shown in \figu{benchmark_spectra}.  However, those limits were derived assuming a fixed value of the UHE neutrino-nucleon cross section while in our analysis the cross section is a free parameter with a wide prior around its standard prediction.  This means that for choices of the cross section that are different from the ones used by IceCube and Auger in computing their limits, those limits need not apply to our results.  Beyond that, we do not attempt to incorporate the lack of observation of UHE neutrinos by IceCube and Auger in our analysis, since doing so would required modeling those two detectors to compute event rates for each of our flux predictions, which is beyond the scope of this work.


\subsection{Results}
\label{sec:discovery_results}


\subsubsection{Baseline analysis choices}
\label{sec:discovery_results_nominal_choices}

\begingroup
\squeezetable
\begin{table*}[tbp]
\begin{ruledtabular}
\centering
\renewcommand{\arraystretch}{1.3}
\begin{tabular}{lclc}
\multicolumn{1}{c}{\multirow{3}{*}{Parameter}}
&
\multicolumn{3}{c}{Baseline analysis choice}
\\
\cline{2-4} 
\multicolumn{1}{c}{}
&
\multirow{2}{*}{Description}
&
\multicolumn{2}{c}{Location in text of results}
\\
\cline{3-4}
\multicolumn{1}{c}{}
&
& \multicolumn{1}{c}{Flux discovery}
& \multicolumn{1}{c}{Flux separation}
\\
\hline\\[-0.3cm]
Atmospheric muon background (Section~\ref{sec:background_mu})
& {\sc Sybill 2.3c}, mitigated by surface veto
&
\begin{tabular}[c]{@{}c@{}}Section~\ref{sec:discovery_results_nominal_results}
\\ \figu{bayes_factor_hard}\end{tabular}
& \begin{tabular}[c]{@{}c@{}}Section~\ref{sec:model_separation_results}
\\ \figu{confusion_hard}\end{tabular}
\\[0.30cm]
\begin{tabular}[c]{@{}l@{}}Background UHE tail of the IceCube high-\\energy neutrino flux, $\mathcal{M}_{\rm HE}$ (Section~\ref{sec:background_nu})\end{tabular}
& \begin{tabular}[c]{@{}c@{}}\textit{Hard flux}: UHE extrapolation of the flux from 9.5-yr\\ IceCube track analysis ($\gamma = 2.37$)\end{tabular}
& \begin{tabular}[c]{@{}c@{}}Section~\ref{sec:discovery_results_nominal_results}\\ \figu{bayes_factor_hard}\end{tabular} 
& \begin{tabular}[c]{@{}c@{}}Section~\ref{sec:model_separation_results}\\ \figu{confusion_hard}\end{tabular} 
\\[0.30cm]
\begin{tabular}[c]{@{}l@{}}Prior on the cut-off energy of the UHE tail of the\\IceCube high-energy $\nu$ flux, $E_{\nu, {\rm cut}}^{\rm HE}$ (Section~\ref{sec:background_nu})\end{tabular} 
& \begin{tabular}[c]{@{}c@{}}Flat prior on $\log_{10}(E_{\nu, {\rm cut}}^{\rm HE}/{\rm GeV})$ between 5 and 12,\\ followed by averaging of Bayes factor over $E_{\nu, {\rm cut}}^{\rm HE}$\end{tabular} 
& \begin{tabular}[c]{@{}c@{}}Section~\ref{sec:discovery_results_nominal_results}\\ \figu{bayes_factor_hard}\end{tabular} 
& \begin{tabular}[c]{@{}c@{}}Section~\ref{sec:model_separation_results}\\ \figu{confusion_hard}\end{tabular}
\\[0.30cm]
\begin{tabular}[c]{@{}l@{}}Prior on the neutrino-nucleon cross section,\\$f_\sigma$ (Section~\ref{sec:propagation})\end{tabular}
& Flat prior on $\log_{10} f_\sigma$ from -1 to 2
& \begin{tabular}[c]{@{}c@{}}Section~\ref{sec:discovery_results_nominal_results}\\ \figu{bayes_factor_hard}\end{tabular}
& \begin{tabular}[c]{@{}c@{}}Section~\ref{sec:model_separation_results}\\ \figu{confusion_hard}\end{tabular}
\\[0.30cm]
\begin{tabular}[c]{@{}l@{}}Detector energy resolution, $\sigma_\epsilon$, and\\angular resolution, $\sigma_{\theta_z}$ (Section~\ref{sec:event_rate_benchmarks})\end{tabular}
& \begin{tabular}[c]{@{}c@{}}$\sigma_\epsilon = 0.1$\\ $\sigma_{\theta_z} = 2^\circ$\end{tabular}
& \begin{tabular}[c]{@{}c@{}}Section~\ref{sec:discovery_results_nominal_results}\\ \figu{bayes_factor_hard}\end{tabular} & \begin{tabular}[c]{@{}c@{}}Section~\ref{sec:model_separation_results}\\ \figu{confusion_hard}\end{tabular}
\end{tabular}
\caption{\label{tab:analysis_choices_base}Baseline analysis choices used to forecast the flux discovery potential (Section~\ref{sec:discovery_potential}) and flux model separation (Section~\ref{sec:model_separation}) in the radio array of IceCube-Gen2, and location in the text of corresponding results.  See Section~\ref{sec:discovery_results_nominal_choices} for an overview and Table~\ref{tab:analysis_choices_alt} for alternative choices.}
\end{ruledtabular}
\end{table*}
\endgroup

\begingroup
\squeezetable
\begin{table*}[tbp]
\begin{ruledtabular}
\centering
\renewcommand{\arraystretch}{1.3}
\begin{tabular}{lclc}
\multicolumn{1}{c}{\multirow{3}{*}{Parameter}}
&
\multicolumn{3}{c}{Alternative analysis choices}
\\
\cline{2-4} 
\multicolumn{1}{c}{}
&
\multirow{2}{*}{Description}
&
\multicolumn{2}{c}{Location in text of results}
\\
\cline{3-4}
\multicolumn{1}{c}{}
&
& \multicolumn{1}{c}{Flux discovery}
& \multicolumn{1}{c}{Flux separation}
\\
\hline\\[-0.3cm]
Atmospheric muon background (Section~\ref{sec:background_mu})
& Baseline $\times 10$, $\times 100$, $\times 1000$
&
\begin{tabular}[c]{@{}c@{}}Section~\ref{sec:discovery_results_impact_muon_bg}
\\ \figu{large_mu_bkg}\end{tabular}
& --
\\[0.30cm]
\begin{tabular}[c]{@{}l@{}}Background UHE tail of the IceCube high-\\energy neutrino flux, $\mathcal{M}_{\rm HE}$ (Section~\ref{sec:background_nu})\end{tabular}
& \begin{tabular}[c]{@{}c@{}}\textit{Soft flux}: UHE extrapolation of the 7.5-yr IceCube\\HESE flux ($\gamma = 2.87$) /  \textit{Intermediate flux}: $\gamma = 2.50$\end{tabular}
& \begin{tabular}[c]{@{}c@{}}Section~\ref{sec:discovery_results_impact_neutrino_bg}\\ Figs~\ref{fig:bayes_other_nu_cases} and \ref{fig:bayes_other_nu_cases_app}\end{tabular} 
& \begin{tabular}[c]{@{}c@{}}Appendix~\ref{sec:appendix_ic_uhe_tail_bg}\\ \figu{confusion_other_nu_cases}\end{tabular} 
\\[0.30cm]
\begin{tabular}[c]{@{}l@{}}Prior on the cut-off energy of the UHE tail of the\\IceCube high-energy $\nu$ flux, $E_{\nu, {\rm cut}}^{\rm HE}$ (Section~\ref{sec:background_nu})\end{tabular} 
& \begin{tabular}[c]{@{}c@{}}Wide Gaussian prior and delta-function prior\\centered on $\log_{10}( E_{\nu, {\rm cut}}^{\rm HE}/{\rm GeV}) = 8$\end{tabular}  
& \begin{tabular}[c]{@{}c@{}}Section~\ref{sec:discovery_results_impact_prior_cut-off}\\ \figu{bayes_factor_ecut_cases}\end{tabular} 
& --
\\[0.30cm]
\begin{tabular}[c]{@{}l@{}}Prior on the neutrino-nucleon cross section,\\$f_\sigma$ (Section~\ref{sec:propagation})\end{tabular}
& \begin{tabular}[c]{@{}c@{}}Wide Gaussian prior and delta-function prior centered \\ on central value of BGR18 prediction, $\log_{10} f_\sigma = 0$\end{tabular} 
& \begin{tabular}[c]{@{}c@{}}Section~\ref{sec:discovery_results_impact_cross_section}\\ \figu{bayes_prior_choices}\end{tabular}
& --
\\[0.30cm]
\begin{tabular}[c]{@{}l@{}}Detector energy resolution, $\sigma_\epsilon$, and\\angular resolution, $\sigma_{\theta_z}$ (Section~\ref{sec:event_rate_benchmarks})\end{tabular}
& \begin{tabular}[c]{@{}l@{}}$\sigma_\epsilon = 0.5, 1.0$\\ $\sigma_{\theta_z} = 5^\circ, 10^\circ$\end{tabular}
& \begin{tabular}[c]{@{}c@{}}Section~\ref{sec:discovery_results_impact_resolution}\\ Figs.~\ref{fig:bayes_factor_res_main}, \ref{fig:bayes_energy_resolution}, \ref{fig:bayes_angular_resolution}\end{tabular} & \begin{tabular}[c]{@{}c@{}}Appendix~\ref{sec:appendix_detector_resolution}\\ Figs.~\ref{fig:confusion_energy_resolution} and \ref{fig:confusion_angular_resolution}\end{tabular}
\end{tabular}
\caption{\label{tab:analysis_choices_alt}Alternative analysis choices used to forecast the flux discovery potential (Section~\ref{sec:discovery_potential}) and flux model separation (Section~\ref{sec:model_separation}) in the radio array of IceCube-Gen2, and location in the text of associated content.  See Section~\ref{sec:discovery_results_nominal_choices} for an overview and Table~\ref{tab:analysis_choices_base} for baseline choices.}
\end{ruledtabular}
\end{table*}
\endgroup

Section~\ref{sec:discovery_results_nominal_results} shows our main results for the flux discovery potential.  To produce them, we adopt baseline analysis choices for the atmospheric muon background, the high-energy neutrino background, including the prior on the value of its cut-off energy, the detector energy and angular resolution, and the prior on the neutrino-nucleon cross section.  (Later, in Section~\ref{sec:model_separation}, when comparing flux models, we keep the same baseline choices.)  {\it Our baseline analysis choices, chosen to be largely conservative, lead to promising results for the flux discovery potential.}  

Sections~\ref{sec:discovery_results_impact_muon_bg}--\ref{sec:discovery_results_impact_cross_section} show results obtained under well-motivated alternative analysis choices.  Different alternative choices may expedite or delay decisive flux discovery, but do not change our main conclusion: fluxes that may be discovered within 20 years of exposure time under the baseline analysis choices remain discoverable.

Tables~\ref{tab:analysis_choices_base} and \ref{tab:analysis_choices_alt}  summarize our baseline and alternative analysis choices, and show where in the text to find associated content.  Below we elaborate on our choices:
\begin{itemize}
 \item
  \textit{Atmospheric muon background}:  Our baseline choice is the event rate computed using the hadronic interaction model {\sc Sybill 2.3c}, mitigated by applying a surface veto~\cite{Glaser:2021hfi}; see Section~\ref{sec:background} for details.  Section~\ref{sec:discovery_results_impact_muon_bg} shows results obtained under a substantially larger atmospheric muon background; with them, the discovery potential shrinks only mildly.
 \item 
  \textit{Background UHE tail of the IceCube high-energy neutrino flux, $\mathcal{M}_{\rm HE}$}:  Our baseline choice is to adopt a flux motivated by the 9.5-year IceCube through-going track analysis~\cite{IceCube:2021uhz} introduced in Section~\ref{sec:background_nu}.  This choice is conservative because, due to its hard spectral index ($\gamma = 2.37$), this flux may extend to higher energies compared to alternative, softer spectra, and so yields a larger background to the discovery of UHE neutrino flux models 3--12.  Section~\ref{sec:discovery_results_impact_neutrino_bg} shows results for the less conservative, softer high-energy neutrino spectra introduced in Section~\ref{sec:background_nu}, motivated by alternative IceCube results; with them, the discovery potential improves appreciably.
 \item
  \textit{Prior on the cut-off energy of the UHE tail of the IceCube high-energy neutrino flux background, $E_{\nu, {\rm cut}}^{\rm HE}$}:  Our baseline choice for $\pi(\log_{10}(E_{\nu, {\rm cut}}^{\rm HE}/{\rm GeV}))$ in \equ{evidence_s_bg} is a flat distribution between $\log_{10}(E_{\nu, {\rm cut}}^{\rm HE}/{\rm GeV}) = 7$ and 12.  In addition, in our baseline analysis we average the mean Bayes factor, \equ{bayes_factor_avg}, over all possible real values of $\log_{10}(E_{\nu, {\rm cut}}^{\rm HE}/{\rm GeV})$, and report the result of that.  These choices of prior and averaging are conservative and reflect our present-day ignorance on the existence and position of a cut-off in the IceCube high-energy neutrino flux.  Section~\ref{sec:discovery_results_impact_neutrino_bg} shows results obtained using alternative, informed priors that reflect possible evidence of a cut-off found in upcoming measurements of the high-energy neutrino flux; with them, the discovery potential improves significantly.
 \item
  \textit{Prior on the neutrino-nucleon cross section, $f_\sigma$}: Our baseline choice for $\pi(\log_{10} f_\sigma)$ in \equ{evidence_s_bg} is a flat distribution between $\log_{10} f_\sigma = -1$ and 2.  This choice is conservative because it ignores the pull from theory towards the central value of the BGR18~\cite{Bertone:2018dse} prediction, \ie, towards $\log_{10} f_\sigma = 0$, when fitting to mock data.  (Nevertheless, we always use $\log_{10} f_\sigma = 0$ as the true value to generate the mock observed event rate against which we fit; see Section~\ref{sec:discovery_stat_analysis}.) Our choice of a baseline wide prior reflects the current lack of direct measurement of the UHE neutrino-nucleon cross section.  Such a wide prior may even encompass new-physics modifications to the cross section (see, \eg, \Refe~\cite{Bustamante:2017xuy}); the fact that it may be possible to discover most flux models even under such a loose prior (\figu{bayes_factor_hard}) is encouraging. Section~\ref{sec:discovery_results_impact_cross_section} shows results for alternative informed priors on $\log_{10} f_\sigma$; their use expedites decisive flux discovery by up to a factor of roughly 3 compared to the baseline expectation, depending on the flux model. 

 \item
  \textit{Detector energy resolution, $\sigma_\epsilon$, and angular resolution, $\sigma_{\theta_z}$}: Our baseline choices for the resolution on the reconstructed shower energy and reconstructed zenith angle are, respectively, $\sigma_\epsilon = 0.1$ and $\sigma_{\theta_z} = 2^\circ$.  These choices are motivated by dedicated simulations~\cite{Glaser:2019rxw, Anker:2019zcx, RNO-G:2021zfm, ARIANNA:2021pzm, Glaser:2022lky, ARA:2021bss, Gaswint:2021smu, Aguilar:2021uzt}; see Section~\ref{sec:event_rate_benchmarks} for details.  The detector energy and angular resolution affect the event rate computed via \equ{spectrum_rec} and determine the energy and angular binning used to compute the likelihood, Eqs.~(\ref{equ:likelihood_s_bg}) and (\ref{equ:likelihood_bg}).  For the baseline energy binning, we use 12 bins equally spaced in logarithmic scale from $E_{\rm sh}^{\rm rec} = 10^7$~GeV to $10^{10}$~GeV.  For the angular binning, we use a single large bin for downgoing events, from $\theta_z^{\rm rec} = 0^\circ$ to $80^\circ$; 10 bins of size $2^\circ$ from $80^\circ$ to $100^\circ$; and two large bins for upgoing events, one from $100^\circ$ to $110^\circ$, and another one from $110^\circ$ to $180^\circ$.  Section~\ref{sec:discovery_results_impact_resolution} shows results for alternative choices of poorer detector resolution, and their associated binning; with them, the discovery potential shrinks for poorer energy resolution---mainly because of features in the energy spectrum become unresolved---and for poorer angular resolution---mainly because the uncertainty in the $\nu N$ cross section is allowed to have a larger impact. 
\end{itemize}

We keep the design of the IceCube-Gen2 radio array fixed to the baseline design of \Refe~\cite{Hallmann:2021kqk}, as described in Section~\ref{sec:ic-gen2}.  We describe the detector response via the energy- and direction-dependent effective volumes generated from dedicated simulations of radio generation, propagation, and detection from \Refe~\cite{Valera:2022ylt}; see Section~\ref{sec:event_rate_benchmarks} and \Refe~\cite{Valera:2022ylt} for details.  We adopt this detector design to make concrete forecasts, but the final design remains under consideration at the time of writing.  

We do not explore alternative detector designs in our forecasts, since doing so requires running intensive simulations for each design choice.  However, we discuss detector-related features that may inform the design of upcoming detectors: the impact of detector energy and angular resolution (Section~\ref{sec:discovery_results_impact_resolution}), the importance of the detector response being sensitive to Earth-skimming neutrinos (Section~\ref{sec:discovery_results_horizontal_events}), and the impact of using an air-shower surface array veto to mitigate the atmospheric muon background (Appendix~\ref{sec:impact_surface_veto}).


\subsubsection{Baseline discovery potential of the benchmark UHE neutrino flux models}
\label{sec:discovery_results_nominal_results}

{\it Even under conservative analysis choices, IceCube-Gen2 should be able to claim decisive evidence for the discovery of most of the benchmark UHE neutrino fluxes models after one decade of operation}. 
\smallskip

Figure~\ref{fig:bayes_factor_hard} shows our main result: the evolution with exposure time of the mean discovery Bayes factor in the radio array of IceCube-Gen2, for the UHE neutrino flux models 1--12 from \figu{benchmark_spectra}, computed as detailed in Section~\ref{sec:discovery_stat_analysis} and under our baseline analysis choices from Section~\ref{sec:discovery_results_nominal_choices}.  The figure reveals promising prospects, in spite of our baseline analysis choices being conservative.  Alternative, less conservative choices of background and priors may hasten discovery; we explore them later.  For flux models 3--12, we include as background of atmospheric muons and the UHE tail of the IceCube high-energy neutrino spectrum; see Section~\ref{sec:discovery_stat_analysis}.  For flux models 1 and 2---unbroken UHE extrapolations of the IceCube TeV--PeV neutrino flux~\cite{IceCube:2020wum, IceCube:2021uhz}---we include only the background of atmospheric muons; see Section~\ref{sec:discovery_results_uhe_ic_tail}. 

Figure~\ref{fig:bayes_factor_hard} sorts flux models 1--12 into three classes, depending on the time it takes for them to be discovered with decisive evidence: models discoverable within 1~year (models 2, 4, 6, 9, 11), models discoverable in 1--10 years  (models 7, 8, 10, 12), and models that are not discoverable within 20 years (models 3 and 5).  Flux model 1, the UHE extrapolation of the 7.5-year IceCube HESE flux~\cite{IceCube:2020wum}, can be discovered with very strong evidence after 20~years.  This classification conveys in detail what \figu{benchmark_spectra} shows roughly: fluxes above the IceCube-Gen2 sensitivity are discoverable within 10 years; fluxes below are not.  Lowering the discovery threshold to ``very strong" or ``strong" evidence expedites flux discovery; however, fluxes that are not discoverable remain as such.

At short exposure times, the predicted event rate, made up of signal plus background events, \equ{event_rate_pred_s_bg}, for most benchmark UHE neutrino flux models is low and observations are compatible with the background-only hypothesis.  (The exception is flux model 4, the highest among all benchmark models, which is compatible with the signal hypothesis even at short exposure times.)  With growing exposure time, the larger predicted event rates enhance the contrast between the alternative hypothesis. Then the observations become more compatible with the signal hypothesis in our statistical analysis; see Section~\ref{sec:discovery_stat_analysis} for details. 

The growth rate of the flux discovery Bayes factor in \figu{bayes_factor_hard} results from the interplay of two factors: the size of the predicted event rate, integrated across all energies and directions, induced by the UHE neutrino flux model---\ie, larger rates lead to larger Bayes factors---and the shape of the event rate induced by the UHE neutrino flux model---\ie, flux models whose energy spectrum peaks at high energies, far from the background concentrated at $E_{\rm sh}^{\rm rec} \lesssim 10^8$~GeV (see \figu{binned_events_all_benchmarks_vs_energy_dep}), lead to larger Bayes factors. This explains the difference in the growth rate of the Bayes factor in \figu{bayes_factor_hard} at short and long exposure times.  We elaborate below.

At short exposure times, \figu{bayes_factor_hard} shows that the Bayes factor grows fast.  There, the growth is dominated by the notable difference between the energy distributions of the event rates induced by the signal and the background; see Fig.~\ref{fig:binned_events_all_benchmarks_vs_energy_dep}.  Differences between their angular distributions are smaller and contribute  weakly to the growth rate; see \figu{binned_events_all_benchmarks_vs_costhzrec}.   Because our analysis is binned in energy and direction (see the likelihood functions in Eqs.~(\ref{equ:likelihood_s_bg}) and (\ref{equ:likelihood_bg})), it is able to resolve differences in the energy distributions of the signal hypothesis {\it vs.}~the background-only hypothesis even if their associated events rate are low, \ie, even at low exposure times.  [For flux models 10 and 12, whose energy spectra peak at lower energies, closer to the background, the Bayes factor grows more slowly because distinguishing between them is harder.  This subtle feature is most clearly seen by comparing flux models 10 and 11: their integrated event rates are similar (see Table~\ref{tab:event_rates_binned}), but the rate for flux model 10 peaks at significantly lower energies than for flux model 11 (see \figu{binned_events_all_benchmarks_vs_energy_dep}).]

At longer exposure times, \figu{bayes_factor_hard} shows that the Bayes factor grows more slowly.  There, the growth rate is dominated by the large difference in integrated event rate induced by the signal and by the background; see Table~\ref{tab:event_rates_binned}.  At large exposure times, the posterior distributions, Eqs.~(\ref{equ:posterior_s_bg}) and (\ref{equ:posterior_bg}), become narrow due to the lessening of the impact of random statistical fluctuations.  As a result, the Bayes factor, \equ{bayes_s_bg}, is dominated by the peak value of the posteriors.  The event rate, $N_{{\rm obs}, ij}$, grows linearly with time for all flux models; see \equ{spectrum_true}.  Therefore, the logarithm of the likelihood function, \equ{likelihood_partial_s_bg}, also grows linearly with time, and so does the logarithm of the discovery Bayes factor, \equ{bayes_s_bg}, as seen in \figu{bayes_factor_hard}.  (For flux model 12, this growth regime is reached beyond 20~years, so it is not seen  in \figu{bayes_factor_hard}.)

\begin{figure}[t!]
 \centering
 \includegraphics[width=\columnwidth]{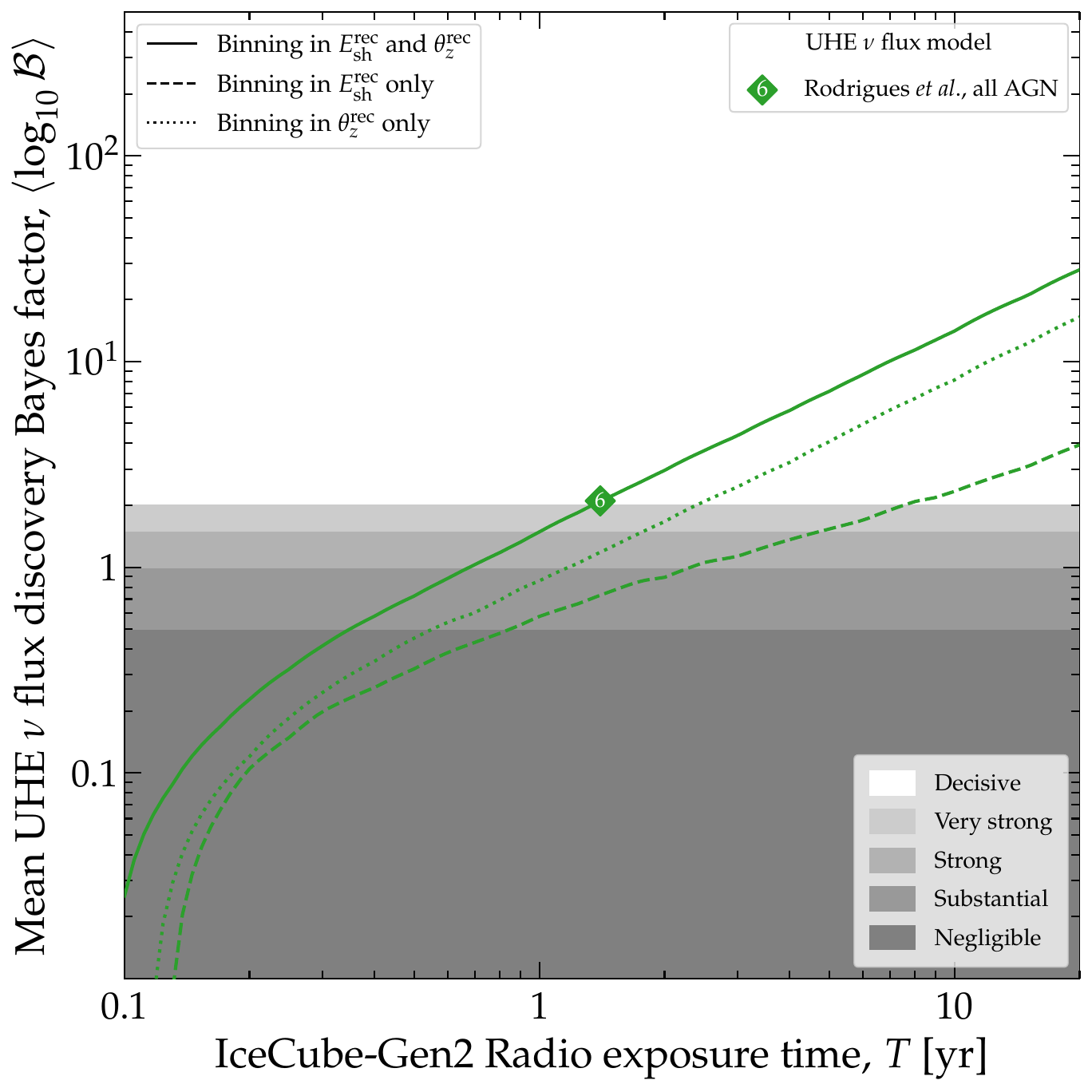}
 \caption{\label{fig:bayes_binning}Impact of binning the detected events only in reconstructed shower energy, $E_{\rm sh}^{\rm rec}$, only in reconstructed direction, $\theta_z^{\rm rec}$, and in both, on the discovery potential of the UHE neutrino flux model 6~\cite{Rodrigues:2020pli}.  Results are for our conservative baseline assumption of a flat prior on the $\nu n$ cross section, $f_\sigma$---representing little to no prior knowledge of the cross section.  \textit{In the absence of precise knowledge of the $\nu N$ cross section, early flux discovery hinges on using both the energy and, especially, angular distribution of events.}  See Section~\ref{sec:discovery_results_nominal_results} for details.}
\end{figure}

Figure~\ref{fig:bayes_binning} illustrates the roles that binning events in reconstructed energy and direction have on the UHE neutrino flux discovery potential.
Binning in energy allows our statistical analysis to distinguish between the energy distributions of events induced by the UHE neutrino flux model {\it vs.}~events induced by the atmospheric muon background---which are concentrated at low energies---and by the UHE tail of the IceCube high-energy neutrino flux---when it has a low cut-off energy, $E_{\nu, {\rm cut}}^{\rm HE}$, compared to the energy at which the UHE neutrino flux model 3--12 peaks.  Binning in direction allows our statistical analysis to break the innate degeneracy between neutrino flux and cross section described in Section~\ref{sec:propagation} and, by doing so, to claim a higher evidence for the discovery of the UHE neutrino flux model 3--12.  This is especially true when using a flat prior on the cross section, \ie, when there is little to no knowledge of the cross section and, thus, a larger degeneracy with the neutrino flux; see Section~\ref{sec:discovery_results_impact_cross_section}.


\subsubsection{Impact of the atmospheric muon background}
\label{sec:discovery_results_impact_muon_bg}

\begin{figure}[t!]
 \centering
 \includegraphics[width=\columnwidth]{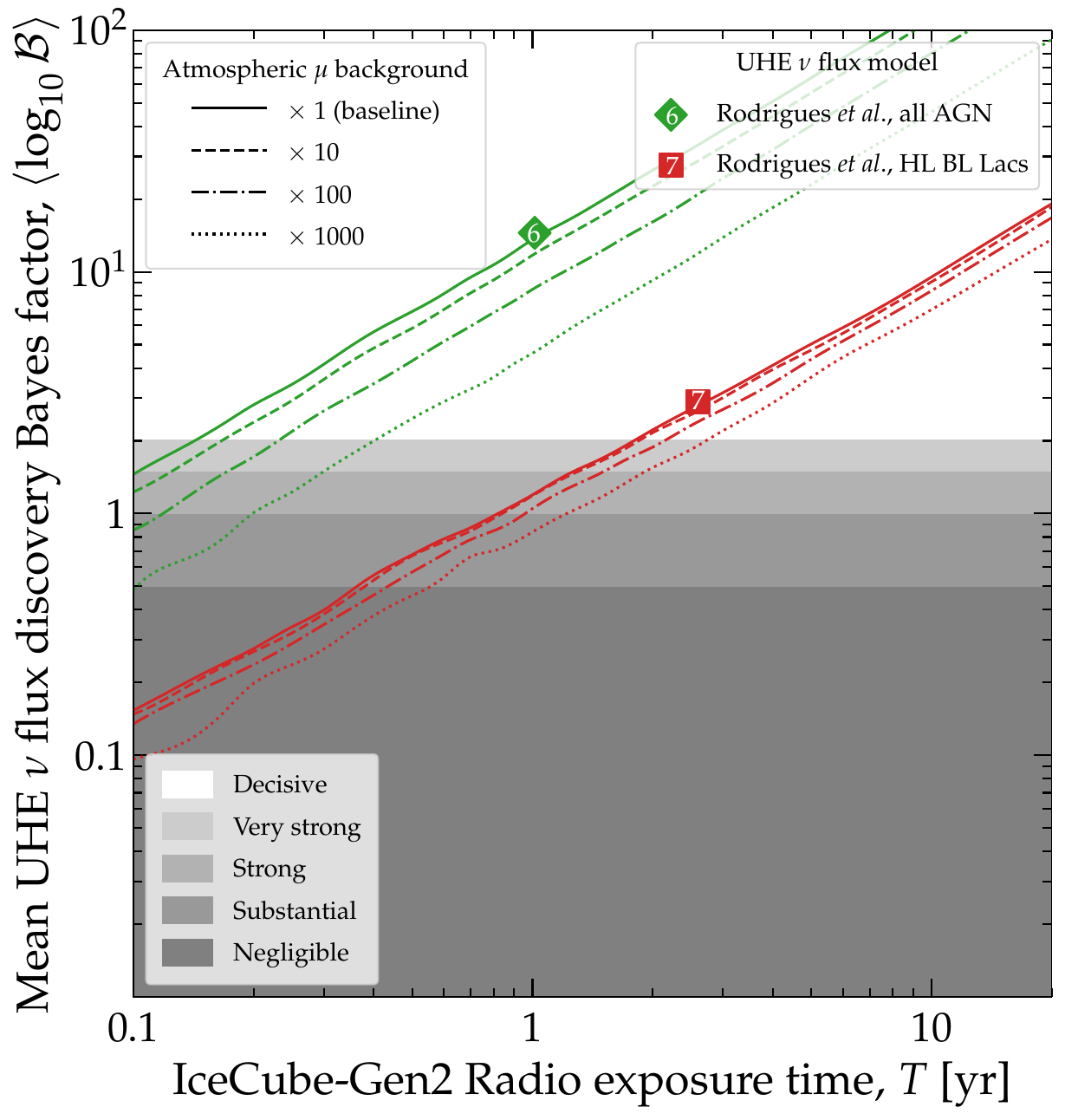}
 \caption{\label{fig:large_mu_bkg}Impact of an enlarged atmospheric muon background on the flux discovery potential of UHE neutrino flux models 6 and 7~\cite{Rodrigues:2020pli}; see \figu{benchmark_spectra}.  Results for other flux models are similar.  For this plot only, the cut-off energy of the high-energy neutrino flux background is fixed at $E_{\nu, {\rm cut}}^{\rm HE} = 10$~PeV, so that the main background is from atmospheric muons.  All other analysis choices are baseline; see Table~\ref{tab:analysis_choices_base} and Section~\ref{sec:discovery_results_nominal_choices}.  {\it Even a thousand-fold increase in the muon background over its baseline expectation reduces the discovery potential only mildly.}  See Section~\ref{sec:discovery_results_impact_muon_bg} for details.}
\end{figure}

{\it The UHE flux discovery potential of IceCube-Gen2 is robust against large uncertainties in the predicted size of the muon background, because it is concentrated mainly at the lowest energies, \ie, $E_{\rm sh}^{\rm rec} \lesssim 10^8$~GeV.}
\smallskip
 
Figure~\ref{fig:large_mu_bkg} shows that artificially increasing the size of the atmospheric muon background only impacts mildly the UHE neutrino flux discovery potential.  (We comment on the impact of changes to the shape of the energy spectrum of the atmospheric muon background later.) To single out the impact of the atmospheric muon background, in \figu{large_mu_bkg} we mitigate the contribution to the background from the UHE tail of the IceCube high-energy flux by fixing its cut-off energy to a relatively low value of $E_{\nu, {\rm cut}}^{\rm HE} = 10$~PeV.  (In most other results, we let the value of $E_{\nu, {\rm cut}}^{\rm HE}$ float generously; see Section~\ref{sec:discovery_results_nominal_choices}.)  Even a hefty thousand-fold increase in the muon background over its baseline expectation (see Section~\ref{sec:background}), which yields an integrated mean yearly rate of fewer than 100 detected muon events, only delays discovery of flux model 6~\cite{Rodrigues:2020pli} by about three months and of flux model 7~\cite{Rodrigues:2020pli} by about one year.  The illustrative choice of $E_{\nu, {\rm cut}}^{\rm HE} = 10$~PeV is conservative: a higher value would increase the contribution of the UHE tail of the IceCube high-energy neutrino flux to the background and reduce the relative contribution of the muon background, weakening further its impact on the flux discovery potential. 

The mild impact that a larger muon background has on the flux discovery potential is due to the difference in the energy and angular distributions of events induced by the muon background and events induced by a UHE neutrino flux model 3--12; see Figs.~\ref{fig:binned_events_all_benchmarks_vs_energy_dep} and \ref{fig:binned_events_all_benchmarks_vs_costhzrec}.  Events induced by the muon background lie at low energies, $E_{\rm sh}^{\rm rec} \lesssim 10^8$~GeV, and above the horizon, but not in downgoing directions, \ie, they lie at $0 \lesssim \cos \theta_z^{\rm rec} \lesssim 0.8$.  In contrast, events induced by the UHE neutrino flux reach higher energies, may be downgoing, but may also come from just below the horizon, especially if the flux is large.  Because our analysis (Section~\ref{sec:discovery_stat_analysis}) is binned in energy and direction, it is sensitive to the above differences in the distributions, which renders the flux discovery potential largely insensitive to increases in the size of the muon background.

The flux discovery potential is robust to changes in the size of the atmospheric muon background, but may not be so to changes in the shape of its event energy distribution.  In particular, a muon-induced event energy distribution that extends to $E_{\rm sh}^{\rm rec} \gtrsim 10^8$~GeV would hinder the discovery of flux models that peak at neutrino energies above $10^8$~GeV that, under our baseline choice for the muon background, are expected to be discoverable.  

We have checked that replacing our baseline muon background, produced using the {\sc Sybill 2.3c}~\cite{Fedynitch:2018cbl} hadronization model, with the central value of the predicted muon backgrounds computed in \Refe~\cite{Garcia-Fernandez:2020dhb} using the {\sc EPOS-LHC}~\cite{Pierog:2013ria} or {\sc QGSJet-II-04}~\cite{Ostapchenko:2010vb} hadronization models, has negligible impact on the flux discovery potential.  However, that exploration is not exhaustive, and may not be representative of all possible variations in the shape of the muon-induced event energy spectrum.  Hence, analyses beyond the scope of this work should account, within their statistical procedures, for the impact that the systematic uncertainties in the hadronization model, and in the size, shape, and mass composition of the parent UHECR energy spectrum have on the energy spectrum of atmospheric muons.  Reference~\cite{IceCube:2020wum} contains a recent implementation of this in an IceCube analysis.

In reality, a factor-of-100 or factor-of-1000 underestimation of the muon background, like the ones in \figu{large_mu_bkg}, is unlikely.  However, such a large contribution could come from other, non-muon and non-neutrino backgrounds, like air-shower cores; see Section~\ref{sec:background_cores} for details.  Its size and shape are presently uncertain, but it may conceivably extend to higher energies than the muon background; see \Refe~\cite{DeKockere:2022bto} for early estimates.  What \figu{large_mu_bkg} suggests is that, even in such a case, a non-neutrino background that is relatively well characterized in energy and direction, even if large in size, may have limited impact on the UHE neutrino flux discovery potential.


\subsubsection{Impact of the background UHE tail of the IceCube high-energy neutrino flux, $\mathcal{M}_{\rm HE}$}
\label{sec:discovery_results_impact_neutrino_bg}

\begin{figure}[t!]
 \centering
 \includegraphics[width=\columnwidth]{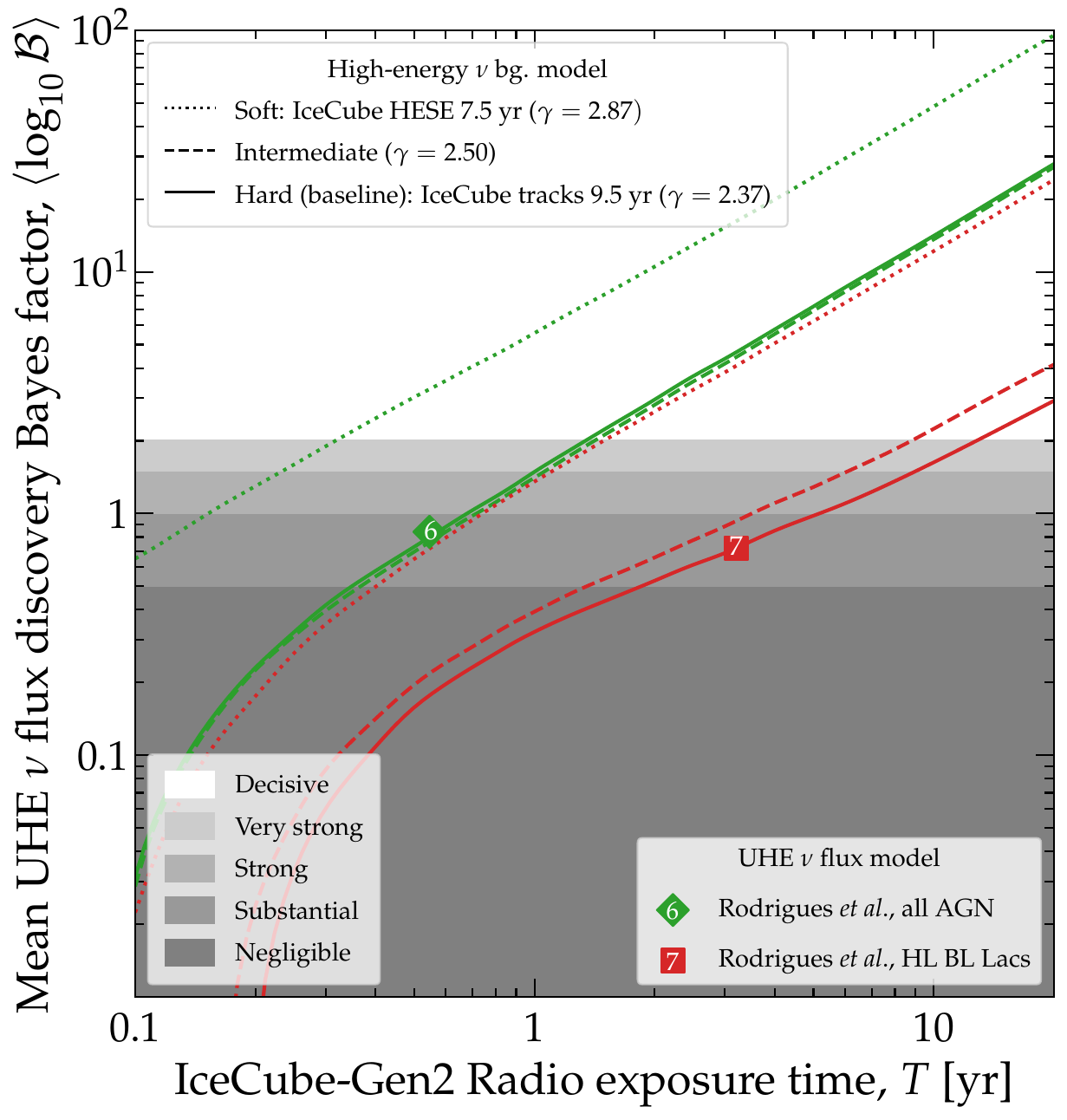}
 \caption{\label{fig:bayes_other_nu_cases}Impact of the choice of high-energy neutrino flux, whose UHE tail constitutes a potential background at ultra-high energies, on the flux discovery potential of UHE neutrino flux models 6 and 7~\cite{Rodrigues:2020pli}; see \figu{benchmark_spectra}.  All other analysis choices are baseline; see Table~\ref{tab:analysis_choices_base} and  Section~\ref{sec:discovery_results_nominal_choices}.  The hard and soft background fluxes are UHE extrapolations of the IceCube through-going track~\cite{IceCube:2021uhz} and HESE~\cite{IceCube:2020wum} results, respectively; we also show an intermediate case.  In all cases, the background flux has a high-energy exponential cut-off; see Section~\ref{sec:background}.  Results for other flux models are similar.  {\it A softer UHE tail of the high-energy neutrino flux background may expedite the discovery of an UHE flux model appreciably, by months or years.}  See Section~\ref{sec:discovery_results_impact_neutrino_bg} for details.}
\end{figure}

{\it The UHE neutrino flux discovery potential of IceCube-Gen2 may be enhanced significantly if the background from the UHE tail of the IceCube high-energy neutrino flux is small, \ie, if its energy spectrum is soft and consistent with the IceCube 9.5-year HESE analysis.}
\smallskip

Figure~\ref{fig:bayes_other_nu_cases} shows the impact on the flux discovery potential of the radio array of IceCube-Gen2 of adopting a different choice for the UHE tail of the IceCube high-energy neutrino background (Section~\ref{sec:background_nu}).  These are extrapolations to ultra-high energies of the power-law neutrino fluxes  measured by IceCube in the TeV--PeV range, suppressed by a high-energy cut-off, \ie, $\propto E_\nu^{-\gamma} e^{-E_\nu/E_{\nu, {\rm cut}}^{\rm HE}}$, following \equ{powerlaw_cutoff}.  We compare the three possibilities introduced in Section~\ref{sec:background_nu}: our baseline choice of a hard spectrum (spectral index of $\gamma = 2.37$) motivated by the 9.5-year IceCube through-going $\nu_\mu$ analysis~\cite{IceCube:2021uhz}, an intermediate spectrum ($\gamma = 2.50$) with the same flux normalization, and a soft spectrum ($\gamma = 2.87$) motivated by the 7.5-year IceCube HESE analysis~\cite{IceCube:2020wum}.  Here we study the impact of the choice of the normalization and spectral index of the background neutrino flux; in Section~\ref{sec:discovery_results_impact_prior_cut-off}, we study the impact of our degree of ignorance of $E_{\nu, {\rm cut}}^{\rm HE}$. 

Figure~\ref{fig:bayes_other_nu_cases} shows that using the softer spectrum may expedite decisive flux discovery significantly: for flux models 6 and 7, decisive discovery is reduced roughly from 1.3 and 13~years, respectively, to 4~months and 1.2 years.  This is because a softer UHE tail of IceCube high-energy neutrino flux corresponds to a lower background.  Figure~\ref{fig:bayes_other_nu_cases} also shows that the improvement stems predominantly from the shape of the spectrum, rather than from its size, {\it viz.}, when comparing the results using the hard and intermediate background spectra, which share the same flux normalization (see Section~\ref{sec:background_nu}).  

Changes to the background UHE tail of the IceCube high-energy neutrino flux differ from changes to the atmospheric muon background (Section~\ref{sec:discovery_results_impact_muon_bg}) mainly in two aspects.  First, if the cut-off energy is known to be $E_{\nu, {\rm cut}}^{\rm HE} \gtrsim 10^8$~GeV or if, as in our baseline treatment, its value is unknown but allowed to be possibly high, then the UHE tail of the IceCube high-energy neutrino flux constitutes the dominant background contribution; see Section~\ref{sec:background}.  Therefore, changes to it naturally affect the flux discovery potential significantly.

Second, unlike the atmospheric muon background, the background from the UHE tail of the IceCube high-energy neutrino flux depends on the $\nu N$ cross section, \ie, on $f_\sigma$.  Because our baseline analysis allows the value of $f_\sigma$ to float (Section~\ref{sec:discovery_results_nominal_choices}), this grants the UHE tail of the IceCube high-energy neutrino flux, when computed under the background-only hypothesis, the freedom to find values of $f_\sigma$ with which it can reproduce closely the observed rate (supplemented, at low energies, by the sub-dominant atmospheric muon background).  (Section~\ref{sec:discovery_results_impact_cross_section} explains in detail how a larger or smaller cross section affects internally the statistical analysis.)  Letting the value of $f_\sigma$ float leads to larger values of the posterior and evidence in the background-only hypothesis, Eqs.~(\ref{equ:posterior_bg}) and (\ref{equ:evidence_bg}), and, consequently, lower values of the Bayes factor, \equ{bayes_s_bg}, and to a longer exposure time required to claim decisively the discovery of a given UHE neutrino flux model 3--12.  In Section~\ref{sec:discovery_results_impact_cross_section} we explore how the impact of the IceCube high-energy neutrino flux is reduced by more precise prior knowledge of $f_\sigma$. 


\subsubsection{Impact of the prior on the cut-off energy of the UHE tail of the IceCube high-energy neutrino flux background, $E_{\nu, {\rm cut}}^{\rm HE}$}
\label{sec:discovery_results_impact_prior_cut-off}

\begin{figure}[t!]
 \centering
 \includegraphics[width=\columnwidth]{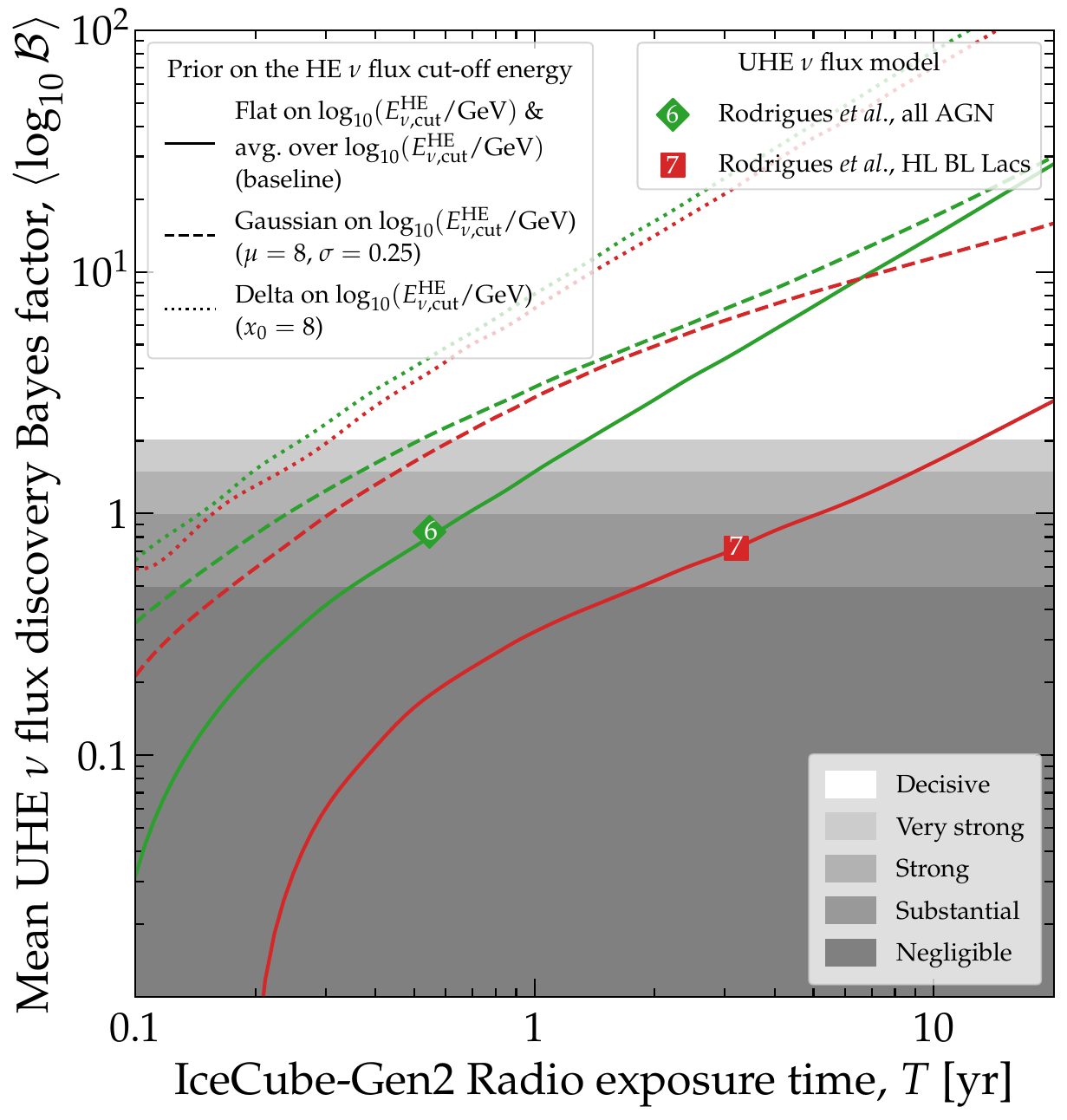}
 \caption{\label{fig:bayes_factor_ecut_cases}Impact of the choice of prior on the cut-off energy of the UHE tail of the background IceCube high-energy neutrino flux, $E_{\nu, {\rm cut}}^{\rm HE}$, on the flux discovery potential of UHE neutrino flux models 6 and 7~\cite{Rodrigues:2020pli}; see \figu{benchmark_spectra}.  All other analysis choices are baseline; see Table~\ref{tab:analysis_choices_base} and  Section~\ref{sec:discovery_results_nominal_choices}.  The baseline choice of a flat prior followed by an average of the Bayes factor over the value of the cut-off energy is conservative and represents complete ignorance of $E_{\nu, {\rm cut}}^{\rm HE}$.  A wide Gaussian prior and delta-function prior, both centered at $E_{\nu, {\rm cut}}^{\rm HE} = 100$~PeV, represent limited and precise knowledge of its value, respectively.  Results for other flux models are similar.  {\it Even limited evidence for the existence and value of a cut-off in the UHE tail of the IceCube high-energy neutrino flux, possibly gathered in upcoming astrophysical TeV--PeV neutrino measurements, may significantly expedite UHE neutrino flux discovery.}  See Section~\ref{sec:discovery_results_impact_prior_cut-off} for details.}
\end{figure}

{\it The UHE neutrino flux discovery potential of IceCube-Gen2 may be significantly enhanced by even limited knowledge of the ultra-high-energy tail end of the high-energy neutrino flux, \ie, of its cut-off energy.}
\smallskip

Figure~\ref{fig:bayes_factor_ecut_cases} shows the impact that the choice of prior  on the cut-off energy of the background UHE tail of the IceCube high-energy neutrino flux, $\pi(\log_{10}(E_{\nu, {\rm cut}}^{\rm HE}/{\rm GeV}))$ in Eqs.~(\ref{equ:posterior_s_bg})--(\ref{equ:evidence_bg}), has on the UHE neutrino flux discovery potential.  We compare results obtained using our conservative baseline analysis choice of a flat prior, followed by an average of the Bayes factor over the real value of the cut-off energy (see Section~\ref{sec:discovery_results_nominal_choices}), against results obtained using two alternative, informed priors: a wide Gaussian prior and a Dirac $\delta$-function prior.  These alternatives reflect, respectively, the possible outcome of limited and precise measurement of the tail end of the high-energy neutrino flux by present and future TeV--PeV neutrino telescopes, \eg, IceCube, IceCube-Gen2~\cite{IceCube-Gen2:2020qha}, via its optical array, KM3NeT~\cite{KM3Net:2016zxf}, Baikal-GVD~\cite{Baikal-GVD:2020xgh}, P-ONE~\cite{P-ONE:2020ljt}, TAMBO~\cite{Romero-Wolf:2020pzh}, TRIDENT~\cite{Ye:2022vbk},  Trinity~\cite{Otte:2019knb}, or a combination of detectors~\cite{Schumacher:2021hhm}.  

In \figu{bayes_factor_ecut_cases}, as illustration, we choose the real value of the cut-off energy to be $E_{\nu, {\rm cut}}^{\rm HE} = 100$~PeV; for the Gaussian prior, we choose a width of 0.25 in $\log_{10}(E_{\nu, {\rm cut}}^{\rm cut}/{\rm GeV})$.  In both cases, the analysis is the same as the one described in Section~\ref{sec:discovery_stat_analysis}, with the exception of using a different prior on $\log_{10}(E_{\nu, {\rm cut}}^{\rm cut}/{\rm GeV})$ and, unlike the baseline analysis choice, of not averaging the Bayes factor over the real value of $\log_{10}(E_{\nu, {\rm cut}}^{\rm cut}/{\rm GeV})$ anymore, since when using informed priors we are no longer in a situation of complete ignorance of the value of the cut-off energy.

Figure~\ref{fig:bayes_factor_ecut_cases} shows that using informed priors on $\log_{10}(E_{\nu, {\rm cut}}^{\rm cut}/{\rm GeV})$ brings significant improvement to the UHE flux discovery potential.  Because the UHE tail end of the IceCube high-energy neutrino flux is the dominant background in our analysis (see Section~\ref{sec:background}), understanding it better, as reflected by using informed priors on $\log_{10}(E_{\nu, {\rm cut}}^{\rm cut}/{\rm GeV})$, significantly improves the separation between the signal and background-only hypotheses.  In \figu{bayes_factor_ecut_cases}, the improvement is striking for flux model 7.  Using our conservative baseline prior on the cut-off energy, we would need about 10~years of exposure time to decisively discover flux model 7.  In contrast, using the wide Gaussian prior reduces the exposure time needed to about 7 months, and using the $\delta$-function prior reduces it to about 4 months.  Similar improvements are achievable for the other benchmark UHE neutrino flux models 1--12.

This significant reduction in the exposure time required for UHE flux discovery highlights the importance of the simultaneous development and deployment of neutrino telescopes that operate in the high-energy (TeV--PeV) and ultra-high-energy ($>$~100~PeV) ranges, and their combined observations~\cite{Ackermann:2019ows, Ackermann:2019cxh, Abraham:2022jse, Ackermann:2022rqc}.


\subsubsection{Impact of the prior on the neutrino-nucleon cross section, $f_\sigma$}
\label{sec:discovery_results_impact_cross_section}

\begin{figure}[t!]
 \centering
  \includegraphics[width=\columnwidth]{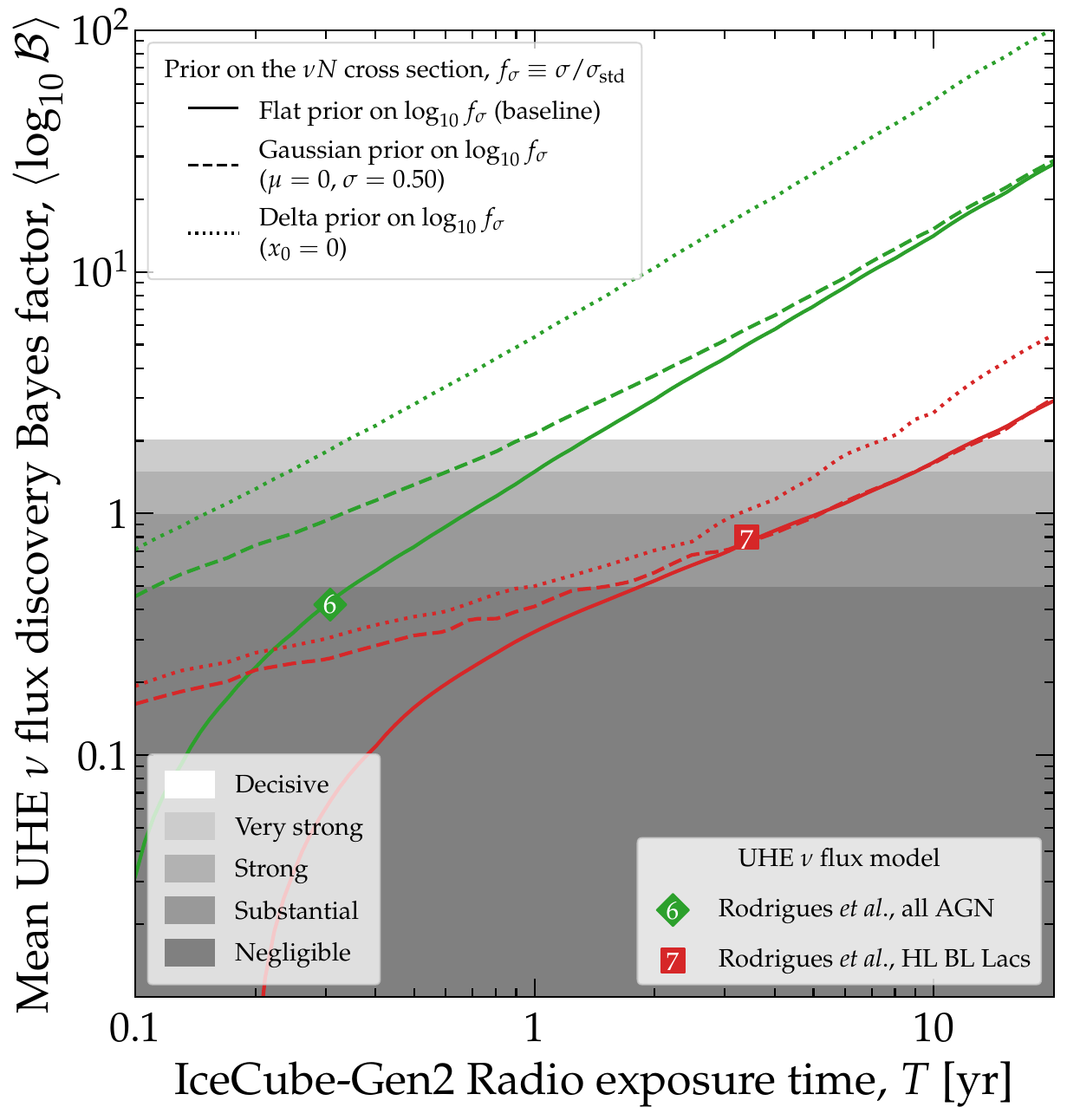}
 \caption{\label{fig:bayes_prior_choices}Impact of the choice of prior on the UHE neutrino-nucleon ($\nu N$) cross section, $f_\sigma \equiv \sigma/\sigma_{\rm std}$, on the flux discovery potential of UHE neutrino flux models 6~\cite{Rodrigues:2020pli} and 7~\cite{Rodrigues:2020pli}; see \figu{benchmark_spectra}.  All other analysis choices are baseline; see Table~\ref{tab:analysis_choices_base} and  Section~\ref{sec:discovery_results_nominal_choices}.  Here, $\sigma$ is the cross section, whose value is allowed to float in the statistical procedure, and $\sigma_{\rm std}$ is its BGR18 prediction~\cite{Bertone:2018dse}.  The baseline choice of a flat prior on $\log_{10 }f_\sigma$ is conservative and represents complete ignorance of $f_\sigma$.  A wide Gaussian prior and delta-function prior, both centered at the central BGR18 prediction of $\log_{10 }f_\sigma = 0$, represent limited and precise knowledge of its value, respectively.  Results for other flux models are similar.  {\it Using even a limited informed prior on $f_\sigma$ expedites flux discovery, especially at low exposure times, where the event rate may be low; at long exposure times, where the event rate is higher, only a precise informed prior helps.}  See Section~\ref{sec:discovery_results_impact_cross_section} for details.}
\end{figure}

{\it The UHE neutrino flux discovery potential of IceCube-Gen2 may be enhanced moderately by limited knowledge of the UHE neutrino-nucleon cross section, especially at low exposure times, and substantially enhanced by precise knowledge of it, for any exposure time.}
\smallskip

Ultra-high-energy neutrinos can be used to measure the UHE deep-inelastic-scattering neutrino-nucleon ($\nu N$) cross section, $\sigma$.  Reference~\cite{Valera:2022ylt} made detailed forecasts of this for IceCube-Gen2, based on the same detector design, effective volume, and calculation framework that we use here; see also \Refes~\cite{Denton:2020jft, Esteban:2022uuw} for complementary forecasts.  In \Refe~\cite{Valera:2022ylt}, the best-fit value and uncertainty on $\sigma$ were determined using a statistical analysis similar to the one we introduced in Section~\ref{sec:discovery_stat_analysis}: after 10~years, as long as at least a few tens of events are detected, the UHE $\nu N$ cross section may be measured to within 50\% of its BGR18 prediction~\cite{Bertone:2018dse}.  Here, while our goal is not to measure the $\nu N$ cross section, we nevertheless account for the important effect that it has on the discovery Bayes factor, via the prior $\pi(\log_{10}f_\sigma)$, where $f_\sigma \equiv \sigma/{\sigma_{\rm std}}$ (Section~\ref{sec:discovery_stat_analysis}).  Below, we show how.

Figure~\ref{fig:bayes_prior_choices} shows the impact that the choice of prior has on the UHE neutrino flux discovery potential.  We compare results obtained using our conservative baseline analysis choice of a flat prior on $\log_{10}f_\sigma$ (see Section~\ref{sec:discovery_results_nominal_choices}), against results obtained using two alternative, informed priors: a wide Gaussian prior, with a half-decade width, and a Dirac $\delta$-function prior, both centered on the central value of the BGR18 prediction of the $\nu N$ cross section, \ie, on $\log_{10}f_\sigma = 0$.

Figure~\ref{fig:bayes_prior_choices} shows that using informed priors expedites flux discovery.  The improvements over the baseline expectations when using the wide Gaussian prior are moderate at short exposure times and negligible at long exposure times.  The improvements when using the $\delta$-function prior are substantial for any exposure time.  For both informed priors, improvements are more evident at short exposure times, where signal event rates are lower, which makes separating them from background event rates more challenging.  At longer exposure times, where event rates are higher and the separation is clearer even when $f_\sigma$ is known uncertainly, there is sizable improvement only when using the $\delta$-function prior.  For example, using our conservative baseline prior on the cross section, we would need roughly 1.3~years of exposure time to decisively discover flux model 6.  In contrast, using the wide Gaussian prior reduces the exposure time needed to about 1~year, and using the $\delta$-function prior reduces it to about 4~months.  Similar improvements are achievable for the other benchmark UHE neutrino flux models.

There is nuanced insight to be gained from how the $\nu N$ cross section affects the flux discovery potential; we describe it below.  Changing the cross section affects the neutrino-induced event rate, \equ{spectrum_true}, in two ways.  (The illustrative simplified event-rate calculation, \equ{event_rate_simple}, also captures these features, as described in Section~\ref{sec:propagation}.)

First, a larger or smaller cross section, respectively, increases or decreases the interaction rate of neutrinos in the detector.  This affects the total event rate, \ie, the rate integrated over all reconstructed energies and directions.  See \Refe~\cite{Valera:2022ylt} for a detailed study.  

Second, a larger or smaller cross section, respectively, strengthens or weakens the attenuation of the neutrino flux as it propagates through the Earth.  This affects the angular distribution of neutrino-induced events: a larger cross section induces a steeper decline in the event rate around the horizon, \ie, at $\theta_z^{\rm rec} \approx 90^\circ$, since upgoing neutrinos are attenuated more strongly.  Reference~\cite{Valera:2022ylt} showed that the sensitivity to the cross section stems from events coming from around the horizon, where in-Earth attenuation is significant, but not overbearing. 

It is from the interplay of the above two effects that more precise prior knowledge of $f_\sigma$ leads to a larger UHE neutrino flux discovery potential. Since information about the cross section is extracted from the angular dependence of the event rate around the horizon, a detector angular resolution that allows us to resolve this accurately is essential (more on this in Section~\ref{sec:discovery_results_impact_resolution}).

Further, because the $\nu N$ cross section grows with neutrino energy, flux models that peak at low energies are less attenuated inside the Earth {\it vs.}~flux models that peak at high energies.  Figure~\ref{fig:binned_events_all_benchmarks_vs_energy_dep} illustrates this: the relative contributions of upgoing events and downgoing events resemble each other more closely for flux models that peak at low energy, \eg, models 10 and 12, than for flux models that peak at high energy, \eg, models 4 and 7.  Thus, for flux models that peak at low energies, the relatively larger number of events at the horizon and below it helps to pin down $f_\sigma$.  This impacts the evolution of the discovery Bayes factor with exposure time: flux models that peak at low energy but have a relatively low integrated event rate reach Bayes factors as high or higher than flux models that predict larger integrated event rates.  For example, \figu{bayes_factor_hard} shows that, after a few years, flux model 12 can be discovered with a significance comparable to flux models 7 and 8, even though it only yields about 70\% and 30\% of their event rates, respectively; see Table~\ref{tab:event_rates_binned}.


\subsubsection{Impact of the detector energy resolution, $\sigma_\epsilon$, and angular resolution, $\sigma_{\theta_z}$}
\label{sec:discovery_results_impact_resolution}

\begin{figure}[t!]
 \centering
 \includegraphics[width=\columnwidth]{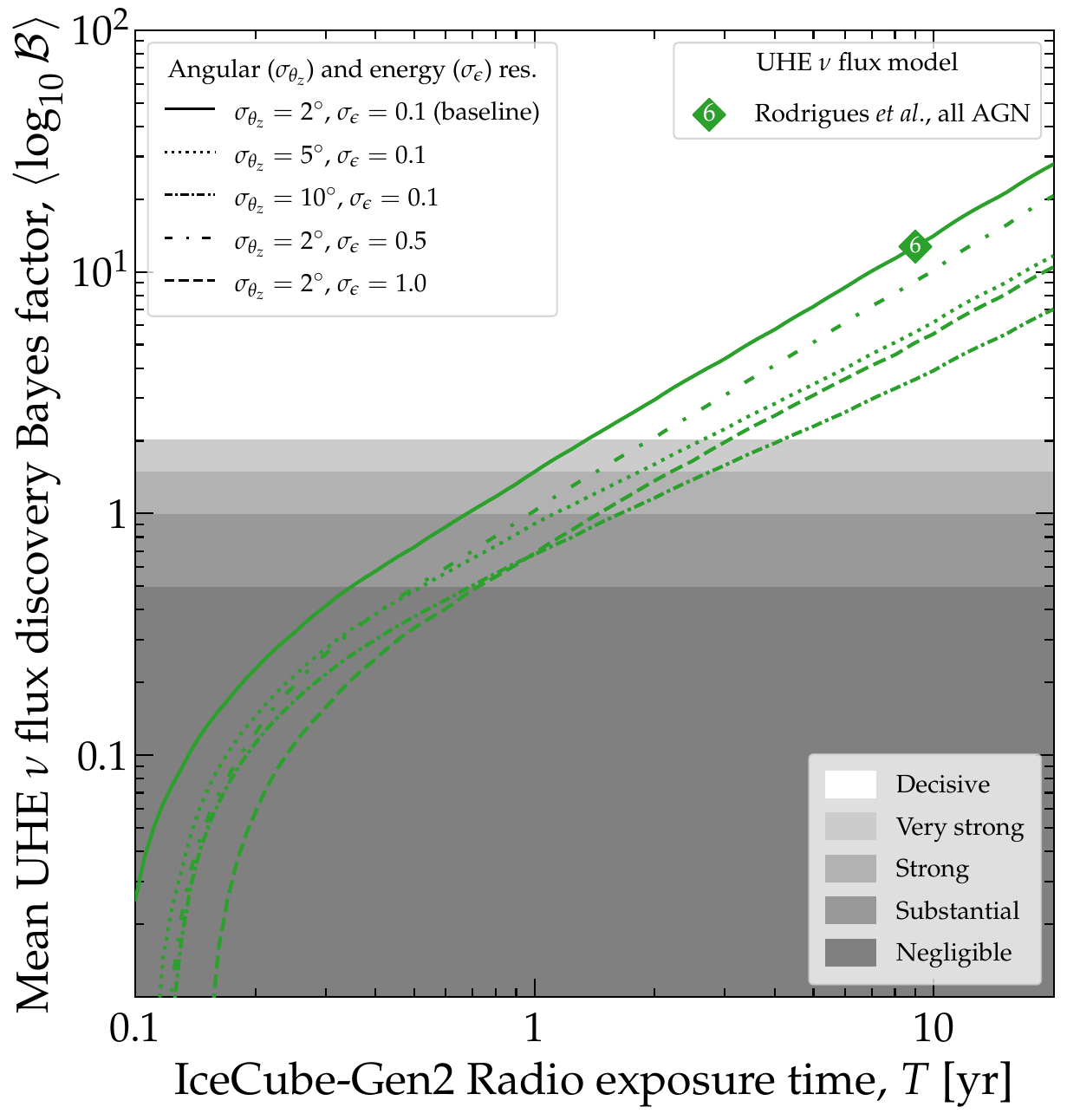}
 \caption{\label{fig:bayes_factor_res_main}Impact of the resolution of the radio array of IceCube-Gen2 in measuring reconstructed energy, $\sigma_\epsilon$, and reconstructed direction, $\sigma_{\theta_z}$, on the flux discovery potential of UHE neutrino flux model 6~\cite{Rodrigues:2020pli}; see \figu{benchmark_spectra}.  All other analysis choices are baseline; see Table~\ref{tab:analysis_choices_base} and  Section~\ref{sec:discovery_results_nominal_choices}.  Results for other flux models are similar; see Figs.~\ref{fig:bayes_energy_resolution} and \ref{fig:bayes_angular_resolution} in Appendix~\ref{sec:appendix_detector_resolution}.  {\it Poorer detector resolution delays flux discovery, but does not prevent it; energy resolution has a stronger impact, because it weakens the distinction between the energy distributions of signal and background.}  See Section~\ref{sec:discovery_results_impact_resolution} for details.}
\end{figure}

{\it The UHE neutrino flux discovery potential of IceCube-Gen2 may be appreciably weakened by poor detector energy resolution---which impairs resolving signal {\it vs.}~background features in the event energy distribution---and poor angular resolution---which preserves the innate degeneracy between neutrino flux and $\nu N$ cross section}
\smallskip

Figure~\ref{fig:bayes_factor_res_main} shows the impact that the detector resolution in reconstructed shower energy, $\sigma_\epsilon$, where $\epsilon \equiv \log_{10}(E_{\rm sh}^{\rm rec}/E_{\rm sh})$, and the reconstructed zenith angle, $\sigma_{\theta_z}$, have on the UHE neutrino flux discovery potential.  The detector resolution affects the computation of event rates via \equ{spectrum_rec}; see Section~\ref{sec:event_rate_benchmarks} for details.  Figure~\ref{fig:bayes_factor_res_main} shows that alternative choices of the energy and angular resolution, poorer than the baseline choices, delay UHE flux discovery, but may not prevent it.

Poorer energy resolution lessens the difference between the distributions in reconstructed energy of the signal and background events (see \figu{binned_events_all_benchmarks_vs_energy_dep} for a comparison {\it vs.}~the muon background); see Section~\ref{sec:discovery_results_nominal_results}.  Thus, a significantly poorer energy resolution may appreciably weaken the UHE flux discovery potential. Figure~\ref{fig:bayes_factor_res_main} shows that, under our baseline choices of detector resolution, with $\sigma_\epsilon = 0.1$, \ie, one-tenth of a decade in shower energy, flux model 6 may be decisively discovered in roughly 1.3~years.  With $\sigma_\epsilon = 0.5$, \ie, a resolution of half a decade in shower energy, decisive discovery is delayed to roughly 2~years.  With $\sigma_\epsilon = 1$, \ie, a resolution of a full decade in shower energy, it is delayed to roughly 3~years.  Similar delays occur for the other benchmark UHE neutrino flux models 3--12; see \figu{bayes_energy_resolution} in Appendix~\ref{sec:appendix_detector_resolution}.  The delays are substantially longer for the flux models with the lowest event rates, for which the separation between signal and background events is more challenging \ie, models 1, 3, 5, 7, and 12; see \figu{binned_events_all_benchmarks_vs_energy_dep} and Table~\ref{tab:event_rates_binned}.  (When changing $\sigma_\epsilon$, we change the binning in reconstructed energy accordingly.  For $\sigma_\epsilon = 0.1$, 0.5, and 1, we use, respectively, 30, 6, and 3 bins equally spaced in logarithmic scale from $E_{\rm sh}^{\rm rec} = 10^7$~GeV to $10^{10}$~GeV.)

Poorer angular resolution lessens the difference between the distributions in reconstructed direction of the signal and background events; see \figu{binned_events_all_benchmarks_vs_costhzrec} for a comparison {\it vs.}~the muon background.  However, this has only a mild impact on the flux discovery potential.  The dominant impact comes instead from the fact that, since our baseline results assume no prior knowledge of the $\nu N$ cross section, \ie, a flat prior on $\log_{10} f_\sigma$, poorer angular resolution preserves the innate degeneracy between the neutrino flux and the cross section, illustrated in \equ{event_rate_simple}.  Ordinarily, the degeneracy would be broken by comparing the angular distribution of events coming from around the horizon, but a poor angular resolution obfuscates this.  For details, see Section~\ref{sec:propagation} and, especially, the discussion in connection to \figu{bayes_binning} in Sections~\ref{sec:discovery_results_nominal_results} and \ref{sec:discovery_results_impact_cross_section}.  (Separately, angular resolution is critical for discovering point sources of UHE neutrinos~\cite{Fang:2016hop, Fiorillo:2022ijt} and important when measuring the UHE neutrino-nucleon cross section~\cite{Connolly:2011vc, Denton:2020jft, Valera:2022ylt, Esteban:2022uuw}). 

Figure~\ref{fig:bayes_factor_res_main} shows that, under our baseline choices of detector resolution, with $\sigma_{\theta_z} = 2^\circ$, flux model 6 may be decisively discovered in roughly 1.3~years.  With $\sigma_{\theta_z} = 5^\circ$ and $10^\circ$, decisive discovery is delayed to roughly 2.5 and 4 years, respectively.  Similar delays occur for the other benchmark UHE neutrino flux models 3--12; see \figu{bayes_angular_resolution} in Appendix~\ref{sec:appendix_detector_resolution}.  As for the case of poorer energy resolution, the delays are longer for the flux models with the lowest event rates.  (When changing $\sigma_{\theta_z}$, we change the binning in reconstructed direction accordingly.  For $\sigma_{\theta_z} = 2^\circ$, $5^\circ$, and $10^\circ$, we use, respectively, 10, 4, and 2 equally spaced bins for events around the horizon, \ie, from $\theta_z^{\rm rec} = 80^\circ$ to $100^\circ$.  We leave the binning of downgoing and upgoing events unchanged).

In our forecasts, we have considered a common detector angular and energy resolution for all of the events.  However, in a real experiment every event will be reconstructed, in general, with a different angular and energy error.  Future, revised versions of our analysis should include this event-by-event treatment~\cite{Valera:InPrep}.


\subsubsection{Importance of Earth-skimming events}
\label{sec:discovery_results_horizontal_events}

\begin{figure}[t!]
 \centering
  \includegraphics[width=\columnwidth]{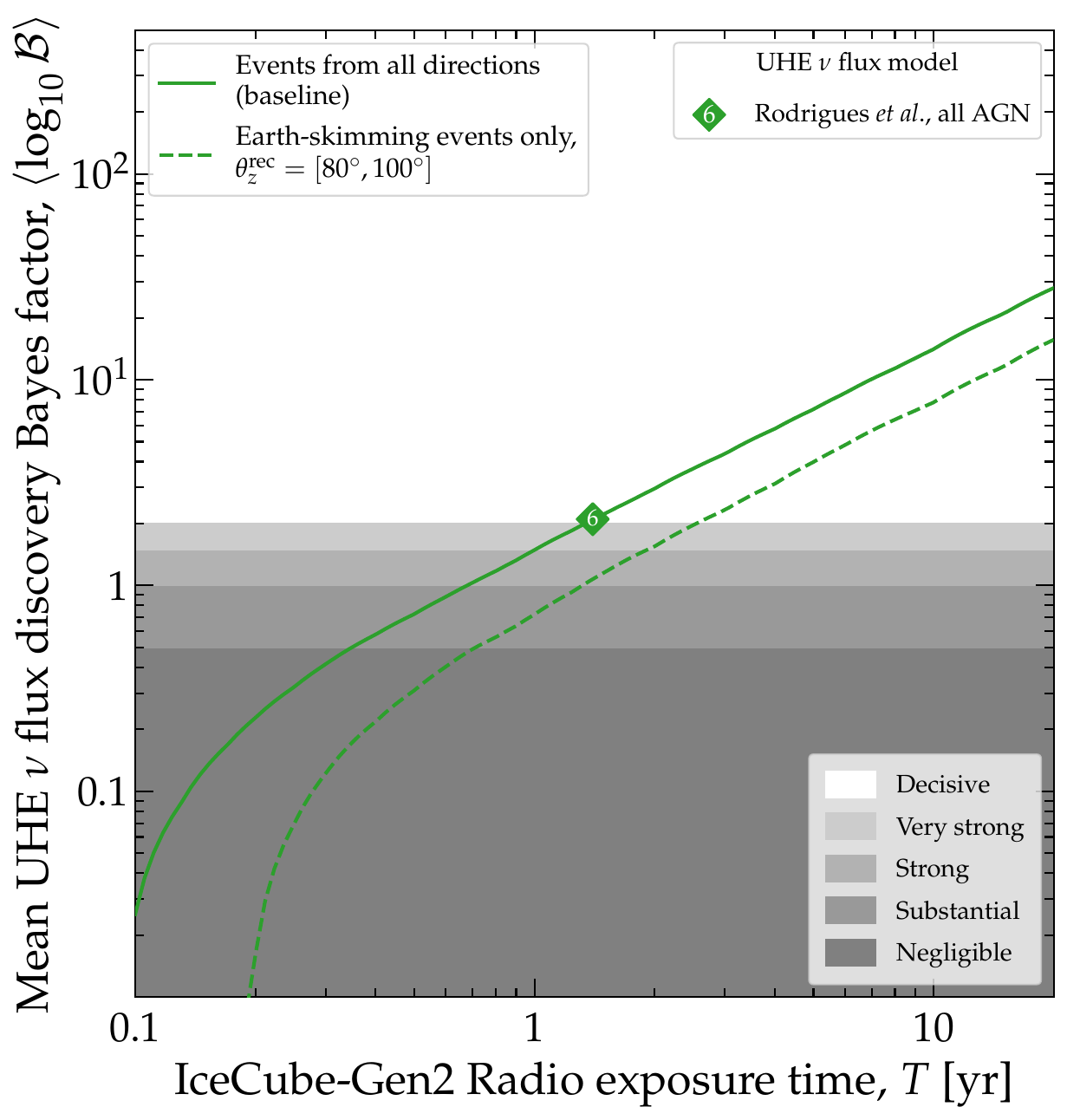}
 \caption{\label{fig:bayes_horizontal}Flux discovery potential of the UHE neutrino flux model 6~\cite{Rodrigues:2020pli} computed using events from all directions {\it vs.}~using only Earth-skimming, or horizontal, events.  All other analysis choices are baseline; see Table~\ref{tab:analysis_choices_base} and  Section~\ref{sec:discovery_results_nominal_choices}.  Results for other flux models are similar.  {\it Earth-skimming events are largely responsible for the flux discovery potential, so optimizing a detector to observe them is a sensible strategy.}  See Section~\ref{sec:discovery_results_horizontal_events} for details.}
\end{figure}

{\it Most of the UHE neutrino flux discovery potential comes from Earth-skimming events, \ie, events that reach the detector from around the horizon ($\theta_z^{\rm rec} \approx 90^\circ$).}
\smallskip

Section~\ref{sec:propagation} showed that the flux of Earth-skimming, or horizontal, \ie, $\theta_z^{\rm rec} \approx 90^\circ$, neutrinos is attenuated by neutrino-matter interactions inside the Earth, but is not obliterated by them, unlike upgoing events, and so they induce a sizable number of events in the detector.  In contrast, downgoing neutrinos reach the detector mostly unattenuated, and the number of neutrinos reaching the detector is much larger.  Figures~\ref{fig:binned_events_all_benchmarks_vs_energy_dep} and \ref{fig:binned_events_all_benchmarks_vs_costhzrec} show these features for the UHE neutrino flux models 1--12.

Because the design of IceCube-Gen2 is still under consideration, its final form might conceivably have a response to downgoing and horizontal events that is different from that of the baseline design~\cite{Hallmann:2021kqk} that we have adopted; see Section~\ref{sec:ic-gen2} for details on it.  Further, other UHE neutrino telescopes presently under planning target mainly Earth-skimming events induced by $\nu_\tau$, \ie, Ashra-NTA~\cite{Sasaki:2014mwa}, AugerPrime~\cite{PierreAuger:2016qzd}, BEACON~\cite{Wissel:2020sec}, EUSO-SPB2~\cite{Adams:2017fjh}, GCOS~\cite{Horandel:2021prj}, GRAND~\cite{GRAND:2018iaj}, POEMMA~\cite{POEMMA:2020ykm}, PUEO~\cite{PUEO:2020bnn}, RET~\cite{RadarEchoTelescope:2021rca}, TAROGE~\cite{Chen:2021egw}, TAx4~\cite{TelescopeArray:2021dri}, TAMBO~\cite{Romero-Wolf:2020pzh}, Trinity~\cite{Otte:2019knb}; see \Refe~\cite{Ackermann:2022rqc} and Fig.~53 in \Refe~\cite{Abraham:2022jse} for an overview.  This prompts us to study the importance that Earth-skimming events have in our forecasts of flux discovery potential.

Figure~\ref{fig:bayes_horizontal} shows, for flux model 6, the extreme case where our forecasts use only Earth-skimming events, with $\theta_z^{\rm rec} = [80^\circ, 100^\circ]$, which make up roughly $35\%$ of the all-sky event rate. To understand this result, consider naively that if events from all directions were equally relevant for flux discovery, then using a subset of only one third of them, with randomly chosen directions, should delay flux discovery by roughly a factor of 3.  However, \figu{bayes_horizontal} shows that using only Earth-skimming events delays the decisive flux discovery of flux model 6, which takes 1.31~years in our baseline predictions (Table~\ref{tab:event_rates_binned}), by only about 1.5~years, rather than by the naive expectation of about 2.6~years.  This is because flux discovery stems largely from resolving the angular distribution of Earth-skimming events, especially if the $\nu N$ cross section is unknown; see  Sections~\ref{sec:discovery_results_impact_cross_section} and \ref{sec:discovery_results_impact_resolution} for details.

To further illustrate the point above, we imagine an extreme alternative design of the radio array of IceCube-Gen2 that detects exclusively Earth-skimming events, but preserves the same total effective volume, integrated over all energies and directions, as our baseline design.  That alternative design should detect roughly three times as many events as are used in the result in \figu{bayes_horizontal} based on Earth-skimming events only.  Increasing the event rate by a factor of 3 is equivalent to increasing the exposure time by the same factor.  Because, at long exposure times, the discovery Bayes factor grows linearly with time (Section~\ref{sec:discovery_results_nominal_results}), the alternative detector design could claim decisive flux discover a factor-of-3 sooner than the baseline result using Earth-skimming events only in \figu{bayes_horizontal}, \ie, a reduction from about 3~years to 1~year.  In turn, this would be a reduction of about 30\% compared to the 1.3~years needed for decisive flux discovery in our baseline forecast in \figu{bayes_horizontal} using events from all directions.  This demonstrates that optimizing the detector response to detect Earth-skimming {\it vs.}~downgoing events is a strategy that merits exploration.


\subsubsection{Discovering the UHE tail end of the IceCube high-energy neutrino flux}
\label{sec:discovery_results_uhe_ic_tail}

{\it The UHE tail of the IceCube high-energy neutrino flux, based the hard-spectrum ($\gamma = 2.37$) flux measured in through-going $\nu_\mu$, may be decisively discovered within 10 years of exposure of the radio array of IceCube-Gen2, even if suppressed by a low-energy cut-off at 50 PeV.}
\smallskip

So far, we have forecast the discovery potential of UHE neutrino flux models 3--12, and considered the UHE tail of the IceCube high-energy neutrino flux as a background to their discovery, together with the background of atmospheric muons; see Section~\ref{sec:discovery_stat_analysis}.  Here, we forecast instead the discovery of the UHE tail of the IceCube high-energy neutrino flux by itself, and consider atmospheric muons as a background to their discovery.  Following the discussion in Section~\ref{sec:background_nu}, we expect that to discover the UHE tail of the high-energy neutrino flux, it must have a hard spectrum, \ie, a value of the spectral index, $\gamma$, not too far from 2, and a cut-off energy, \ie, $E_{\nu, {\rm cut}}^{\rm HE}$, in the tens of PeV, or no cut-off at all.  If these conditions are met, the UHE tail of the high-energy neutrino flux could induce a sizable number of neutrino events in the radio array of IceCube-Gen2 and in other UHE neutrino telescopes, and become detectable over the atmospheric muon background.  Below we quantify this.

\begin{figure}[t!]
 \centering
 \includegraphics[width=\columnwidth]{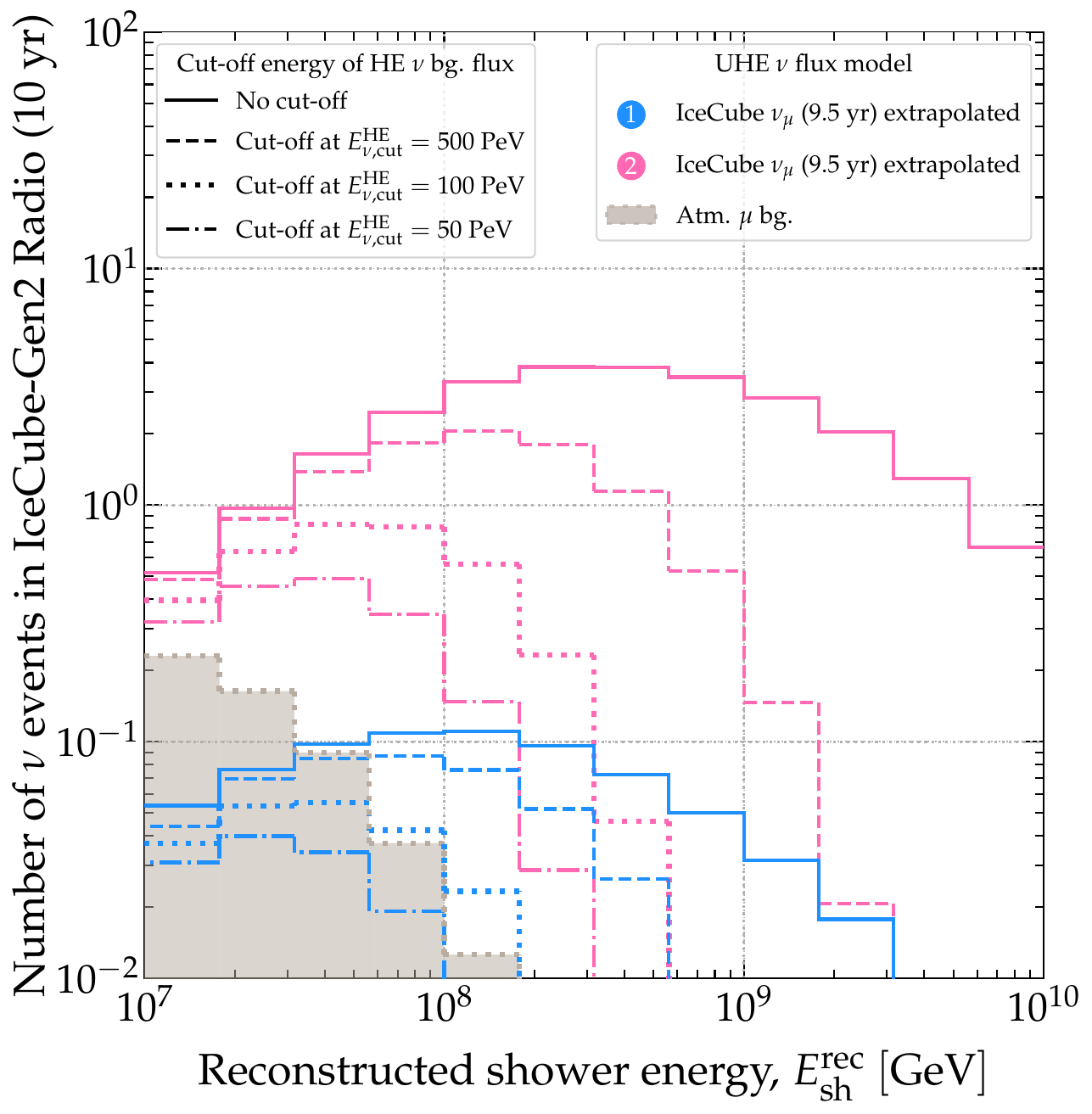}
 \caption{\label{fig:binned_events_ic_hard}Mean event distribution in reconstructed shower energy, $E_{\rm sh}^{\rm rec}$, expected in the radio array of IceCube-Gen2 after 10 years of exposure for flux models 1 and 2, \ie, the UHE extrapolation of the IceCube high-energy neutrino flux from the 7.5-year HESE~\cite{IceCube:2020wum} and 9.5-year through-going muon analyses~\cite{IceCube:2021uhz}, respectively (see \figu{benchmark_spectra}), augmented with a high-energy exponential cut-off at energy $E_{\nu, {\rm cut}}^{\rm HE}$; see \equ{powerlaw_cutoff}.  Event rates are computed using the procedure from Section~\ref{sec:event_rate_benchmarks}, and under our baseline analysis choices; see Section~\ref{sec:discovery_results_nominal_choices}.  See Table~\ref{tab:event_rates_binned} for integrated event rates, \figu{bayes_factor_no_marg_unbinned_ic_hard_atm_mu_bg_low} for the associated flux discovery potential, and Section~\ref{sec:discovery_results_uhe_ic_tail} for details.}
\end{figure}

\begin{figure}[t!]
 \centering
 \includegraphics[width=\columnwidth]{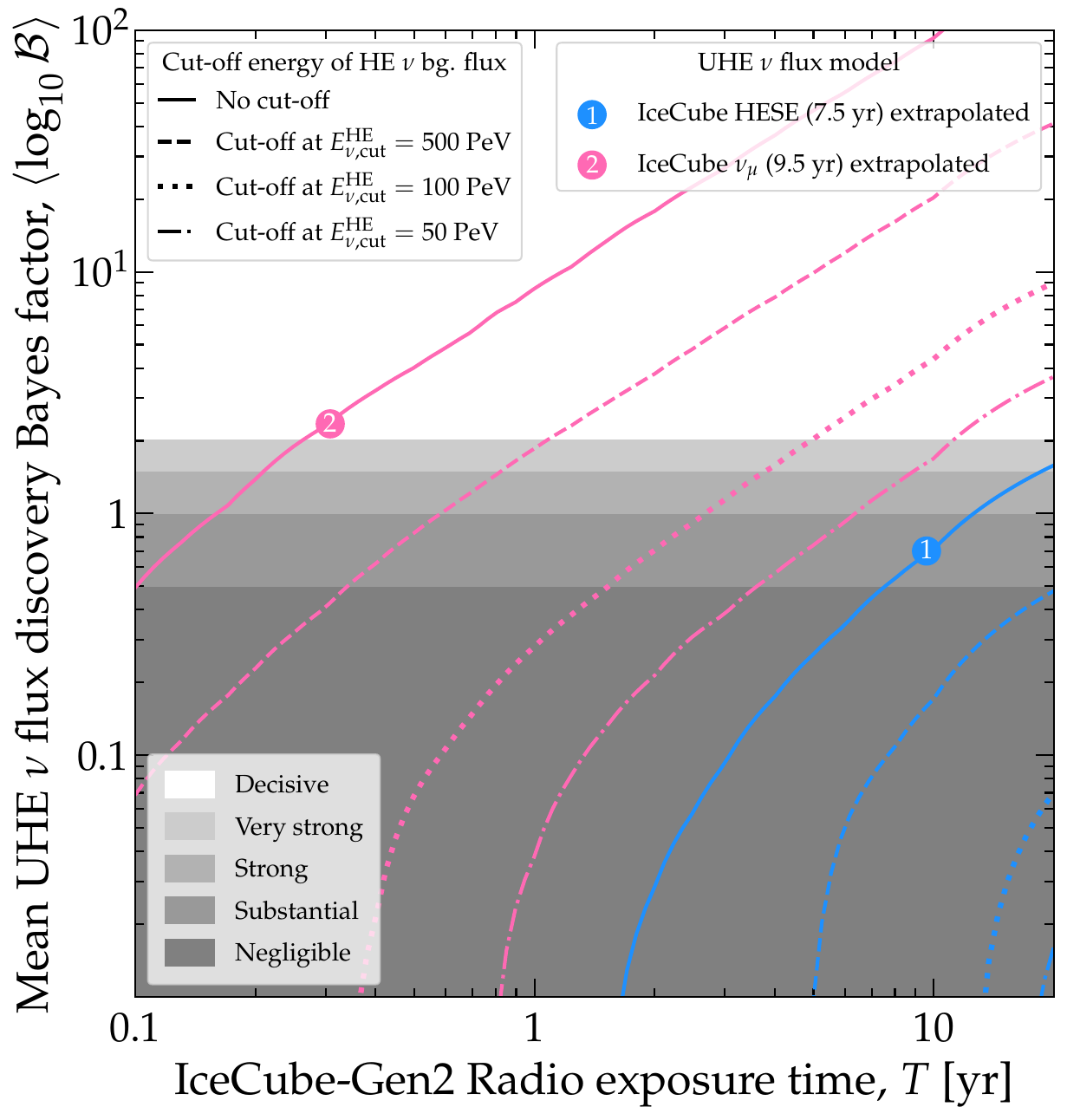}
 \caption{\label{fig:bayes_factor_no_marg_unbinned_ic_hard_atm_mu_bg_low}Impact of the cut-off energy, $E_{\nu, {\rm cut}}^{\rm HE}$, on the discovery potential of flux model 2, \ie, the UHE extrapolation of the high-energy flux from the IceCube 9.5-year through-going muon analysis~\cite{IceCube:2021uhz}, augmented by a high-energy exponential cut-off at energy $E_{\nu, {\rm cut}}^{\rm HE}$; see \equ{powerlaw_cutoff}.  Unlike our baseline procedure, in this plot we do not average the Bayes factor over the real value of $E_{\nu, {\rm cut}}^{\rm HE}$, since it is assumed to be known; however, we still use our baseline flat prior on $\log_{10} (E_{\nu, {\rm cut}}^{\rm HE}/{\rm GeV})$ when computing the posterior, \equ{posterior_s_bg}.  For this plot only, the sole background to flux discovery is from atmospheric muons.  All other analysis choices are baseline; see Table~\ref{tab:analysis_choices_base} and  Section~\ref{sec:discovery_results_nominal_choices}.  See \figu{binned_events_ic_hard} and Table~\ref{tab:event_rates_binned} for the event rates and Section~\ref{sec:discovery_results_uhe_ic_tail} for details.}
\end{figure}

Figure~\ref{fig:binned_events_ic_hard} compares, for different choices of the cut-off energy, the mean expected event rate induced by our alternative soft-spectrum choice and our baseline hard-spectrum choice for the UHE tail of the IceCube high-energy neutrino flux introduced in Section~\ref{sec:background_nu}.  They are based respectively, on the flux measured in the 7.5-year HESE analysis~\cite{IceCube:2020wum}, with $\gamma =2.87$, and in the 9.5-year through-going $\nu_\mu$ analysis~\cite{IceCube:2021uhz} by the IceCube Collaboration, with $\gamma = 2.37$, both augmented by a high-energy exponential cut-off, \equ{powerlaw_cutoff}.  When without a cut-off, \ie, with $E_{\nu, {\rm cut}}^{\rm HE} \to \infty$, they correspond to our benchmark UHE neutrino flux models 1 and 2, respectively.  

Figure~\ref{fig:binned_events_ic_hard} shows that lower values of $E_{\nu, {\rm cut}}^{\rm HE}$ reduce the integrated event rate and, especially, the event rate at high energies, thus concentrating events at low energies, and making their energy distribution resemble that of the atmospheric muon background.  Below we show how this erodes their prospects of being discovered.  Table~\ref{tab:event_rates_binned} shows event rates for flux models 1 and 2, for different choices of the cut-off energy.  Flux model 1 yields, on average, fewer than one event in 10~years, and versions of it with a cut-off yield even less.  Flux model 2 yields, on average, about 27 events in 10~years, and about 2 events even with an early cut-off at 50~PeV.  

Figure~\ref{fig:bayes_factor_no_marg_unbinned_ic_hard_atm_mu_bg_low} shows the impact that the value of $E_{\nu, {\rm cut}}^{\rm HE}$ has on the discovery potential of the UHE tail of the IceCube high-energy neutrino flux.  As expected from \figu{binned_events_ic_hard}, lower values of $E_{\nu, {\rm cut}}^{\rm HE}$ hinder discovery.  To compute the flux discovery Bayes factor in this case, we follow the same procedure introduced in Section~\ref{sec:discovery_stat_analysis}, but using only atmospheric muons for the background, \ie, using $N_{{\rm pred}, ij}^{({\rm s+bg})}(\boldsymbol\theta) = N^{\rm HE}_{\nu, ij}(\boldsymbol\theta) + N_{\mu, ij}$ instead of \equ{event_rate_pred_s_bg} for the signal hypothesis and $N_{{\rm pred}, ij}^{({\rm bg})} = N_{\mu, ij}$ instead of \equ{event_rate_pred_bg} for the background-only hypothesis.  In addition, unlike our baseline prescription (Section~\ref{sec:discovery_results_nominal_choices}), when the value of $E_{\nu, {\rm cut}}^{\rm HE}$ is fixed at 50, 100, or 500~PeV, we no longer average the Bayes factor over it.  Figure~ \ref{fig:bayes_factor_no_marg_unbinned_ic_hard_atm_mu_bg_low}, and also \figu{bayes_factor_hard}, show that in the absence of a cut-off flux model 2 could be discovered decisively within 4~months.  The presence of a cut-off delays its discovery, but does not preclude it: for a high cut-off at $E_{\nu, {\rm cut}}^{\rm HE} = 500$~PeV, the flux may be discovered decisively within 1~year, and even for a low cut-off at 50~PeV, it may still be discovered after roughly 10 years.  Flux model 2, without a cut-off, may be discovered with very strong evidence after 20~years, but versions of it with a cut-off are undiscoverable. 

Figure~\ref{fig:bayes_factor_no_marg_unbinned_ic_hard_atm_mu_bg_low} posits the intriguing possibility of using the IceCube high-energy neutrino flux to calibrate the response of UHE neutrino telescopes, which is known uncertainly.  If the value of the cut-off energy of the IceCube flux can be measured or constrained by complementary measurements in detectors with high sensitivity in the energy range of 1--10~PeV, like the optical array of IceCube-Gen2~\cite{IceCube-Gen2:2020qha}, TAMBO~\cite{Romero-Wolf:2020pzh}, or Trinity~\cite{Otte:2019knb}, then it may be possible to make informed predictions about the contribution of its UHE tail to the event rate in UHE neutrino telescopes.   However, there is an unavoidable trade-off: a low cut-off energy would be easier to characterize with 1--10~PeV telescopes, but it would also imply a low event rate in the UHE range (that is, in the absence of other contributions, like UHE neutrino flux models 3--12).


\subsubsection{Summary}

Our forecasts---Figs.~\ref{fig:bayes_factor_hard} and \ref{fig:bayes_factor_no_marg_unbinned_ic_hard_atm_mu_bg_low}, and in Table~\ref{tab:event_rates_binned}---tempered by design by important nuance from theory and experiment, reveal promising prospects for the discovery of an UHE neutrino flux in the first decade of operation of IceCube-Gen2. Several of our benchmark UHE neutrino flux models (Section~\ref{sec:fluxes}) may even be decisively discovered within 5 years of detector exposure.

Less conservative, but still reasonable and well-motivated alternative analysis choices may hasten or delay decisive flux discovery within a decade, but are unlikely to preclude discovery.  {\it This renders our forecasts robust against analysis choices.}  In summary, the impact of the different analysis choices on the UHE neutrino flux discovery potential is as follows:
\begin{itemize}
 \item
  The size of the atmospheric muon background has only a mild impact---as long as it only affects the lowest energy bins, as predicted by current hadronic models.  See Section~\ref{sec:discovery_results_impact_muon_bg}.
 \item
  The normalization and, especially, the spectral index of the UHE tail of the background IceCube high-energy neutrino flux has a large impact; a softer spectrum yields a smaller background, which hastens the discovery of UHE neutrino flux models 3--12.  See Section~\ref{sec:discovery_results_impact_neutrino_bg}.
 \item
  Using an informed prior on the cut-off energy of the background UHE tail of the IceCube high-energy neutrino flux, $E_{\nu, {\rm cut}}^{\rm HE}$, may significantly hasten flux discovery, even if the prior is based on limited knowledge.  See Section~\ref{sec:discovery_results_impact_prior_cut-off}.
 \item
  Using an informed prior on the UHE $\nu N$ cross section, $f_\sigma$, may hasten flux discovery moderately, if the prior is based on limited knowledge, or substantially, if it is based on precise knowledge.  See Section~\ref{sec:discovery_results_impact_cross_section}.
 \item
  Poor detector resolution on shower energy, $\sigma_\epsilon$, and zenith angle, $\sigma_{\theta_z}$, may appreciably delay flux discovery.  See Section~\ref{sec:discovery_results_impact_resolution}.
 \item 
  Because Earth-skimming events, with $\theta_z^{\rm rec} = [80^\circ, 100^\circ]$, provide most of the UHE neutrino flux discovery potential, a detector with a total effective volume equivalent to that of our baseline design, but focused on the horizontal directions, could enhance discovery opportunities.  See Section~\ref{sec:discovery_results_horizontal_events}.
\end{itemize}

Finally, we also found that the UHE tail of the IceCube high-energy neutrino flux, augmented with a high-energy cut-off, \equ{powerlaw_cutoff}---which is typically a background for the discovery of other flux models---may itself be discovered.  Depending on the value of the cut-off energy, which determines the UHE event rate induced by this flux, discovery may occur within months, if the cut-off energy is high, or years, if it is low.  See Section~\ref{sec:discovery_results_uhe_ic_tail}.


\section{Flux model separation}
\label{sec:model_separation}

In Section~\ref{sec:discovery_potential}, we discussed the UHE neutrino flux discovery potential of benchmark flux models 1--12.  Here we tackle a related question: how well can two UHE neutrino flux models be experimentally distinguished from each other?  To answer it, we consider two hypotheses: the {\it true signal hypothesis}, built assuming knowledge of which is the ``true" neutrino flux model, and the {\it test signal hypothesis}, built for alternative, ``test" models.  Below, we forecast how well these hypotheses can be experimentally distinguished in the radio array of IceCube-Gen2.  We focus on benchmark flux models 3--12.  Like in Section~\ref{sec:discovery_potential}, we account for the background from atmospheric muons and the UHE tail of the IceCube high-energy neutrino flux, for random statistical fluctuations in the event rate, and for the uncertainty in analysis parameters.  We adopt the same baseline analysis choices as for the flux discovery potential (Table~\ref{tab:analysis_choices_base} and Section~\ref{sec:discovery_results_nominal_choices}), but limit our exploration of alternative analysis choices to the effect of different choices for the background UHE tail of the IceCube high-energy neutrino flux (Appendix~\ref{sec:appendix_ic_uhe_tail_bg}) and of the detector energy and angular resolution (Appendix~\ref{sec:appendix_detector_resolution}).


\subsection{Statistical analysis}
\label{sec:model_separation_stat_analysis}

We model the statistical analysis used for model separation closely after the analysis used for flux discovery introduced in Section~\ref{sec:discovery_stat_analysis}.  For a given choice of the true UHE neutrino flux, $\mathcal{M}_{\rm UHE}^{\rm true}$, of the test UHE neutrino flux, $\mathcal{M}_{\rm UHE}^{\rm test}$, and of the background UHE tail of the IceCube high-energy neutrino flux, $\mathcal{M}_{\rm HE}$, we compute the likelihood function under the true and test hypotheses, $\mathcal{L}_{\mathcal{M}_{\rm UHE}^{\rm true}, \mathcal{M}_{\rm HE}}^{({\rm s}+{\rm bg})}$ and $\mathcal{L}_{\mathcal{M}_{\rm UHE}^{\rm test}, \mathcal{M}_{\rm HE}}^{({\rm s}+{\rm bg})}$, respectively, using \equ{likelihood_s_bg}.  The true and test flux is any of the benchmark flux models 3--12; see Section~\ref{sec:fluxes} and \figu{benchmark_spectra}.  When computing these likelihood functions we sample the value of the observed event rate in each energy and angular bin, $N_{{\rm obs}, ij}$, at random from a Poisson distribution with central value $N_{{\rm pred}, ij}^{({\rm s}+{\rm bg})}$ equal to the event rate predicted by the true UHE neutrino flux model, using \equ{event_rate_pred_s_bg}, which includes the background of atmospheric muons and the UHE tail of the IceCube flux.  We use the same random realization to compute the likelihood functions under the true and test hypotheses.  (Like before, we repeat this procedure using many random realizations; we explain this below.)

We compute the corresponding statistical evidence, $\mathcal{Z}_{\mathcal{M}_{\rm UHE}^{\rm true}, \mathcal{M}_{\rm HE}}^{({\rm s}+{\rm bg})}$ and $\mathcal{Z}_{\mathcal{M}_{\rm UHE}^{\rm test}, \mathcal{M}_{\rm HE}}^{({\rm s}+{\rm bg})}$, using \equ{evidence_s_bg}, and, with them, the {\it model separation} Bayes factor,
\begin{equation}
 \label{equ:bayes_factor_model_separation}
 \mathcal{B}_{\mathcal{M}_{\rm UHE}^{\rm true}, \mathcal{M}_{\rm UHE}^{\rm test}, \mathcal{M}_{\rm HE}}^{\rm sep}
 = 
 \frac{\mathcal{Z}_{\mathcal{M}_{\rm UHE}^{\rm true}, \mathcal{M}_{\rm HE}}^{({\rm s}+{\rm bg})}}
 {\mathcal{Z}_{\mathcal{M}_{\rm UHE}^{\rm test}, \mathcal{M}_{\rm HE}}^{({\rm s}+{\rm bg})}} \;,
\end{equation}
via which we report the model separation potential between two UHE neutrino flux models.  It represents the preference for the true signal hypothesis over the test signal hypothesis, given an observed event rate.  

Like we did for the flux discovery Bayes factor in Section~\ref{sec:discovery_stat_analysis}, we use {\sc UltraNest}~\cite{Ultranest} to find the statistical evidence under the true and test hypotheses, compute the model separation Bayes factor for $N_{\rm samples} = 10^4$ different random realizations of the observed event rate for each choice of $\mathcal{M}_{\rm UHE}^{\rm true}$, $\mathcal{M}_{\rm UHE}^{\rm test}$, and $\mathcal{M}_{\rm HE}$, and report only the mean Bayes factor, averaged over all the realizations, $\langle \mathcal{B}_{\mathcal{M}_{\rm UHE}^{\rm true}, \mathcal{M}_{\rm UHE}^{\rm test}, \mathcal{M}_{\rm HE}}^{\rm sep} \rangle$, computed similarly to \equ{bayes_factor_avg}.


\subsection{Results}
\label{sec:model_separation_results}

\begin{figure*}[t!]
 \centering
 \includegraphics[width=\textwidth]{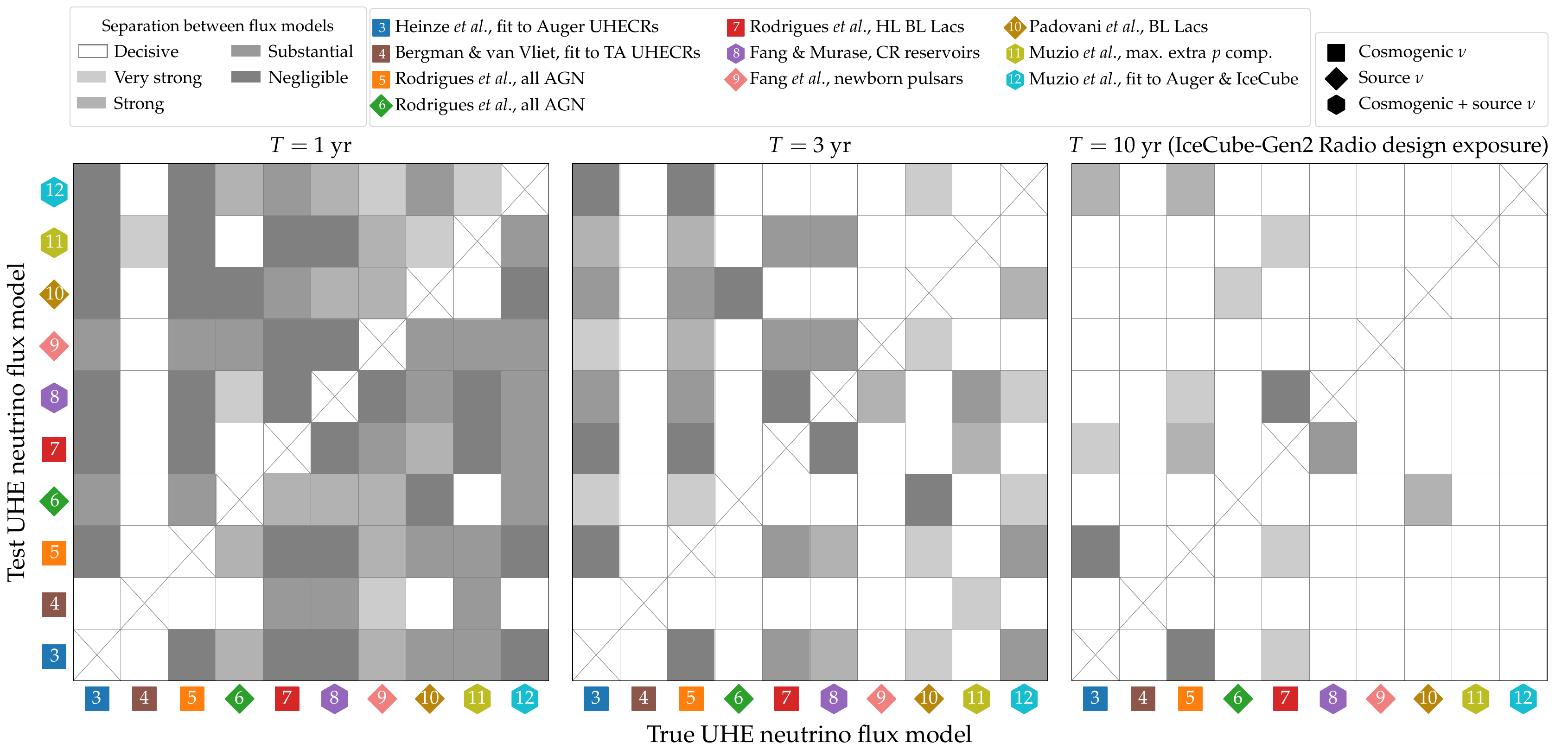}
 \caption{\label{fig:confusion_hard}Confusion matrix showing the experimental separation between true and test UHE neutrino flux models in the radio array of IceCube-Gen2, after an exposure time $T$.  The true flux model determines the observed event rate, and they are contrasted against event-rate predictions from the test models.  The color coding shows the mean model separation Bayes factor, $\langle \mathcal{B}_{\mathcal{M}_{\rm UHE}^{\rm true}, \mathcal{M}_{\rm UHE}^{\rm test}, \mathcal{M}_{\rm HE}}^{\rm sep} \rangle$, accounting for the background from atmospheric muons and from the UHE tail of the high-energy neutrino spectrum, and interpreted qualitatively using Jeffreys' table~\cite{jeffreys1998theory} (Section~\ref{sec:discovery_stat_analysis}).  All analysis choices are baseline; see Table~\ref{tab:analysis_choices_base} and  Section~\ref{sec:discovery_results_nominal_choices}.  {\it After 10~years, the majority of UHE neutrino flux models that are discoverable (see Section~\ref{sec:discovery_results_nominal_results}), can also be distinguished from each other; yet, some discoverable flux models with similar energy spectra can still be confused with each other.}  See Section~\ref{sec:model_separation} for details, and Appendices~\ref{sec:appendix_ic_uhe_tail_bg} and \ref{sec:appendix_detector_resolution} for results obtained under alternative analysis choices.}
\end{figure*}

Figure~\ref{fig:confusion_hard} shows the \textit{confusion matrix} for 1, 3, and 10 years of detector exposure time.  In each matrix, the horizontal axis shows the true UHE neutrino flux models, $\mathcal{M}_{\rm UHE}^{\rm true}$, which we use to generate the observed rate, and the vertical axis shows the test UHE neutrino flux models, $\mathcal{M}_{\rm UHE}^{\rm test}$.  Each entry in the confusion matrix represents the value of the model separation Bayes factor, \equ{bayes_factor_model_separation}, interpreted qualitatively according to Jeffreys' table~\cite{jeffreys1998theory}; see Section~\ref{sec:discovery_stat_analysis} for details.

Figure~\ref{fig:confusion_hard} shows that, as expected, at short exposure times most UHE neutrino flux models cannot be distinguished from each other.  (The exceptions are flux model 4 and, to a lesser extent, flux model 6, which have the highest event rates; see Table~\ref{tab:event_rates_binned}.)  This is because at short exposure times the observed event rate is generally low, so features in the energy and angular distribution of the observed events are resolved poorly, or not at all.  Model separation is further marred by the relatively large random statistical fluctuations that affect low event rates.

At longer exposure times, the observed event rate grows, features in the energy and angular event distribution become better resolved and more robust against random fluctuations.  Consequently, the observed features can be more cleanly contrasted against the features predicted by different flux models, and the true model may be more easily distinguished from others.  Accordingly, \figu{confusion_hard} shows that after 10~years most of the flux models that can be discovered (see Section~\ref{sec:discovery_results_nominal_results}) can also be distinguished from each other with at least strong evidence, and many with decisive evidence.  There are a few exceptions.  For instance, flux models 7 and 8 remain easy to confuse even after 10 years, because the energy spectrum of model 7, supplemented by the background of the UHE tail of the IceCube high-energy neutrino flux,  resembles the energy spectrum of model 8.  Naturally, flux models that yield low event rates and that are not expected to be discovered within a decade, \ie, models 1, 3, and 5 (see \figu{bayes_factor_hard}), cannot be distinguished from each other.  

A subtle feature of the confusion matrix is that it is nearly, but not exactly, symmetric along its diagonal.  This is because, when comparing a pair of true and test flux models, the observed event rate and the size of its  statistical fluctuations are computed using the true flux model, while the predicted event rate is computed using the test flux model.  As a result, at short exposure times, when event rates are low and affected significantly by fluctuations, swapping the roles of the true and test flux models impacts the size of the fluctuations significantly.  This, in turn, impacts the model separation Bayes factor appreciably, and is reflected in the asymmetry of the confusion matrix.  At longer exposure times, when event rates are higher and more robust to fluctuations, swapping the true and test flux models does not impact the Bayes factor as much.  Consequently, the confusion matrix becomes more symmetric with exposure time.

The results for flux model separation above were obtained using our baseline analysis choices (see Table~\ref{tab:analysis_choices_base} and Section~\ref{sec:discovery_results_nominal_choices}); specifically, using a hard spectrum for the UHE tail of the IceCube high-energy neutrino flux, detector angular resolution of $\sigma_{\theta_z} = 2^\circ$, and detector energy resolution of $\sigma_\epsilon = 0.1$.  Appendix~\ref{sec:appendix_ic_uhe_tail_bg} contains confusion matrices generated instead using our soft and intermediate choices of the spectrum for the background UHE tail of the IceCube high-energy flux, introduced in Section~\ref{sec:background_nu}.  Figure~\ref{fig:confusion_other_nu_cases} shows appreciable improvement in the flux model separation when switching from the baseline hard to the soft background flux.  

Appendix~\ref{sec:appendix_detector_resolution} contains confusion matrices generated using poorer choices of detector angular and energy resolution.  Figure~\ref{fig:confusion_angular_resolution} shows that  poorer angular resolution has little effect on flux model separation.  This is because the angular distributions of events for all flux benchmark models are comparable.  Figure~\ref{fig:confusion_energy_resolution} shows, in contrast, that poorer energy resolution strongly erodes flux model separation.  This is because most of the model separation power stems from the differences between the event energy distributions of the different flux models.  For an energy resolution of a decade in shower energy, \ie, $\sigma_\epsilon = 1.0$, model separation is largely unfeasible, except for flux model 4 due to its high event rate.  This reveals that, while good detector energy resolution is important for flux discovery (see Section~\ref{sec:discovery_results_impact_resolution}), it is essential for flux model separation.


\section{Future directions}
\label{sec:limits_improvements}

To produce our forecasts of the UHE neutrino flux discovery potential and model separation above, we used state-of-the-art detector simulations and theoretical input.  Yet, there are potential improvements that could be implemented in future revisions.  None of them represents a fundamental limitation of our present analysis.  

Since our calculation framework is similar to that of \Refe~\cite{Valera:2022ylt}, it shares potential directions of future improvement identified in that work: raising the maximum neutrino energy in in-Earth propagation, including the Landau–Pomeranchuk–Migdal effect in the relation between neutrino and shower energies, including the contribution of secondary leptons in the detector effective volumes for $\nu_\mu$- and $\nu_\tau$-initiated CC showers, improving the modeling of angular resolution, using an unbinned likelihood analysis rather than a binned one, including the background of air-shower cores, including nuclear effects in the cross section, using flavor identification, and jointly measuring the $\nu N$ cross section, flux normalization and spectral shape.  See \Refe~\cite{Valera:2022ylt} for details.  Work is ongoing on several of these fronts.  

Below, we present additional potential future improvements that are directly relevant to the present work, listed roughly in order of implementation challenge.

\smallskip
{\it Characterizing the UHE tail of the IceCube high-energy neutrino flux.---}  In our forecasts, we found that the main background to the discovery of flux model 3--12, and the separation between them, may be from the UHE tail of the IceCube high-energy neutrino flux, especially if it has a hard energy spectrum and a high cut-off energy; see Section~\ref{sec:background}.  In our results, we factored in already the large uncertainty in the position of the cut-off energy, but not the uncertainties on the flux normalization and spectral index.  Instead, we fixed them to their current best-fit values, which is reasonably motivated: by the time that the radio array IceCube-Gen2 gathers sufficient UHE data to perform the above analyses, it is likely that the TeV--PeV range of the flux will have been precisely characterized, by IceCube, the optical array of IceCube-Gen2, or by a combination of neutrino telescopes~\cite{Schumacher:2021hhm}.  Even then, it is foreseeable that the UHE tail of the IceCube high-energy neutrino flux will be known imprecisely due to the paucity of UHE events.  Thus, factoring in the full uncertainty on the shape of the UHE tail, not only from the cut-off energy, but also from its normalization and spectral index, may weaken future revised forecasts.

\smallskip
{\it Reconstructing the UHE neutrino energy spectrum.---}  When computing the UHE neutrino flux discovery potential of benchmark flux models 1--12, we did not quantify how well their neutrino energy spectrum could be reconstructed.  Yet, doing this is critical to being able to claim flux discovery and model separation without the theory bias that comes from working with a limited collection of flux models that, while representative of the model parameter space, is evidently not exhaustive.  Work in reconstructing the normalization and spectral shape of the UHE neutrino flux using a generic parametrization of the UHE neutrino flux, and in addition to jointly measure the $\nu N$ cross section, is ongoing~\cite{Valera:InPrep}. 

\smallskip
{\it Applying our analysis methods to other UHE neutrino telescopes.---}  To make concrete forecasts that represent realistic experimental nuance, we geared them to IceCube-Gen2, presently in advanced planning stages.  Yet, it is straightforward to apply our analysis methods to compute event rates, flux discovery potential, and flux model separation to other UHE neutrino telescopes~\cite{Ackermann:2022rqc}, radio-based or otherwise, without large alteration.  Section~\ref{sec:event_rates} shows that the particulars of the detector affect the calculation via the detector geometry, \ie, when computing where neutrinos hit the detector after propagating inside the Earth, and in the modeling of the detector response via the energy- and direction-dependent effective volume.  Given the same information for a different UHE neutrino telescope, our methods can be repeated.

\smallskip
{\it Informing the design of the radio array of IceCube-Gen2 and of other detectors.---}  Conversely, our methods can be used to optimize the design of the IceCube-Gen2 detector---or any future detector---based on its potential to discover the UHE neutrino flux, in terms of effective volume, energy resolution, and angular resolution. 

\smallskip
{\it Combining the optical and radio arrays of IceCube-Gen2.---}  We focused our forecasts exclusively on the radio array of IceCube-Gen2.  Yet, the planned design of IceCube-Gen2~\cite{IceCube-Gen2:2020qha} includes also a large extension of the optical array that is expected to characterize in detail the TeV--PeV neutrino flux beyond the capabilities of IceCube.  Because the optical array will be sensitive to neutrinos with up to roughly 10~PeV, using it in combination with the radio array could enhance the discovery potential of UHE neutrino flux models that peak at relatively low energies, and the separation between flux models that differ primarily at low energies.  References~\cite{vanSanten:2022wss, toise} have early results in this direction.


\section{Summary and outlook}
\label{sec:summary}

Ultra-high-energy (UHE) neutrinos, with EeV-scale energies, represent the ultimate high-energy neutrino frontier~\cite{Berezinsky:1969erk, Stecker:1978ah}.  Sought unsuccessfully for the past half-century, there is a real chance of finally discovering them in the next 10--20 years, thanks to new large-scale UHE neutrino telescopes presently under development~\cite{Ackermann:2022rqc}.  Their discovery would reveal key insight into extant questions in astrophysics and particle physics~\cite{Ahlers:2018fkn, Ahlers:2018mkf, Ackermann:2019cxh, Ackermann:2019ows,  AlvesBatista:2019tlv, Arguelles:2019rbn, AlvesBatista:2021gzc, Abraham:2022jse, Ackermann:2022rqc, Adhikari:2022sve}. 

So far, existing forecasts of the discovery of UHE neutrinos, and general-purpose methods to produce them, while pioneering, have of necessity lacked detail.  Still, further work from theory and phenomenology is needed to accurately forecast the discovery potential of upcoming detectors.  In dialogue with ongoing experimental development, these forecasts will help map out near-future capabilities and may inform design choices and science programs of upcoming detectors.

Our work addresses this need.  We have produced detailed forecasts of the discovery of a diffuse flux of UHE neutrinos, aimed at upcoming UHE neutrino telescopes.  (The discovery of point sources of UHE neutrinos is addressed elsewhere, \eg, in \Refes~\cite{Fang:2016hop, Fiorillo:2022ijt}.)  To make our forecasts realistic, robust, and useful, we factor in nuance that previous works either considered partially or not at all.  By design, and inasmuch as possible, our forecasts are anchored in detailed theory and experimental considerations.  Despite being tempered by them, we have found encouraging prospects.

On the theory front, since the diffuse flux of UHE neutrinos is predicted uncertainly, we have considered a large number of competing benchmark predictions~\cite{Fang:2013vla, Padovani:2015mba, Fang:2017zjf, Heinze:2019jou, Muzio:2019leu, Rodrigues:2020pli, Anker:2020lre, IceCube:2020wum,  Muzio:2021zud, IceCube:2021uhz}, built on diverse assumptions, that span the allowed parameter of flux models presently allowed, from optimistic to pessimistic; see \figu{benchmark_spectra}.  The models include extrapolations of the TeV--PeV IceCube neutrino flux to ultra-high energies~\cite{IceCube:2020wum, IceCube:2021uhz}, cosmogenic neutrinos~\cite{Heinze:2019jou, Anker:2020lre, Rodrigues:2020pli}, neutrinos made in astrophysical sources~\cite{Rodrigues:2020pli, Fang:2013vla, Padovani:2015mba}, and combinations of the latter two~\cite{Fang:2017zjf, Muzio:2019leu, Muzio:2021zud}. 

On the experimental front, we have used state-of-the-art ingredients to compute the propagation of neutrinos through the Earth (Section~\ref{sec:propagation}), of neutrino-induced event rates at the detector (Section~\ref{sec:event_rate_benchmarks}), including dedicated simulations of the detector response, and of neutrino and atmospheric muon backgrounds (Section~\ref{sec:background}).  We factored in the uncertainty in the UHE neutrino-nucleon cross section, which affects neutrino in-Earth propagation and detection.  We made our forecasts concrete by focusing on UHE neutrino detection in the envisioned radio array of IceCube-Gen2~\cite{IceCube-Gen2:2020qha}, whose target sensitivity is among the best~\cite{Ackermann:2022rqc}.  We produced forecasts using a Bayesian statistical approach, and reported them via Bayes factors that account for random statistical fluctuations in the predicted event rates (Section~\ref{sec:discovery_stat_analysis}).

In our baseline results, we adopted conservative analysis choices for the detector capabilities, backgrounds, and neutrino-nucleon cross section (Table~\ref{tab:analysis_choices_base} and Section~\ref{sec:discovery_results_nominal_choices}).  With them, and even after accounting for the above experimental nuance, {\it we found (\figu{bayes_factor_hard}) that most of our benchmark UHE neutrino diffuse flux models may be  discovered decisively (\ie, with a Bayes factor larger than 100) within ten years of operation of the radio array of IceCube-Gen2; most of them, within a handful of years and some, within a few months.}  On average, discoverable flux models are expected to induce roughly 10--300 events with energies from 10~PeV to 10~EeV per decade (Table~\ref{tab:event_rates_binned}).  Discovery may be claimed sooner at a lower statistical significance, or by adopting alternative analysis choices (Table~\ref{tab:analysis_choices_alt}).  Flux models with less than one event per decade will remain undiscovered; these include, \eg, some cosmogenic neutrino flux models fit to the heavy UHECR mass composition measured by Auger~\cite{Heinze:2019jou}.

We found that the potential UHE tail of the IceCube TeV--PeV neutrino flux may be the dominant background to discovering the benchmark flux models.  In some cases, knowing the precise energy where the background IceCube neutrino flux cuts off may reduce the time needed for the discovery of an UHE neutrino flux model from several years to a few months (Section~\ref{sec:discovery_results_impact_prior_cut-off}).  This stresses the need for a precise understanding of the size and shape of the high-energy tail of the IceCube neutrino flux.  Detectors that will target the tens-of-PeV range, like TAMBO~\cite{Romero-Wolf:2020pzh}, Trinity~\cite{Otte:2019knb}, and the optical array of IceCube-Gen2~\cite{IceCube-Gen2:2020qha}, should prove valuable.  The UHE tail of the IceCube flux may itself be discovered in the radio array of IceCube-Gen2 within 10~years, provided its spectrum is hard~\cite{IceCube:2021uhz} and its cut-off is beyond 50~PeV (Section~\ref{sec:discovery_results_uhe_ic_tail}).  This opens up the possibility that the IceCube flux span the TeV--EeV range and that it could be used as a calibration flux bridging TeV--PeV-scale telescopes and EeV-scale telescopes.

Finally, we have found that, in the event of UHE neutrino flux discovery, it should be possible within 10~years to distinguish between nearly all our competing benchmark flux models (Section~\ref{sec:model_separation}).  The power to separate competing flux predictions stems from the differences in the energy distribution of the events that they induce. 

We provide our forecasts and methods in the hope of complementing ongoing work in the planning and building of UHE neutrino telescopes.  We encourage experimental collaborations to adopt our methods, or similar ones; or, alternatively, to make publicly available the simulated response function of their detectors.  In light of our findings, the coming decades have a real chance to bring transformative progress to astroparticle physics.


\section*{Acknowledgements}

We thank Douglas Bergman, Ke Fang, Alfonso Garc\'ia, Kohta Murase,  Foteini Oikonomou, and  Juan Rojo for valuable discussion and input and, especially, Jonas Heinze, Kumiko Kotera, Marco Muzio, Paolo Padovani, Xavier Rodrigues, and Arjen van Vliet for providing or helping to generate the detailed neutrino fluxes used in this work.  VBV is grateful to Olga Mena and the Instituto de F\'isica Corpuscular (IFIC), Universidad de Valencia, for their hospitality during part of the developement of this work.  MB and VBV are supported by the {\sc Villum Fonden} under project no.~29388.  This work used resources provided by the High Performance Computing Center at the University of Copenhagen.  This work was made possible by Institut Pascal at Universit\'e Paris-Saclay during the Paris-Saclay Astroparticle Symposium 2021, with the support of the P2IO Laboratory of Excellence (programme ``Investissements d’avenir" ANR-11-IDEX-0003-01 Paris-Saclay and ANR-10-LABX-0038), the P2I research departments of the Paris-Saclay University, as well as IJCLab, CEA, IPhT, APPEC, the IN2P3 master project UCMN, and EuCAPT.  The computations and data handling were enabled by resources provided by the Swedish National Infrastructure for Computing (SNIC) at UPPMAX partially funded by the Swedish Research Council through grant agreement no.~2018-05973.


\newpage
\onecolumngrid
\appendix


\section{Impact of the background UHE tail of the IceCube high-energy neutrino flux\\on all UHE neutrino flux models}
\label{sec:appendix_ic_uhe_tail_bg}
\renewcommand{\thefigure}{A\arabic{figure}}
\setcounter{figure}{0}

Figure~\ref{fig:bayes_other_nu_cases} in the main text showed, for benchmark UHE neutrino flux models 6 and 7~\cite{Rodrigues:2020pli}, the impact on the flux discovery potential of using our three choices of background UHE tail of the IceCube high-energy neutrino flux: hard-spectrum---our conservative baseline choice---intermediate, and soft-spectrum. See Section~\ref{sec:background_nu} for details.  

\begin{figure*}[t!]
 \centering
 \includegraphics[width=\textwidth]{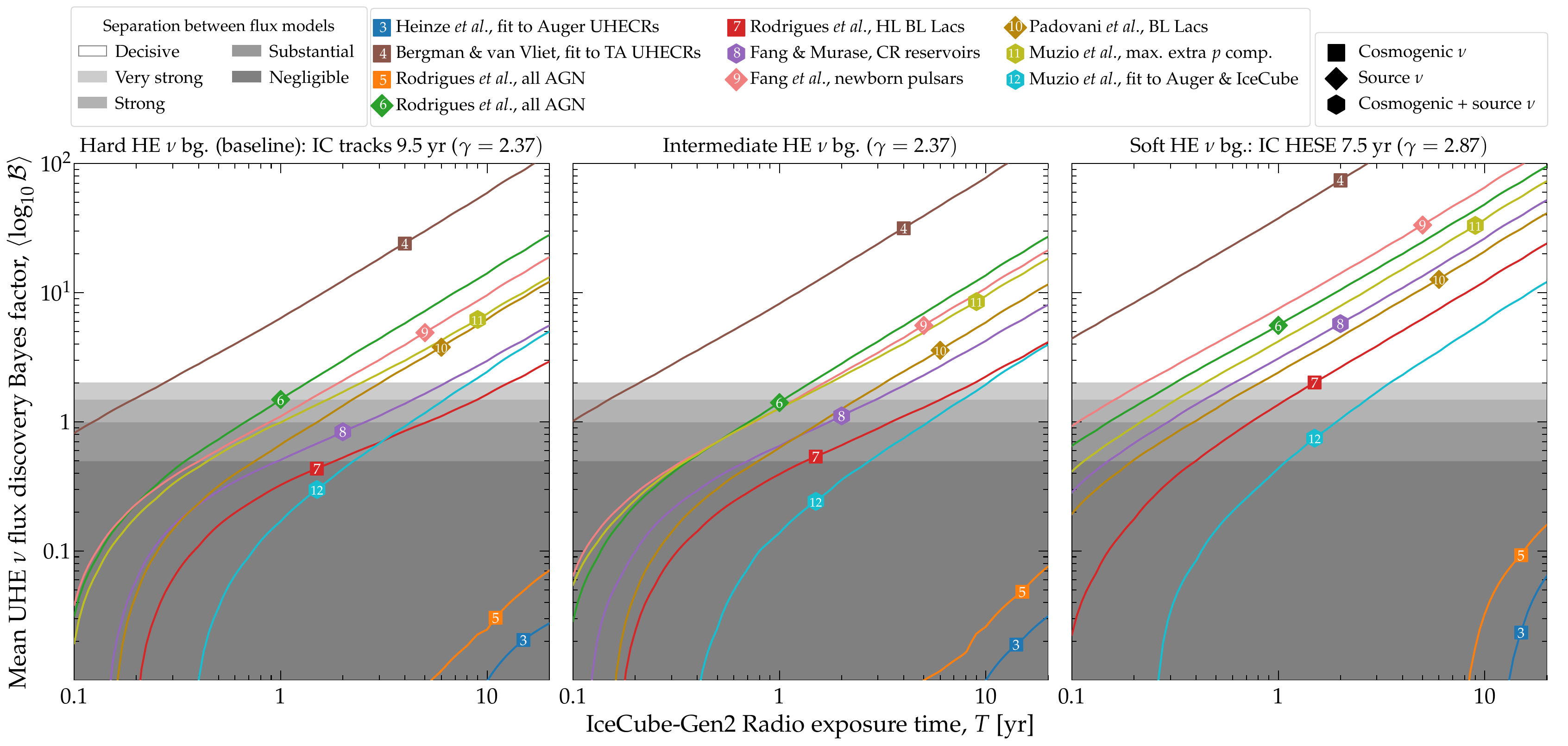}
 \caption{\label{fig:bayes_other_nu_cases_app}Discovery potential of benchmark UHE neutrino flux models 3--12~\cite{Fang:2013vla, Padovani:2015mba, Fang:2017zjf, Heinze:2019jou, Muzio:2019leu, Rodrigues:2020pli, Anker:2020lre, Muzio:2021zud} (see \figu{benchmark_spectra}) in the radio array of IceCube-Gen2, for three different choices of the background UHE tail of the IceCube high-energy neutrino flux; see Section~\ref{sec:background_nu}. All other analysis choices are baseline and conservative; see Table~\ref{tab:analysis_choices_base} and Section~\ref{sec:discovery_results_nominal_choices}.  UHE neutrino flux models 1~\cite{IceCube:2020wum} and 2~\cite{IceCube:2021uhz}---the UHE extrapolation of the IceCube high-energy neutrino flux---are not included in this figure because in their analysis the sole background is from atmospheric muons; see Section~\ref{sec:discovery_results_uhe_ic_tail}. \textit{Left:} Baseline choice of a hard-spectrum background, with spectral index $\gamma = 2.37$, motivated by the IceCube 9.5-year through-going $\nu_\mu$ analysis~\cite{IceCube:2021uhz}.  Results match \figu{bayes_factor_hard} in the main text. \textit{Center:} Intermediate background, with the same normalization, but $\gamma = 2.50$. \textit{Right:} Soft-spectrum background, with $\gamma = 2.87$, motivated by the IceCube 7.5-year HESE analysis~\cite{IceCube:2020wum}.  See Appendix~\ref{sec:appendix_ic_uhe_tail_bg} and Section~\ref{sec:discovery_results_impact_neutrino_bg} for details.}
\end{figure*}

Figure~\ref{fig:bayes_other_nu_cases_app} extends the result to all flux models 3--12.  The same conclusions as in Section~\ref{sec:discovery_results_impact_neutrino_bg} hold: using a softer background UHE tail of the IceCube high-energy neutrino flux may expedite flux discovery significantly.  However, UHE neutrino flux models with meager event rates, like flux models 3 and 5 (see Table~\ref{tab:event_rates_binned}) remain undiscoverable regardless of the choice of background.

Figure~\ref{fig:confusion_hard} in the main text showed, for all benchmark UHE neutrino flux models, the confusion matrix representing the degree of separation between models achievable, after 1, 3, and 10 years of detector exposure.  Those results were computed under our baseline choice of a hard-spectrum UHE tail of the IceCube high-energy neutrino flux.

\begin{figure*}[t!]
 \centering
 \includegraphics[width=\textwidth]{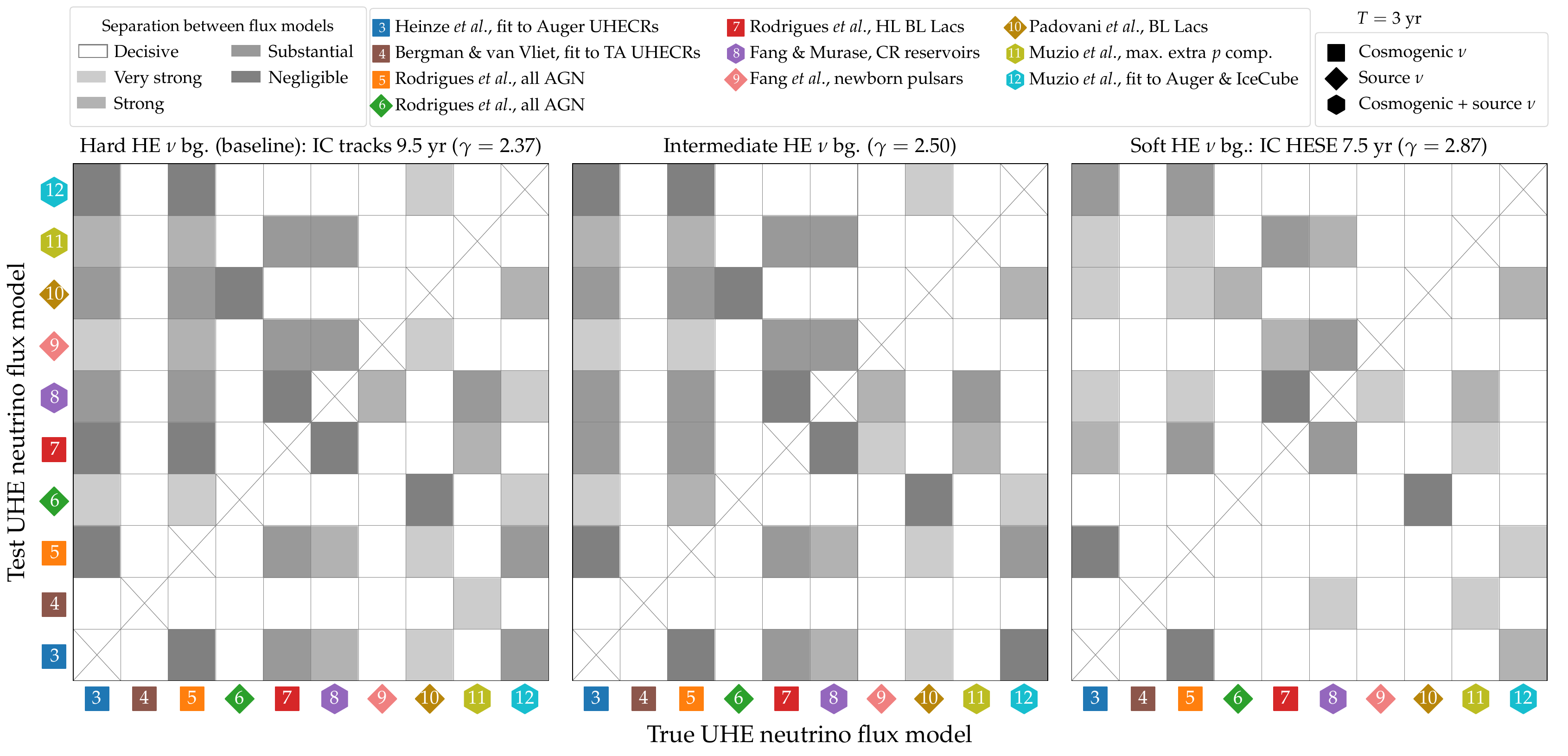}
 \caption{\label{fig:confusion_other_nu_cases}Confusion matrix showing the experimental separation between true and test UHE neutrino flux models 3--12 (see \figu{benchmark_spectra}) in the radio array of IceCube-Gen2, after $T = 3$~years of detector exposure, computed using the same three choices for the background UHE tail of the IceCube high-energy neutrino flux as in \figu{bayes_other_nu_cases_app}; see Section~\ref{sec:background_nu}.  All other analysis choices are baseline and conservative; see Table~\ref{tab:analysis_choices_base} and Section~\ref{sec:discovery_results_nominal_choices}. \textit{Left:} Baseline choice of a hard-spectrum background.  Results match \figu{confusion_hard} in the main text.  \textit{Center:} Intermediate background.  \textit{Right:} Soft-spectrum background.  See Appendix~\ref{sec:appendix_ic_uhe_tail_bg} and Section~\ref{sec:model_separation_results} for details.}
\end{figure*}

Figure~\ref{fig:confusion_other_nu_cases} shows the confusion matrix computed using also the intermediate- and soft-spectrum choices of the UHE tail of the IceCube high-energy neutrino flux, for a fixed exposure time of 3 years.  Using a softer background appreciably improves the separation between UHE neutrino flux models 3--12, since it allows the features of their event energy distributions to be resolved more cleanly.


\section{Impact of the detector angular and detector resolution on all UHE neutrino flux models}
\label{sec:appendix_detector_resolution}
\renewcommand{\thefigure}{B\arabic{figure}}
\setcounter{figure}{0}

Figure~\ref{fig:bayes_factor_res_main} in the main text showed, for benchmark UHE neutrino flux model 6, the impact on the flux discovery potential of alternative choices of the detector energy and angular resolution.  In connection to \figu{bayes_factor_res_main}, in Section~\ref{sec:discovery_results_impact_resolution} we found that poorer energy resolution hinders flux discovery by diluting the features of the event energy spectrum and poorer angular resolution, by preserving the innate degeneracy between the UHE neutrino flux and cross section (see Section~\ref{sec:propagation}).  Here we extend these results to all the benchmark UHE neutrino flux models 1--12.  As for \figu{bayes_factor_res_main}, when changing the energy and angular resolution, we change the size of the bins of reconstructed shower energy and reconstructed zenith angle commensurately; see Section~\ref{sec:discovery_results_impact_resolution}.

\begin{figure*}[t!]
 \centering
 \includegraphics[width=\textwidth]{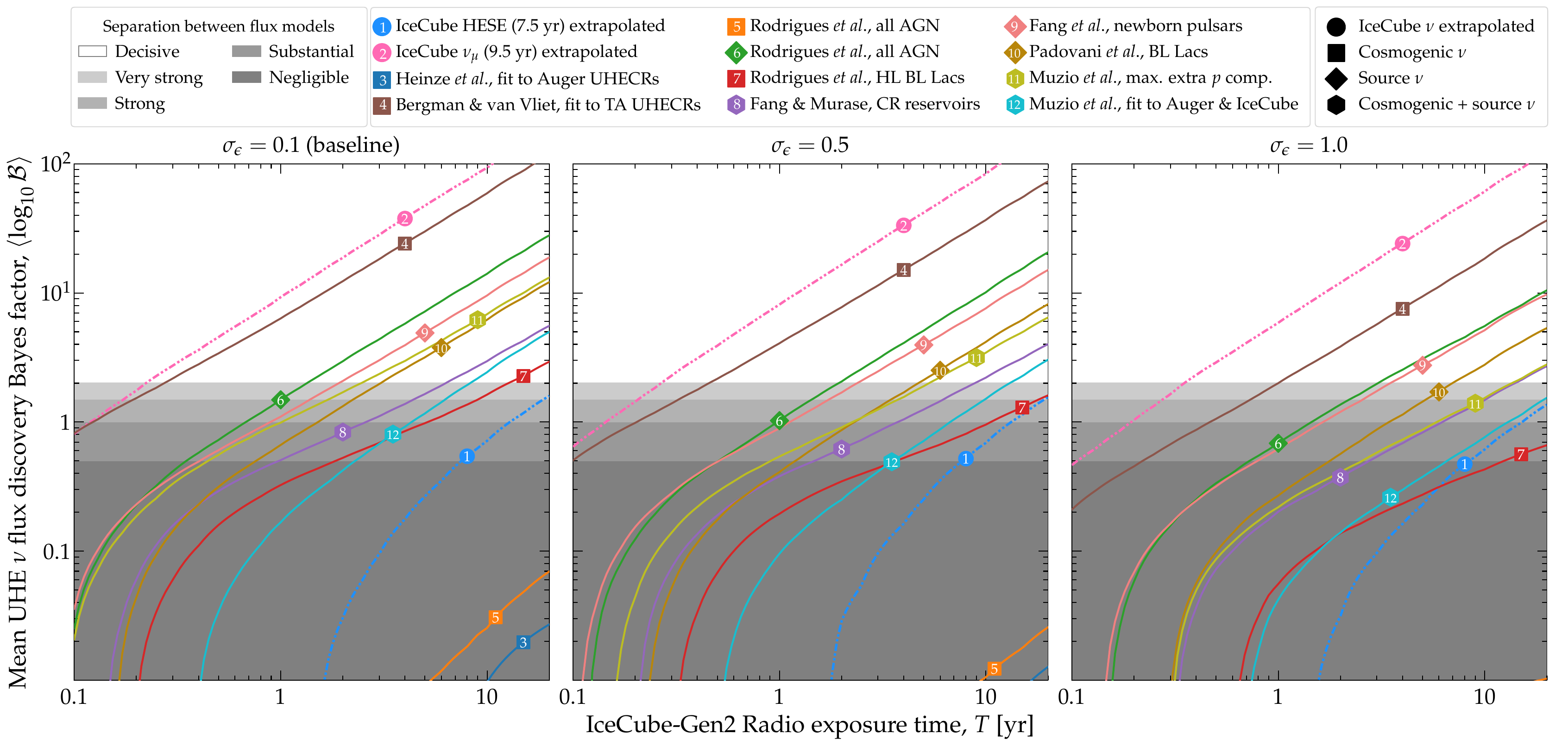}
 \caption{\label{fig:bayes_energy_resolution}Impact of the resolution of the radio array of IceCube-Gen2 in measuring the reconstructed energy, $\sigma_\epsilon = 0.1$ ({\it left}, our baseline choice), $0.5$ ({\it center}), and $1.0$ ({\it right}), on the flux discovery potential of UHE neutrino flux models 1--12.  All other analysis choices are baseline; see Table~\ref{tab:analysis_choices_base} and  Section~\ref{sec:discovery_results_nominal_choices}.   See Appendix~\ref{sec:appendix_detector_resolution} and Section~\ref{sec:discovery_results_impact_resolution} for details.}
\end{figure*}

\begin{figure*}[t!]
 \centering
 \includegraphics[width=\textwidth]{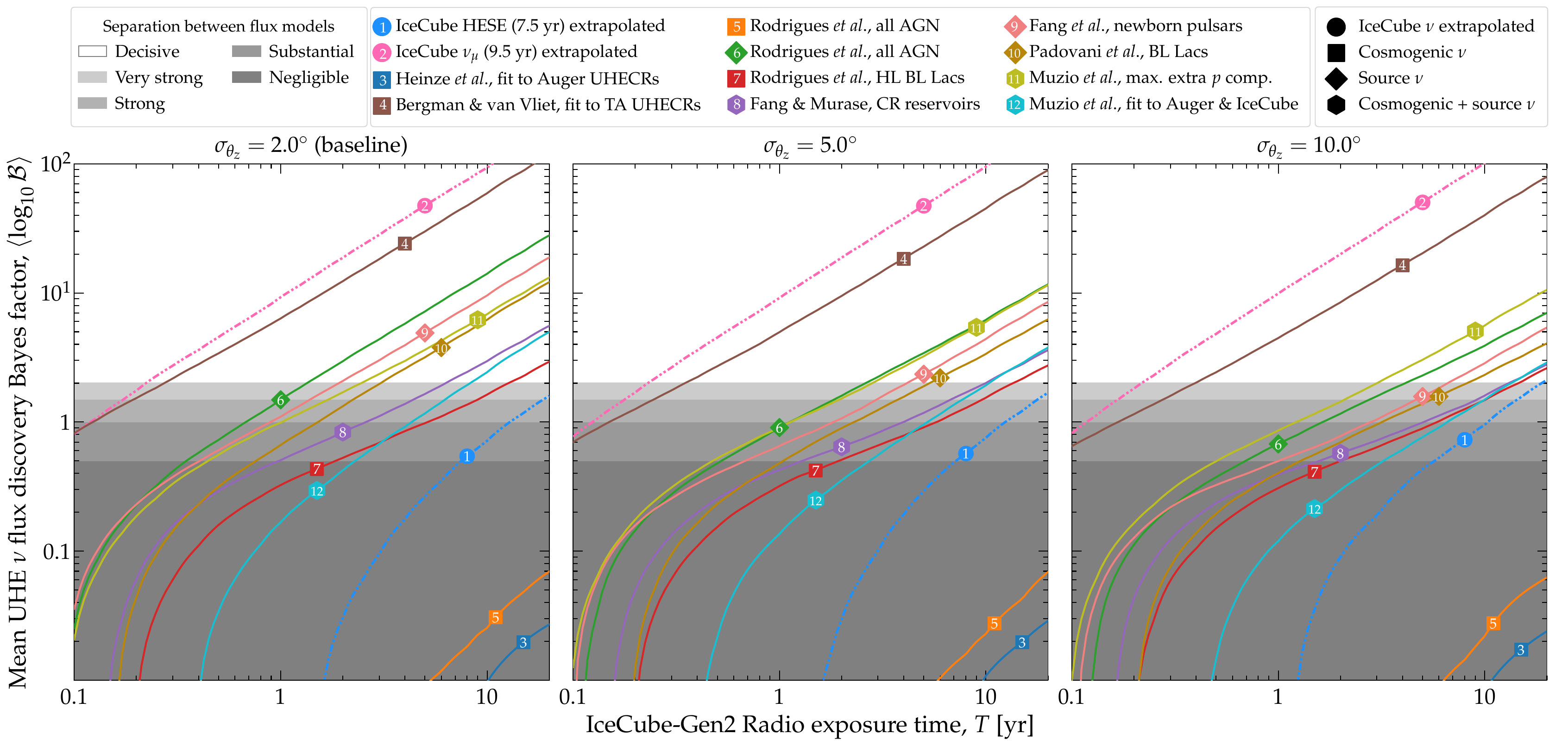}
 \caption{\label{fig:bayes_angular_resolution}Impact of the resolution of the radio array of IceCube-Gen2 in measuring the reconstructed zenith angle, $\sigma_{\theta_z} = 2^\circ$ ({\it left}, our baseline choice), $5^\circ$ ({\it center}), and $10^\circ$ ({\it right}), on the flux discovery potential of UHE neutrino flux models 1--12.  All other analysis choices are baseline; see Table~\ref{tab:analysis_choices_base} and  Section~\ref{sec:discovery_results_nominal_choices}.   See Appendix~\ref{sec:appendix_detector_resolution} and Section~\ref{sec:discovery_results_impact_resolution} for details.}
\end{figure*}

Figures~\ref{fig:bayes_energy_resolution} and \ref{fig:bayes_angular_resolution} show, respectively, the flux discovery potential for detector energy resolution of $\sigma_\epsilon = 0.1$ (our baseline choice), 0.5, and 1.0, and for detector angular resolution of $\sigma_{\theta_z} = 2^\circ$ (our baseline choice), $5^\circ$, and $10^\circ$, for the benchmark UHE neutrino flux models 3--12.  The results are similar as for flux model 6 in \figu{bayes_factor_res_main}.

\begin{figure*}[t!]
 \centering
 \includegraphics[width=\textwidth]{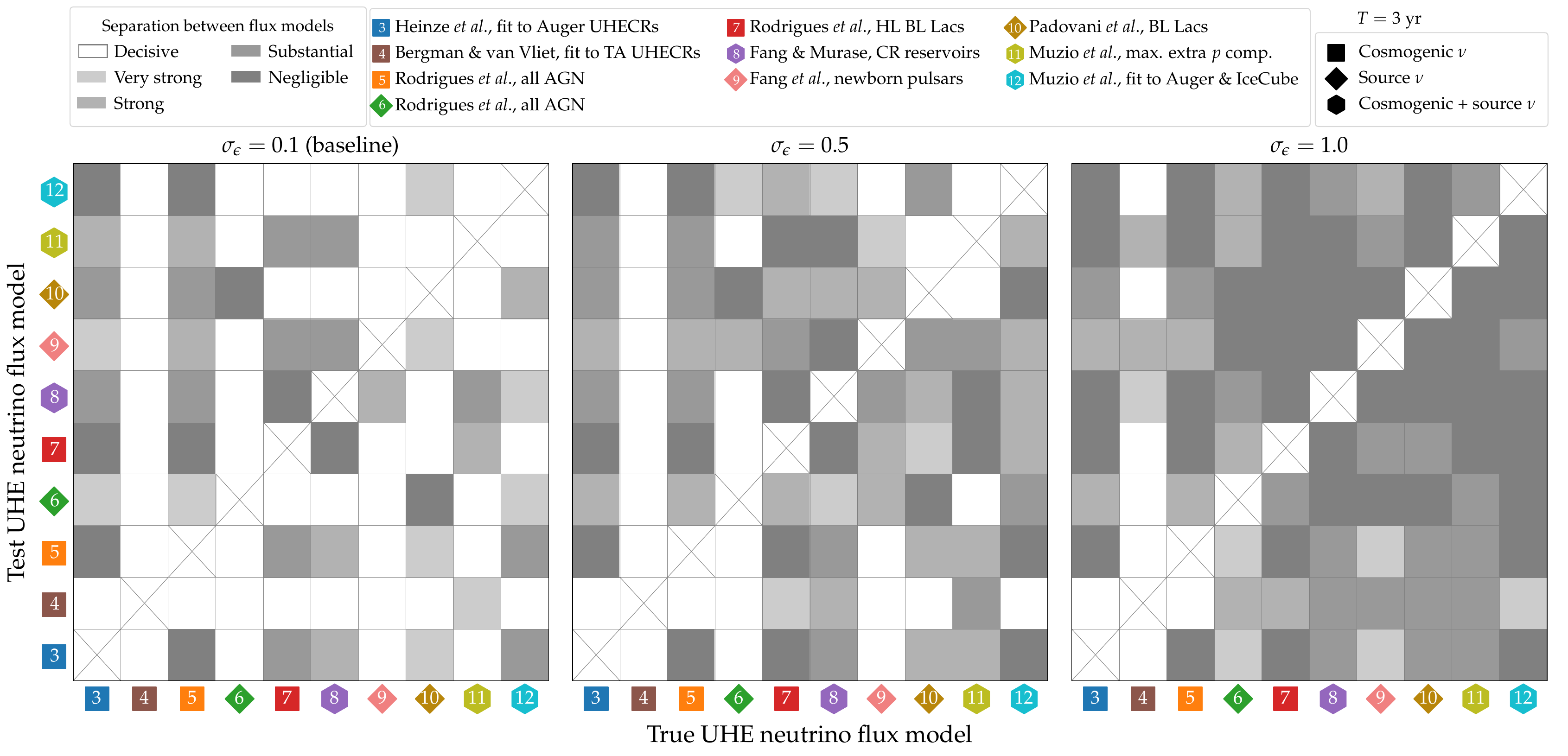}
 \caption{\label{fig:confusion_energy_resolution} Impact of the resolution of the radio array of IceCube-Gen2 in measuring the reconstructed energy, $\sigma_\epsilon = 0.1$ ({\it left}, our baseline choice), $0.5$ ({\it center}), and $1.0$ ({\it right}), on the confusion matrix that represents the separation between true and test UHE neutrino flux models 3--12, after $T = 3$~years of detector exposure.   All other analysis choices are baseline; see Table~\ref{tab:analysis_choices_base} and  Section~\ref{sec:discovery_results_nominal_choices}.  The left panel coincides with \figu{confusion_hard} in the main text.  See Appendix~\ref{sec:appendix_detector_resolution} and Section~\ref{sec:model_separation} for details. }
\end{figure*}

\begin{figure*}[t!]
 \centering
 \includegraphics[width=\textwidth]{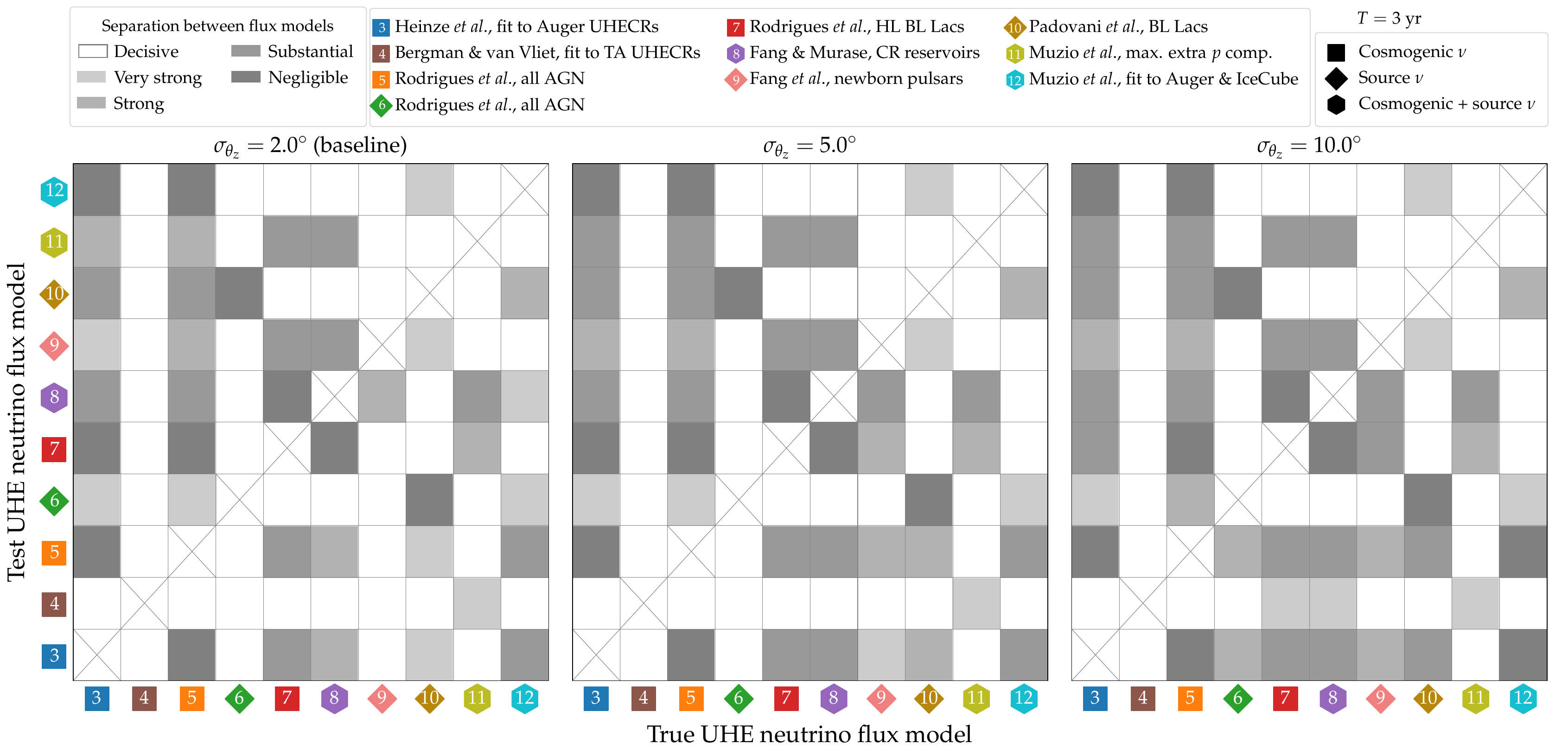}
 \caption{\label{fig:confusion_angular_resolution} Impact of the resolution of the radio array of IceCube-Gen2 in measuring the reconstructed zenith angle, $\sigma_{\theta_z} = 2^\circ$ ({\it left}, our baseline choice), $5^\circ$ ({\it center}), and $10^\circ$ ({\it right}), on the confusion matrix that represents the separation between true and test UHE neutrino flux models 3--12, after $T = 3$~years of detector exposure.   All other analysis choices are baseline; see Table~\ref{tab:analysis_choices_base} and  Section~\ref{sec:discovery_results_nominal_choices}.  The left panel coincides with \figu{confusion_hard} in the main text.  See Appendix~\ref{sec:appendix_detector_resolution} and Section~\ref{sec:model_separation} for details. }
\end{figure*}

Figures~\ref{fig:confusion_angular_resolution} and \ref{fig:confusion_energy_resolution} show the confusion matrix representing flux model separation, after 3~years of detector exposure, for the same choices of detector angular and energy resolution, respectively, as Figs.~\ref{fig:bayes_angular_resolution} and \ref{fig:bayes_energy_resolution}.  Since most of the power to separate between flux models comes from resolving the differences between their event energy distributions, model separation is affected more severely by poorer energy resolution than by poorer angular resolution.

Figure~\ref{fig:bayes_angular_resolution} shows a subtle feature: for some flux models, notably, for flux model 1, poorer angular resolution improves the discovery potential, which seems counter-intuitive.  The reasons behind this behavior expose limitations associated to using a binned likelihood to compute the flux discovery potential (Section~\ref{sec:discovery_stat_analysis}).  In our prescription, the event rates of low UHE neutrino fluxes, like flux model 1, that predict fewer than one event in a decade of detector exposure (Table~\ref{tab:event_rates_binned}), are plagued by a large number of unpopulated event-rate bins in many of the random realizations of their mock observed event rates that we use to compute the mean discovery Bayes factor (Section~\ref{sec:discovery_stat_analysis}).  (High UHE neutrino fluxes are unaffected because they do not have unpopulated event bins.)  When using poorer detector resolution in our forecasts, as in \figu{bayes_angular_resolution}, our prescription changes to using coarser event bins (Section~\ref{sec:discovery_results_impact_resolution}).  This reduces the number of unpopulated bins by merging formerly unpopulated with populated bins and this, in turn, improves the flux discovery potential of the low UHE neutrino flux models, as seen in \figu{bayes_angular_resolution} for flux model 1.

Figure~\ref{fig:bayes_energy_resolution} shows that, in contrast, when using poorer energy resolution the discovery potential of flux model 1 remains unchanged.  There are two competing effects responsible for this. On the one hand, like with poorer angular resolution, using a poorer energy resolution induces coarser binning and improves discovery prospects.  On the other hand, unlike with poorer angular resolution, using poorer energy resolution reduces the integrated event rate due to some events leaking out of the energy range of interest to our analysis, $E_{\rm sh}^{\rm rec} = 10^7$--$10^{10}$~GeV, on account of broader Gaussian energy resolution function; see Eq.~(15) in \Refe~\cite{Valera:2022ylt}.  These two effects balance each other out, leaving the discovery potential of flux model 1 unchanged.

Evidently, the above features are not physical, but rather limitations that stem from using a binned likelihood for scenarios of low event rates.  In revised versions of our analysis~\cite{Valera:InPrep}, these limitations will be overcome by switching to an unbinned likelihood analysis.

\section{Impact of the surface veto on the flux discovery potential}
\label{sec:impact_surface_veto}
\renewcommand{\thefigure}{C\arabic{figure}}
\setcounter{figure}{0}

\begin{figure*}[t!]
 \centering
 \includegraphics[width=0.5\textwidth]{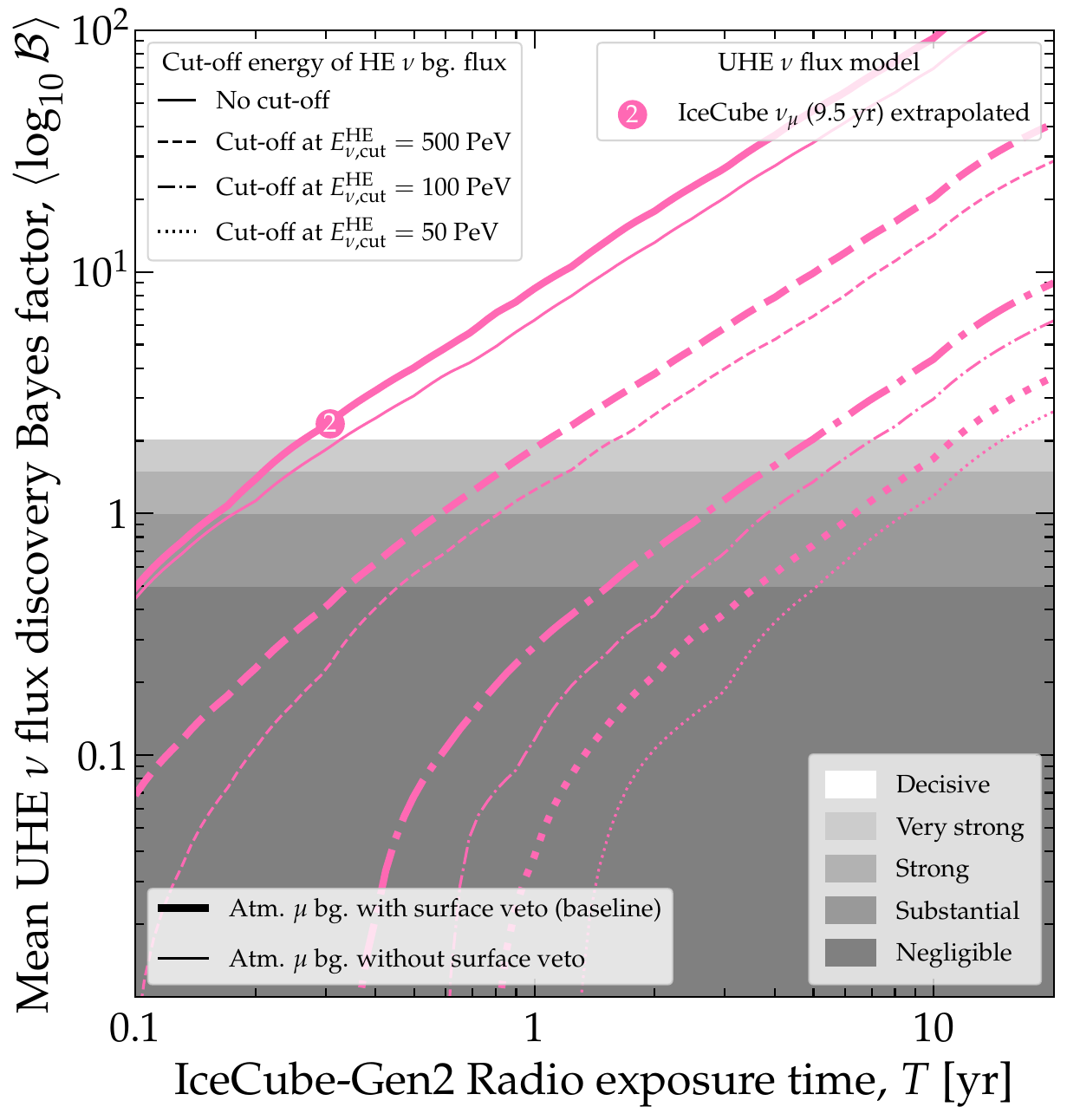}
 \caption{\label{fig:no_veto}Same as \figu{bayes_factor_no_marg_unbinned_ic_hard_atm_mu_bg_low} in the main text, for UHE neutrino flux model 2~\cite{IceCube:2021uhz} only, but showing the impact of using the veto of air-shower surface detectors to mitigate the background of atmospheric muons.  All other analysis choices are baseline and conservative; see Table~\ref{tab:analysis_choices_base} and Section~\ref{sec:discovery_results_nominal_choices}.  See Appendix~\ref{sec:impact_surface_veto} for details.}
\end{figure*}

Section~\ref{sec:discovery_results_impact_muon_bg} found that increasing the normalization of the background of atmospheric muons has little impact on the flux discovery potential.  (The same may not be true of changing the energy spectrum of the background.)  Our baseline analysis choice for the atmospheric muon background  includes applying a veto from air-shower surface detectors to mitigate it, following \Refe~\cite{Hallmann:2021kqk}; see Section~\ref{sec:background_mu} for details.  Removing the surface veto increases the background rate, especially at high energies, though not greatly; see Fig.~15 in \Refe~\cite{Valera:2022ylt}.  Here we study the impact of the surface veto on our results; we focus on its impact on the discovery of flux model 2~\cite{IceCube:2021uhz}, for which the only background is from atmospheric muons.

Figure~\ref{fig:no_veto} shows that switching off the surface veto shifts the discovery Bayes factor to longer exposure times.  This delays the discovery of flux model 2, but only slightly.  This weak impact is expected, since flux model 2---and our other benchmark UHE neutrino flux models (Section~\ref{sec:fluxes})---reach higher energies than the atmospheric muon background, even without the surface veto, and so can be clearly separated from it; see \figu{binned_events_all_benchmarks_vs_energy_dep} and Fig.~15 in \Refe~\cite{Valera:2022ylt}.  

Thus, the discovery of flux model 2 with a high cut-off energy is only delayed by a few months, since in that case events induced by it reach $E_{\rm sh}^{\rm rec} \gg 10^8$~GeV, far from the atmospheric muon background that is concentrated at $E_{\rm sh}^{\rm rec} \lesssim 10^8$~GeV.  The discovery of flux model 2 with a low cut-off energy is delayed longer; at worst, by a handful of years, for $E_{\nu, {\rm cut}}^{\rm HE} = 50$~PeV.  Other UHE neutrino flux models with meager associated event rates, like flux models 1, 3, and 5, remain undiscoverable even if the surface veto is switched off.  Overall, these results suggest that the surface veto, while helpful, might not be determinant for UHE neutrino flux discovery (barring a change in the atmospheric muon background that stretches it to higher energies; see Section~\ref{sec:discovery_results_impact_muon_bg}).


\section{Statistical significance of the mean Bayes factor}
\label{sec:stat_sig_bayes_factor}
\renewcommand{\thefigure}{D\arabic{figure}}
\setcounter{figure}{0}

\begin{figure*}[t!]
 \centering
 \includegraphics[width=\textwidth]{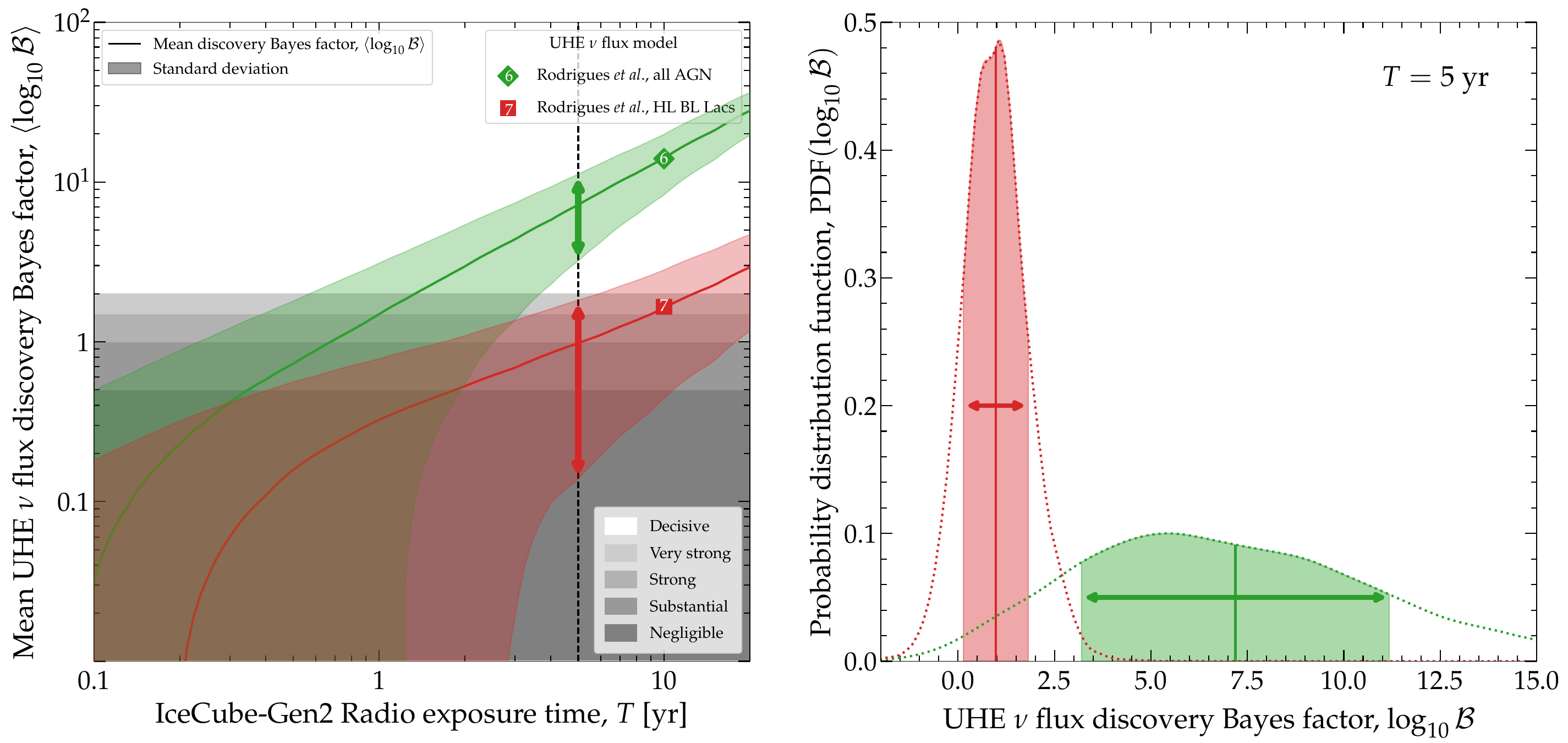}
 \caption{\label{fig:statistical_spread}Mean and standard deviation of the discovery Bayes factor of benchmark UHE neutrino flux models 6 and 7~\cite{Rodrigues:2020pli} (see \figu{benchmark_spectra}).  The mean is computed using \equ{bayes_factor_avg}; the standard deviation is built from many random realizations of the observed event rate in the radio array of IceCube-Gen2.  {\it Left:} Evolution of the mean and standard deviation with detector exposure time.  {\it Right:} Probability distribution functions of the Bayes factor after $T = 5$~years of detector exposure. See Appendix~\ref{sec:stat_sig_bayes_factor} and Section~\ref{sec:discovery_stat_analysis} for details.}
\end{figure*}

In the main text and other appendices, we reported our results in terms of the mean Bayes factor, \equ{bayes_factor_avg}, averaged over $N_{\rm samples} = 10^4$ random realizations of the observed event rate, following the prescription in Section~\ref{sec:discovery_stat_analysis}.  Here, we examine the distribution of values of the discovery Bayes factor obtained in those random realizations, in order to assess how representative the mean value is of the underlying distribution. 

Figure~\ref{fig:statistical_spread} shows the mean value and standard deviation of the discovery Bayes factor for two representative UHE neutrino flux models, models 6 and 7~\cite{Rodrigues:2020pli}.  For both models, the standard deviation is broad, especially at low exposure times, where event rates are low and more severely affected by random Poissonian fluctuations; see Section~\ref{sec:discovery_stat_analysis}. 

Figure~\ref{fig:statistical_spread}, left, shows that within the standard deviation of the Bayes factor the detector exposure time needed for flux discovery can be significantly longer or shorter compared to the man, by up to years.  In the case of flux model 7, decisive discovery may be unfeasible within 20~years if the observed Bayes factor lies close to the bottom of its standard deviation.  UHE neutrino flux models with larger event rates have a smaller spread of their discovery Bayes factor, and so their discovery is more robust to statistical fluctuations, {\it viz.}~flux models 6 {\it vs.}~7 in \figu{statistical_spread}, left.

Figure~\ref{fig:statistical_spread}, right, shows the corresponding probability distribution functions of the Bayes factor after 5~years of exposure.  The distributions span several orders of magnitude in $\mathcal{B}$.  This justifies our averaging procedure, introduced in Section~\ref{sec:discovery_results_nominal_choices}, of reporting our results via the arithmetic mean of $\log_{10} \mathcal{B}$ or, equivalently, the geometric mean of $\mathcal{B}$, to avoid biasing the mean Bayes factor towards large, unrepresentative values. At longer exposure times, where event rates are higher, the probability distribution functions become more symmetric and narrower around the mean, mitigating our original need to avoid the above bias.


\pagebreak
\twocolumngrid
\bibliography{refs.bib}


\end{document}